\documentclass[useAMS,usenatbib]{mnras}
\usepackage{amsmath}
\usepackage{amssymb}
\usepackage{color}
\usepackage{graphicx}
\usepackage{hyperref}
\usepackage[compatibility=false]{caption}
\usepackage{subfig}
\usepackage{lscape}
\usepackage{longtable}

\newcommand{\hii}{H\textsc{ii}}
 
\newcommand{\mewm}{$\mu$m} 
\newcommand{\kms}{kms$^{-1}$}
\newcommand{\asec}{$^{\prime\prime}$}
\newcommand{\exOH}{ex-OH} 
\newcommand{\meth}{CH$_3$OH}

\title[Excited-state hydroxyl maser catalogue]{Excited-state hydroxyl maser catalogue from the methanol multibeam survey -- I. Positions and Variability}
\author[A.Avison et al.]{A.~Avison,$^{1,2}$ \thanks{Email: adam.avison@manchester.ac.uk}, L.J. Quinn,$^{2,3}$, G.A.~Fuller,$^{1,2}$, J.L.~Caswell$^{4}$\thanks{Deceased 2015 January 14.}, J.A.~Green,$^{4,5}$, S.L.~Breen$^4$, 
\newauthor
S.P.~Ellingsen$^6$, M.D.~Gray,$^2$, M.~Pestalozzi,$^7$, M. A.~Thompson$^8$ and M. A.~Voronkov$^4$\\
 $^1$UK ALMA Regional Centre Node\\ $^2$Jodrell Bank Centre for Astrophysics, Alan Turing Building, School of Physics and Astronomy, The University of Manchester,\\ Manchester, M13 9PL, UK\\  $^3$Bury College, Millennium Centre, Market Street, Bury BL9 0DB, UK \\ $^4$CSIRO Astronomy and Space Science, Australia Telescope National Facility, PO Box 76, Epping, NSW 2121, Australia\\ $^5$SKA Organization, Jodrell Bank Observatory, Lower Withington, Macclesfield, Cheshire SK11 9DL, UK\\ $^6$School of Physical Sciences, University of Tasmania, Private Bag 37, Hobart, TAS 7001, Australia\\ $^7$ Istituto di Astrofisica e Planetologia Spaziali IAPS - INAF, via del Fosso del cavaliere 100, I-00133 Roma, Italy\\ $^8$Centre for Astrophysics Research, Science and Technology Research Institute, University of Hertfordshire, College Lane, Hatfield AL10 9AB}
\date{Released 2016 Xxxxx XX}

\pagerange{\pageref{firstpage}--\pageref{lastpage}} 
\pubyear{XXXX}

\def\LaTeX{L\kern-.36em\raise.3ex\hbox{a}\kern-.15em T\kern-.1667em\lower.7ex\hbox{E}\kern-.125emX}

\DeclareRobustCommand{\VAN}[2]{#2}
\begin{document}

\label{firstpage}

\maketitle

\begin{abstract}\\
We present the results of the first complete unbaised survey of the Galactic Plane for 6035-MHz excited-state hydroxyl masers undertaken as part of the Methanol Multibeam Survey. These observations cover the Galactic longitude ranges $186^{\circ}< l < 60^{\circ}$ including the Galactic Centre. We report the detection of 127 excited-state hydroxyl masers within the survey region, 47 being new sources. The positions of new detections were determined from interferometric observations with the Australia Telescope Compact Array. We discuss the association of 6035-MHz masers in our survey with the 6668-MHz masers from the MMB Survey, finding 37 likely \meth$-$\exOH\ maser pairs with physical separations of $\leq$ 0.03pc and 55 pairings separated by $\leq$ 0.1pc. Using these we calculate for the first time an excited-state hydroxyl maser life time of between 3.3$\times 10^3$ and 8.3$\times 10^3$ years. We also discuss the variability of the 6035-MHz masers and detection rates of counterpart 6030-MHz excited-state hydroxyl masers (28\% of our sample having detection at both frequencies).

\end{abstract}

\section{Introduction}
\label{intro:sec}
The Methanol Multibeam (MMB) survey is an unbiased Galactic plane survey searching for the 6668-MHz methanol maser line \citep{GreenTech}. The MMB survey region covers the whole southern portion of the Galactic plane (longitudes $186^{\circ} < l < 60^{\circ}$, latitude $|b|\leq2^{\circ}$).

The 6668-MHz methanol maser transition is uniquely associated with high-mass star formation \citep{Minier03, Xu08, Breen13} making it a clear signpost of regions of high-mass star formation and their distribution throughout the Milky Way. Ground-state and excited-state hydroxyl (hereafter ex-OH) maser transitions are often associated with the 6668-MHz methanol maser \citep[e.g.][]{Caswell97,Caswell98}. First observed in the late 1960s and early 1970s \citep[e.g][]{Yen69,Rydbeck70} the $^2\Pi_{3/2}$ $J=5/2, F=3-3$ \exOH\ maser transition at 6035-MHz and the nearby, often weaker, 6030-MHz maser line ($^2\Pi_{3/2}$ $J=5/2, F=2-2$) have been extensively studied with multiple targeted surveys, usually toward other maser species (e.g. 6668-MHz \meth\ or ground-state OH, see \citealt[][]{Knowles75,CaswellVaile95,Baudry97,Caswell97,Caswell01}). The \exOH\ maser transitions are primarily observed toward star forming regions, with only two known examples observed toward evolved stars  \citep[Vy 2-2 and K 3-35, see][]{Desmurs10} making \exOH\ masers in evolved stars extremely rare and a potentially transient phenomenon \citep{Richards12}. These sources are not present in this current work as both exist at Galactic latitudes outside of the MMB survey range. 

When observed in association with the 6668-MHz \meth\ maser, the rarer \exOH\ maser gives important additional information on the environment of the star forming region it inhabits. This maser transition appears to exist in pre-ionising high-mass protostellar objects as well as sources with ultra compact \hii\ regions \citep[and references therein]{Fish07}. 

\begin{figure}
\begin{center}
\includegraphics[scale=0.3]{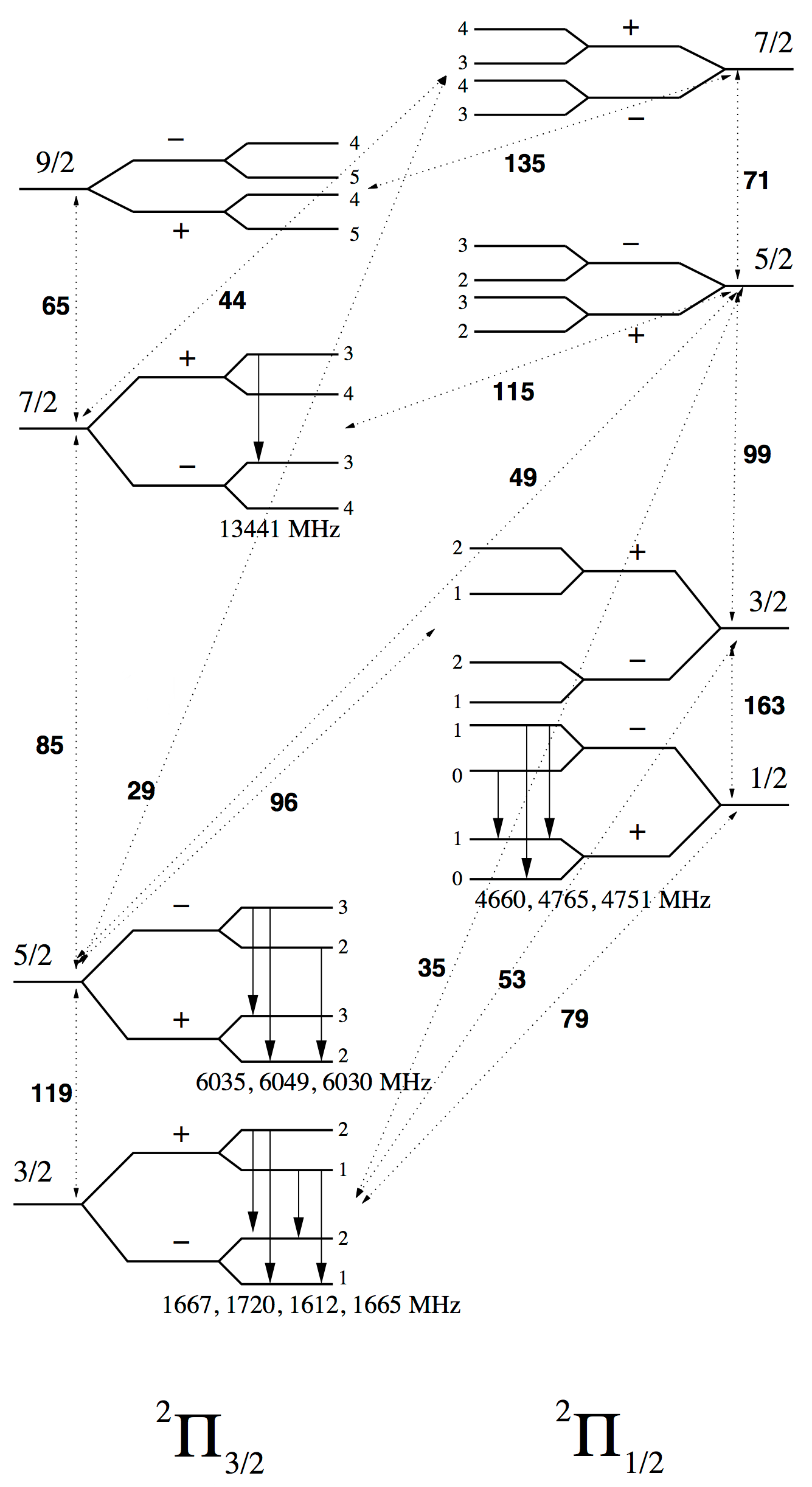}
\caption{Rotational energy level diagram of OH up to J= 9/2. The level are split in to the left-hand $^2\Pi_{3/2}$ and right-hand $^2\Pi_{1/2}$ ladders by spin-orbit coupling.  Each rotational level is split first into halves of opposite parity and then again by hyperfine splitting. The frequencies in MHz of emission are shown for various transitions below solid arrows for each J level. The values in bold next to dotted arrows give approximate wavelengths in microns. Figure and caption adapted from \citet{GrayBook}.}
\label{OHtrans:fig}
\end{center}
\end{figure}

Ex-OH masers are predominantly radiatively pumped requiring a nearby source of infrared emission at wavelengths of the order $\sim$30-120\mewm\ to generate the population inversion, \citep[e.g][]{Cesaroni91,Gray92,Gray01}. Figure \ref{OHtrans:fig} shows rotational energy level diagram for both ground state and \exOH\ transitions. Models of maser pumping by \citet{Cragg02} find that both the 6030 and 6035-MHz transitions occur in zones of low gas temperatures ($T_k < 70$K) and high densities (up to $n_H = 10^{8.5}$cm$^{-3}$), covering a parameter space coincident with, but not identical to, both the 6668-MHz \meth\ and ground-state OH (1612, 1665, 1667, 1720-MHz) masers \citep[see also][]{Gray92}. As such, coincident detection of multiple maser species can place constraints on the environment of a high-mass star forming region \citep[e.g.][for methanol]{Cragg01,Sutton01}. The \exOH\ maser also provides important information about the magnetic fields in its local environment as OH has a relatively large Lande g-factor resulting in a significant Zeeman splitting of the maser lines \citep[][and references therein]{Fish07}.

In this paper we present the results from a survey for the 6035-MHz line of \exOH\ carried out in parallel with the MMB survey. We present high signal-to-noise spectra at both 6035-MHz and 6030-MHz (the latter from targeted follow-up observations) for all sources and accurate interferometric positions for the newly detected masers, as well as discussing \exOH\ maser source variability and association with 6668-MHz \meth\ maser counterparts. In a companion paper we will explore the results of the survey in relation to the magnetic fields probed by the masers (Avison et al. \textit{in prep}).

\section{Methanol Multibeam Survey Parameters and Equipment}

The full details of the techniques used by the MMB survey are presented in \citet{GreenTech}, therefore here we present only a summary of those points relevant to the \exOH\ observations. The MMB survey was conducted as a blind survey using a custom built seven-beam receiver on the Parkes Radio Telescope. Using the receiver's 1-GHz bandwidth both the 6668-MHz \meth\ line and the 6035-MHz \exOH\ lines were surveyed simultaneously, to a typical RMS noise level of 0.17Jy \citep{GreenTech}. The Parkes data included both Left-hand (LHCP) and Right-hand (RHCP) circular polarisation data. 

Any new \exOH\ detections or sources without previous high-resolution positions were then followed up with the Australia Telescope Compact Array (ATCA) to acquire a position (to within $\sim$ 0.4\asec). Finally, to obtain high signal-to-noise spectra of these sources, they were observed again at Parkes using the ATCA position with an observing setup termed `MX' \citep[see][]{GreenTech}. `MX' observations at 6030-MHz were also obtained at positions of 6035-MHz emission to detect potential counterpart \exOH\ masers at this frequency. The typical RMS noises in the MX observations at 6035-MHz and 6030-MHz are 0.10Jy and 0.11Jy respectively.

\subsection{Ex-OH ATCA observations}
The positioning observations of forty-seven new \exOH\ sources were made with ATCA in January 2009 over three days.  Observations were conducted with the ATCA in the 6C configuration (maximum baseline 6.0km). The observations were made centred on the \exOH\ 6035-MHz transition, with 2048 channels over a bandwidth of 4-MHz. The data were flux density and bandpass calibrated using the ATCA standard sources PKS B1934-638 and PKS B1921-293 respectively. Phase calibration was interleaved between on source observations of three to four maser sources which were close in position and velocity, using a calibration source local (typical offset $<10^{\circ}$) to that area of sky.

Further ATCA observations were taken in October 2013 and February 2014 to complete the source positioning for new sources. These observing runs were each taken in blocks over one (Oct. 2013) and two days (Feb. 2014) in the H214 and 6D configurations (maximum baselines 274m and 6.0km respectively). They focused on the more northerly sources from the MMB survey. The CABB correlator system \citep{CABBpaper} had replaced the correlator system used in the 2009 observations, the new observing setup providing 10,240 channels over a 5.0MHz bandwidth (giving 0.488 kHz channels). These observations provided full Stokes polarisation (XX, YY, XY, YX) and also included observations of the 6030-MHz line. For the Oct 2013 observations in the H214 array configuration antenna 6 (at 6.0km) was excluded from the data reduction meaning these data were taken with five antennas leading to higher RMS values. The positional accuracy of all ATCA observations for the MMB survey is approximately 0.4\asec\ \citep{MMB345to006}. 

For all the ATCA observing epochs data reduction was conducted using the \textsc{MIRIAD} software package following  standard ATNF reduction strategies for either pre- or post-CABB data.

\section{Survey Results}
\label{SurRes:ref}
A total of 127 6035-MHz and 32 6030-MHz \exOH\ masers are found within the MMB survey region, Galactic longitudes $186^{\circ} < l < 60^{\circ}$ and latitudes $|b|\leq2^{\circ}$. We present in Table \ref{Results:tab} the properties of the \exOH\ masers found within this survey.  Column 1 gives the source name defined as the source position in Galactic coordinates (rounded at the third decimal point) prefixed with MMBOH-G, columns 2 and 3 the J2000 Right Ascension and Declination co-ordinates. Column 4 the maser transition frequency, column 5 the literature or observational position reference for the RA and Dec. Columns 6, 7 and 8 give the maser peak flux density, peak velocity and a guideline velocity range from the Parkes MX of the LHCP maser emission, with columns 9, 10 and 11 the same for RHCP. The exceptions being sources \textit{49.046$-$0.290} and \textit{326.447$-$0.749} for which the Stokes I values from the ATCA data are presented in columns 6, 7 and 8 in italics.

Figure \ref{abcd} shows the Parkes MX spectra for each \exOH\ maser with the exception of sources \textit{49.046$-$0.290} and \textit{326.447$-$0.749} for which the Stokes I spectra from ATCA is used as these sources are not present in the nearest Parkes MX. The full version of this figure is available in the online version only.

Finally, there remains one \exOH\ maser, \textit{189.030+0.783} clearly present in the Parkes MX data but for which we have been unable to attain an ATCA position. This source is at high positive declination where the beam of the ATCA becomes elongated and positioning becomes difficult. This source is included in Table \ref{Results:tab}, and Figures \ref{GalacticDist:fig} and \ref{abcd} but is not included in any analysis beyond $\S$\ref{PaperIV:sec}.

\subsection{Remarks on sites of ex-OH maser emission}
\label{NotesObj:sec}
Here we provide additional information to that given in Table \ref{Results:tab} on each of the 47 newly detected \exOH\ sources, or those which have changed significantly since their last published observation. This additional information primarily focuses on association with 6030-MHz emission from the MMB MX observations and any nearby MMB \meth\ source. We also note if there are previous detections at 1665-MHz OH masers and then include any further comments. It should be noted that many of the new 6035-MHz detections are low flux density sources and as such there are only a small number of corresponding 6030-MHz detections (see \S \ref{sixohthreeoh:sec} for these sources). 

For new detections we systematically comment on the nearest detections in the Bolocam Galactic Plane Survey (BGPS;  \citealt{BGPSII, Aguirre11}) and the APEX Telecsope Large Area Survey of the GALaxy (ATLASGAL, with sources prefixed AGAL; \citealt{Schuller09,Contreras13,Urquhart14}) data. Both surveys trace dense clumps and cores in the mm/sub-mm indicative of regions of star formation which may be associated with the \exOH\ in our sample. The BGPS covers a region of the Galactic plane from $349.5^{\circ}\leq l \leq90.5^{\circ}$, $|b|\leq0.5^{\circ}$; with the latitude range extended to  $|b|\leq1.5^{\circ}$ in some regions see \citealt{Aguirre11} at 1.1mm with an effective resolution of 33\asec\, with ATLASGAL covering a region of $280^{\circ} \leq l \leq60^{\circ}$, $-2^{\circ}<b<1^{\circ}$ ( see \citealt[see][]{Urquhart14}) at 870\mewm\  with a beam full width half-maximum of 19.2\asec. We extend our search for nearby BGPS or ATLASGAL sources out to 30\asec, approximately the resolution of BGPS \citep{Schuller09} and the median clumps size reported by \citet{Urquhart14} for ATLASGAL. Additionally, we include the name of the nearest infrared to radio wavelength continuum source or star formation tracer to our \exOH\ detection up to an angular separation of 5.0\asec\ equivalent to 0.1pc, the typical size of a star forming core at the fiducial distance to a high-mass star forming region of 5kpc. 

For brevity some references to previous OH (both excited and ground-state) detection are abbreviated, these are as follows: FC89 is \citet{ForesterCaswell89} (1665-MHz, reported RMS noise $\sim$0.1Jy), CV95 is \citet{CaswellVaile95} (6035-MHz, typical detection limit $\sim$0.3Jy), C98 is \citet{Caswell98} (1665/7-MHz, detection $\sim$0.16Jy), A00 is \citet{Argon00} (1665/7-MHz, detection limit 0.9-2.7Jy) and C03 is \citet{Caswell03} (6035/0-MHz, reported RMS noise $\sim$0.03-0.05Jy).  Finally `ND' stands for new detection, to differentiate them from known sources with additional comments.\\

\begin{figure}
\begin{center}
\includegraphics[scale=0.33, trim= 00 00 00 00]{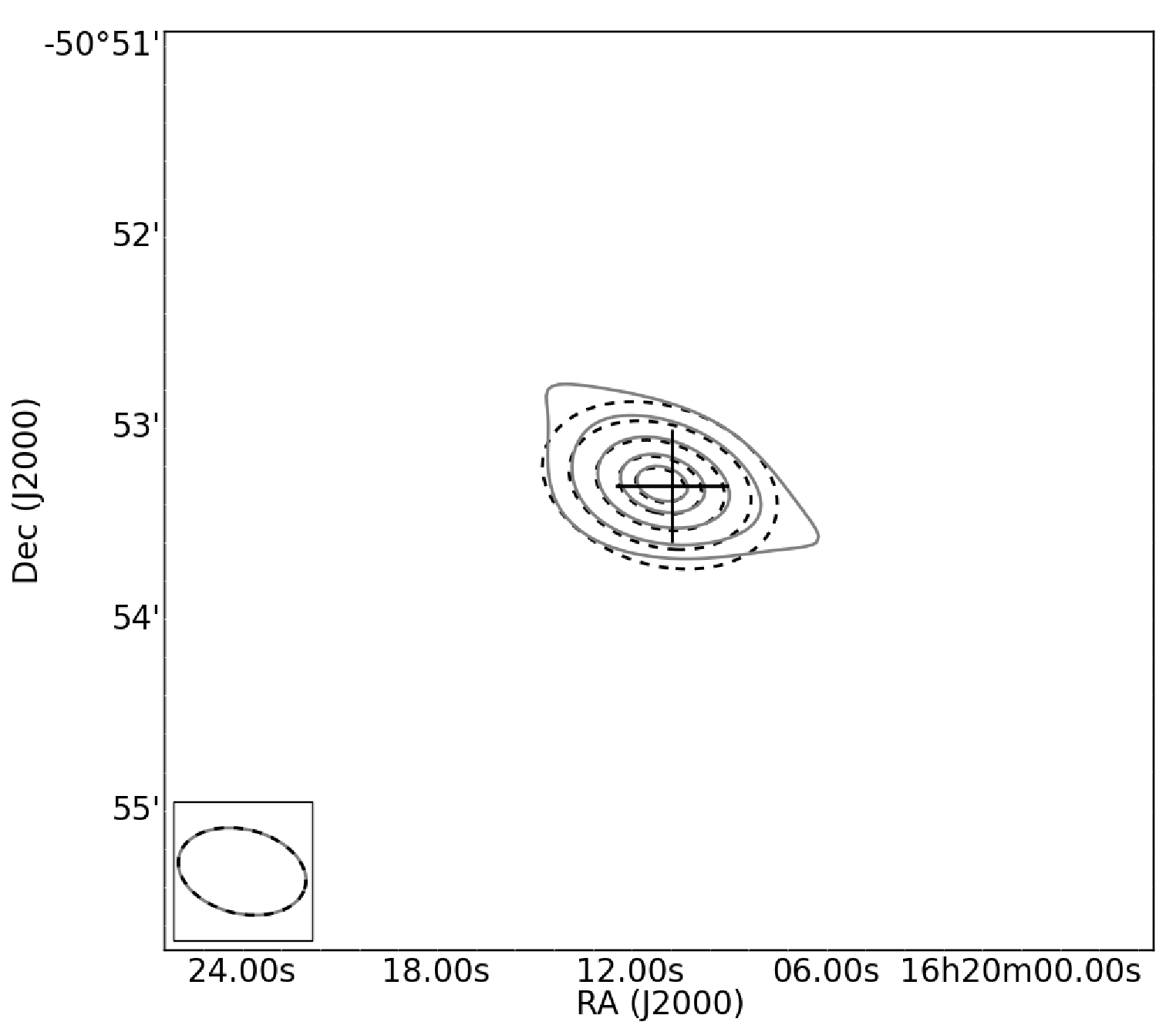}
\caption{Continuum emission detected around source G332.824$-$0.548 at 6035-MHz (black dashed contour), and 6030-MHz (grey solid contour). Contours denote 10, 25, 50, 75 and 90\% of the peak emission at each frequency (see text). The `+' indicates the 6035-MHz \exOH\ maser as observed in the ATCA-2013 data. The synthesised beam of each observation is given in the box at the lower left of the image.}
\label{332p824cont:fig}
\end{center}
\end{figure}

\noindent \textit{4.682+0.278:} ND, position from ATCA-2009 data. No 1665-MHz from C98. No 6030-MHz detection. BGPS G004.681+00.277 mm source within 3.9\asec\ \citep{BGPSII} and AGAL 004.681+00.277 sub-mm source within 4.16\asec. The nearest MMB \meth\ maser, G04.676+0.276, is offset by an angular distance of 21.3\asec.
\\
\\
\textit{6.882+0.094:} ND, position from ATCA-2009 data. 6030-MHz detection in MMB MX both hands of polarisation. No counterpart 1665-MHz in C98, and a MMB \meth\ maser, G06.881+0.093,nearby at 4.8\asec\ separation. No other radio/mm source within 5\asec. (The source BGPS G006.885+00.091 source is at a distance of 6\asec\ and no ATLASGAL source is reported nearby).
\\
\\
\textit{8.352+0.478:} ND, Positions from ATCA-2013 data. No known 1665-MHz or 6030-MHz counterparts. No radio/mm source within 5\asec. This source is one of the `isolated' \exOH\ sources discussed in $\S$\ref{isolated:sec}.  No mm/sub-mm sources are detected nearby by ATLASGAL or BGPS.
\\
\\
\textit{9.620+0.194 and 9.622+0.196:} NDs, Positions from ATCA-2009 data, the pair are separated by 12.0\asec. Each target has a 1665-MHz maser detected in C98, with a single detection at this site at 1665-MHz from FC89. No 6030-MHz detection from either source. These sources are small angular distances from the 6668-MHz maser pair \textit{G09.619+0.193 and G09.621+0.196} (offsets 5.23\asec\ and  4.75\asec) respectively. The 6668-MHz maser \textit{G09.621+0.1$-$96} is the brightest methanol maser detected in the MMB survey with a peak flux density of 5239.85Jy \citep{MMB006to020}, whereas the 6035-MHz peak flux density from the ATCA-2009 data is 0.2Jy and from the MMB MX data 0.24/0.25Jy (LHCP/RHCP, respectively). The methanol maser is also known to periodically vary on timescales of 244 days \citep{vanderWalt09}. \textit{9.620+0.194} is coincident with BGPS G009.620+00.194 (within our positional uncertainties) and is likely associated with AGAL 009.621+00.194 (offset $\sim$ 4.1\asec), whilst \textit{9.622+0.196} sits at greater separation from both these sources at 12.2\asec and 9.6\asec respectively.
\\
\\
\textit{10.322$-$0.258:} ND, Position from ATCA-2009 data. Nearby is the radio source GPSR5 10.322$-$0.259 source (2.63\asec) \citep{GPRS5} and an MMB \meth\ maser, G10.320$-$0.259, at an offset of 4.70\asec. At greater separation lie AGAL 010.321$-$00.257 and  BGPS G010.320-00.258 at 4.5\asec\ and 7.0\asec\ respectively.
\\
\\
\textit{10.960+0.022:} ND, Positions from ATCA-2013 data. No 6030-MHz detection in the MMB MX data nor a 1665-MHz detection in FC89, C98 or A00. The MMB \meth\ maser source G10.958+0.022, is offset from this position by 7.1\asec. Nearby mm/sub-mm cores are BGPS G010.959+00.020 (offset at 7.0\asec) and AGAL 010.957+00.022 (offset at 11.6\asec).
\\
\\
\textit{12.681$-$0.182:} 1665-MHz maser seen in FC89, C98 and A00. ND from ATCA-2009. No 6030-MHz source detected in MX data. Within the W33 star-forming region. At an offset of 2.9\asec\ is the \meth\ maser G12.681$-$0.182. BGPS G012.681$-$00.182 is offset at angular separation of 0.5\asec\ and AGAL 012.679$-$0.181 is offset by $\sim$9.2\asec.
\\
\\
\textit{18.460$-$0.005:} ND, Position from ATCA-2009. Outside of C98 range, no detections of 1665-MHz in FC89. No 6030-MHz source detected in MX data. Offset from Compact / UC \hii\ region by 3.89\asec\ from e.g. \citet{Walsh98} and the MMB \meth\ maser, G18.460$-$0.004, by 1.6\asec. Nearby are the dense mm/sub-mm cores of BGPS G018.462-00.002 at 12.7\asec\ and AGAL 018.461-00.002 at 9.8\asec.
\\
\\
\textit{18.836$-$0.299:} ND, position from ATCA-2009 data. Outside of C98 range, no detections of 1665-MHz in FC89. No 6030-MHz source detected in MMB MX data. The nearest MMB \meth\ maser is G18.834$-$0.300 at an offset of 7.2\asec. There are mm/sub-mm detections made by both BGPS and ATLASGAL with sources AGAL 018.833-00.301 and BGPS G018.834-00.299 offset from the maser peak by 10.6\asec\ and 5.1\asec\ respectively.
\\
\\
\textit{19.752$-$0.191:} ND, positioning from the MMB `piggyback' data (see \citealt{GreenTech} and Ellingsen \textit{et al. in prep.}). 6030-MHz data unavailable for this source. This source is an `isolated' \exOH\ sources (discussed in $\S$\ref{isolated:sec}) with no MMB \meth\ detection within $>$200\asec\ and the nearest BGPS and ATLASGAL detections over an arcminute away.
\\
\\
\textit{24.147$-$0.010:} ND, position from ATCA-2009 data. No detection FC89 at 1665-MHz. Non detection of 6030-MHz in the MMB MX data. MMB \meth\ maser source G24.148$-$0.009 is offset for this source position by 3.5\asec. Nearby sub-mm detection of AGAL 024.148$-$00.009 (offset 4.1\asec), whilst the nearest BGPS sources is G024.154-00.008 offset by 24.0\asec.
\\
\\
\textit{25.509$-$0.060:} ND, position from ATCA-2009 data. No detection FC89 at 1665-MHz nor a 6030-MHz source in MMB MX data. This source is an `isolated' \exOH\ sources (discussed in $\S$\ref{isolated:sec}) with no MMB \meth\ detection within $>$400\asec. No radio/mm sources within 5\asec. There are no BGPS or ATLASGAL sources within  2 arcminutes.
\\
\\
\textit{25.648+1.050:} ND, position from ATCA-2013. No 1665- or 6030-MHz counterpart. No mm- or radio source within 5\asec. The MMB \meth\ maser G25.650+1.049 is offset by 5.8\asec. AGAL 025.649+01.051 is the nearest sub-mm detection at an angular offset of 3.9\asec, there is no nearby BGPS source.
\\
\\
\textit{28.819+0.366:} ND, position from ATCA-2014 data. Not observed at 6030-MHz as part of the Parkes MMB survey, nor was a 6030-MHz maser seen in the ATCA-2014 data. No radio/mm sources within 5\asec, with the nearest ATLASGAL and BGPS sources  028.816+00.366 and G028.817+00.363 at angular offsets of 9.9 and 9.7\asec\ respectively.
\\
\\
\textit{30.778$-$0.801:} ND, position from ATCA-2009. No radio/mm sources within 5\asec\ and no nearby ATLASGAL detection. The BGPS sources G030.772-00.801 and G030.782-00.795 are at angular offsets of 23.2\asec\ and 24.0\asec\ respectively. The nearest MMB \meth\ maser G30.771-0.804 is at an angular separation of 29.0\asec\ making it very unlikely these are a true pair excited by the same source.
\\
\\
\textit{34.261$-$0.213:} ND, position from ATCA-2009. No 1665-MHz detection from previous studies or 6030-MHz source in MMB MX data. No nearby radio or mm- continuum source reported within 5\asec. At a separation of 25.1\asec\ the nearest MMB \meth\ maser G34.267$-$0.210 is unlikely associated with this \exOH\ source. The nearest mm/sub-mm sources are at similarly large separations and unlikely associated with the \exOH\ maser, these sources being BGPS G034.264$-$00.210 (separation of 15.6\asec) and  AGAL 034.266$-$00.209 (separation of 23.5\asec).
\\
\\
\textit{34.258+0.153:} Known source. Here the 55\kms\ feature in left hand circularly polarised (LHCP) now dominates the source whereas in C03 the feature at 62.1\kms\ (LHCP) was dominant. Source detected at 6030-MHz. For this maser  the nearest mm/sub-mm detections from BGPS and ATLASGAL are BGPS G034.258+00.154 and AGAL 034.258+00.154 with angular offsets between their reported positions and the \exOH\ maser peak of 1.4\asec\ and 2.5\asec\ respectively.
\\
\\
\textit{35.133$-$0.744:} ND, position from ATCA-2013. No source seen in 6030-MHz MMB MX data. MMB \meth\ maser G35.132$-$0.744 is located at a 2.0\asec\ separation from this \exOH\ maser. Whilst there is no nearby BGPS source,  ATLASGAL source AGAL 035.132$-$00.744 is at an angular separation of 3.3\asec.
\\
\\
\textit{35.198$-$0.743:} Known source at 1665-MHz (FC89) and 6035-MHz (C03) seen in absorption in the MMB 6030-MHz data. Over three epochs the 6035-MHz maser source has been increasing (by a factor 3.99 and 2.44 between its reported flux density in CV95 and C03 and the MMB result respectively, see Table \ref{VariabilityTab:tab}). ATLASGAL detects a source (AGAL 035.197$-$00.742) offset from the maser peak position by 4.0\asec. There is no nearby BGPS detection.
\\
\\
\textit{35.200$-$1.736:} Known source at both 1665-MHz and 6035-MHz (FC89 and C03 respectively), see in absorption from MMB MX data at 6030-MHz. This maser source is seen to decrease in flux density between the two previous observations as reported by CV95 and C03 (by factors of -6.60 and -8.37 respectively). Interestingly in this time the polarization which displays peak emission at each epoch has switched handedness (Table \ref{VariabilityTab:tab}). No ATLASGAL or BGPS  detections within an arcminute.
\\
\\
\textit{40.282$-$0.220:} ND, position and 6030-MHz detection from ATCA-2014. Not observed at 6030-MHz as part of the Parkes MMB survey.  HC\hii\ region at offset of 1.21\asec\ observed with the EVLA by \citet{SanchezMonge11} and the MMB \meth\ maser G40.425+0.700 is observed at an offset of 3.0\asec. Sub-mm detection AGAL 040.283$-$00.219 has angular offset from the maser of 5.4\asec, there is no BGPS source detected within and arcminute.
\\
\\
\textit{48.988$-$0.300:} ND, position from ATCA-2009. Source BGPS G48.989$-$0.299 (mm source) offset by 4.47\asec\ and AGAL 048.991$-$00.299 (sub-mm) offset by 11.6\asec. The MMB \meth\ maser G48.990-0.299 by 8.9\asec. No 6030-MHz detection. 
\\
\\
\textit{49.046$-$0.290:} ND, position from ATCA-2009. No 6030-MHz in MMB MX data and no detection at 1665-MHz in FC89, C98 or A00. No nearby radio or mm- continuum source reported within 5\asec. This source is one of the `isolated' \exOH\ sources with no nearby \meth\ detection discussed in $\S$\ref{isolated:sec}, the nearest mm/sub-mm sources from the ATLASGAL and BGPS surveys are offset by 59.0\asec\ and 56.6\asec\ respectively, meaning there is likely no association between these detection and the \exOH\ maser.
\\
\\
\textit{49.490$-$0.388:} Known source, seen at 6035-MHz in C03 and 1665-MHz in FC89. Seen in absorption in the MMB 6030-MHz MX data. Nearby mm/sub-mm sources from ATLASGAL and BGPS are 049.489$-$00.389 (offset by 6.3\asec) and G049.489$-$00.386 (offset by 7.7\asec) respectively.
\\
\\
\textit{50.478+0.705:} ND, ATCA-2009 position. Associated with IRAS 19194+1548 (IR 7.84 \asec). Source outside of range of previous studies at 1665-MHz. No nearby radio or mm- continuum source reported within 5\asec\ and no ATLASGAL or BGPS detection within an arcminute. An `isolated' \exOH\ sources with the nearest \meth\ maser over 595\asec\ away.
\\
\\
\textit{51.683+0.714:} ND, ATCA-2014 position, this maser was not observed as part of the MMB MX at 6030-MHz and no maser was observed at 6030-MHz with our ATCA observations. Source outside of range of previous studies at 1665-MHz. No nearby radio or mm- continuum source reported within 5\asec, no BGPS detection within an arcminute however the nearest ATLASGAL source AGAL 051.678+00.719 has an angular separation from the \exOH\ maser peak of 22.5\asec. The nearest MMB \meth\ maser G51.679+0.719 is offset by 21.9\asec.
\\
\\
\textit{284.016$-$0.856} ND, position from ATCA-2009. No 1665-MHz counterpart in C98 or FC89, no 6030-MHz counterpart in MMB MX spectra. There is no BGPS within an arcminute and the ATLASGAL source at the nearest angular separation is AGAL 284.016$-$00.857, 4.8\asec\ away. This source is an `isolated' \exOH\ source (discussed in $\S$\ref{isolated:sec}) with no \meth\ maser detected nearby by the MMB.
\\
\\
\textit{298.723$-$0.086} ND, ATCA-2009 position with maser covering very small velocity range, no 1665-MHz or 6030-MHz counterparts. Nearby is MMB \meth\ maser G298.723$-$0.086 at an offset of 3.05\asec. ATLASGAL source AGAL 298.724$-$00.086 is offset from the \exOH\ maser peak by 3.6\asec, no other nearby radio or mm- continuum source reported within 5\asec.
\\
\\
\textit{305.208+0.206} ND, ATCA-2013 position, source previously detected at 1665-MHz (C98) where the source was significantly stronger (15.9Jy), than either hand of 6035-MHz detection (0.98(LHCP) and 1.25(RHCP) Jy). No 6030-MHz source in MMB MX data. SIMBA detection at 1.2-mm, G305.21+0.21, from \citet{Hill05} with the 1.2mm peak offset by 4.42\asec from the maser peak position and an MMB \meth\ maser, G305.208+0.206, separated by 0.57\asec. AGAL 305.209+00.206 is offset by 2.0\asec. There is no BGPS detected within an arcminute.
\\
\\
\textit{305.362+0.150} ND, ATCA-2009 position. Counterpart at 1665-MHz reported by C98, significantly brighter 10.6 (7.6)Jy than the 6035-MHz detection (0.41(LHCP) and 0.36(RHCP) Jy). No 6030-MHz seen in the MMB MX data. Candidate ultra-compact \hii\ region, \textit{G305.362+00.150}, in the G305 star-forming complex reported by \citet{Hindson12} at a reported position offset from the \exOH\ maser of $\sim$1.4\asec.  ATLASGAL detects source AGAL 305.362+00.151 at a similar separation of 2.0\asec. There also exists the MMB \meth\ maser G305.362+0.150 at an angular offset of 0.78\asec. 
\\
\\
\textit{308.056$-$0.396} Brightest newly detected source at 6035-MHz with a flux density of 15.93Jy (LHCP) and 25.46Jy (RHCP); interestingly no 6030-MHz counterpart seen in MMB MX data. No 1665-MHz detections C98. MMB \meth\ maser G308.056$-$0.396 is offset for this \exOH\ source by 1.2\asec. Sub-mm source AGAL 308.057-00.397 is at an angular separation of 4.3\asec.
\\
\\
\textit{308.651$-$0.507} ND, ATCA-2009 position, with no detected counterpart at 6030-MHz in MMB MX. There is no 1665-MHz detections in C98. MMB \meth\ maser G308.651$-$0.507 is offset by 1.5\asec\ and sub-mm core AGAL 308.652$-$00.507 at an offset of 2.1\asec.
\\
\\
\textit{309.384$-$0.135} ND at 6035-MHz from ATCA-2009. Weak detections at 1665-MHz (C98) of 0.23Jy. No 6030-MHz seen in MMB MX data. \meth\ maser G309.384$-$0.135 from the MMB data is offset by 1.5\asec\ and Extended Green Object (EGO) G309.38-0.13(a) present within 2.05\asec\ \citep{Cyganowski08}.  At a similar angular offset (2.7\asec) is sub-mm AGAL 309.384-00.134.
\\
\\
\textit{309.901+0.231} ND, position from ATCA-2009. No 6030-MHz detection in MMB MX data and no 1665-MHz detection in C98. The MMB detected the nearby \meth\ maser G309.901+0.231 which is offset from this \exOH\ maser by 1.34\asec, this is a similar separation to AGAL 309.901+00.231 (offset 1.3\asec), the nearest sub-mm detection. The EGO G309.90+0.23 is within 4.4\asec\ \citep{Cyganowski08}.
\\
\\
\textit{312.598+0.045} ND, position from ATCA-2009. Known source at 1665-MHz (C98). No 6030-MHz detection in MMB MX data. Offset from sub-mm source G312.598+00.044 in the ATLASGAL survey by 4.46\asec\ \citep{Contreras13}.
\\
\\
\textit{320.427+0.103} ND with position from ATCA-2009. No 1665-MHz in FC98 or  C98, no 6030-MHz detection in MMB MX data. Sub-mm source AGAL 320.427+00.102 is at an angular offset of 3.6\asec. The nearest MMB \meth\ maser,  G320.424+0.089, is at an angular offset of 50.5\asec\ so is highly unlikely to be associated with the \exOH\ detection.
\\
\\
\textit{326.447$-$0.749} and \textit{326.448$-$0.749} Newly detected maser pair separated by 3.4\asec\ with positions from ATCA-2009. Single 6030-MHz counterpart with similar velocity profile to \textit{326.448$-$0.749}. Both sources within $\leq3.5$\asec\ of Young Stellar Object Candidate from the RMS survey \citep{MottramHoare07}. The pair are at angular separations of 6.5\asec and 3.5\asec from sub-mm source AGAL 326.449-00.749. The nearest MMB \meth\ source to the pair is G326.448$-$0.748, offset by 5.5 and 2.5\asec from \textit{326.447$-$0.749} and \textit{326.448$-$0.749} respectively.
\\
\\
\textit{327.944$-$0.115} ND, position from ATCA-2009. Beyond FC89 and A00 range and no detection of a 1665-MHz maser in C98. No 6030-MHz source in MMB MX data. The nearest MMB \meth\ maser, G327.945$-$0.115, is at a separation of 3.7\asec\ with no other nearby radio or mm- continuum source reported within 5\asec, the nearest sub-mm source in ATLASGAL (AGAL 327.948-00.117)  is at an angular offset of 14.3\asec, meaning the maser and dense core are unlikely to be associated.
\\
\\
\textit{329.184$-$0.314} ND, position from ATCA-2009. No 6030-MHz MX data available for this source. 1665-MHz source from C98. \meth\ maser G329.183$-$0.314 from the MMB is offset by 2.8\asec. Sub-mm source AGAL 329.184-00.314 is offset from the maser peak by 1.8\asec.
\\
\\
\textit{332.824$-$0.548:} Known source. In both the 6035-MHz and 6030-MHz ATCA data there is a very strong continuum source (integrated flux density of 3.71$\pm$0.03Jy at 6035-MHz and 4.1$\pm$0.3Jy at 6030-MHz) at the same position as the 6035-MHz maser. There is no maser detected at 6030-MHz, instead this source appears in absorption. The nearest 6668-MHz \meth\ maser is offset from the \exOH\ detection by 7.0\asec. Figure \ref{332p824cont:fig} shows the continuum emission as contours along with the 6035-MHz maser position. This source is likely associated with the nearby \hii\ candidate G332.8256-00.5498 seen in the RMS survey data \citep{Mottram07}, which is offset by 9.4\asec\ and the sub-mm source AGAL 332.826-00.549 is at a separation of 4.2\asec. 
\\
\\
\textit{332.964$-$0.679:} ND with ATCA-2009 position. No 6030-MHz emission detected in the MMB MX. No 1665-MHz detection in C98. Separated by 2.1\asec\ from EGO G332.96-0.68 \citep{Cyganowski08}. The nearest MMB \meth\ maser is G332.963$-$0.679 which is offset by 2.4\asec. The sub-mm source AGAL 332.962-00.679 is at an angular offset of 6.9\asec.
\\
\\
\textit{333.068$-$0.447:} ND with ATCA-2009 position. No 6030-MHz or 1665-MHz detections from the MMB MX data and C98, respectively.  G333.068$-$0.447 at an offset of 1.9\asec\ is the closest MMB \meth\ maser detected.  The young stellar object (YSO) candidate G333.0682-00.4461(1)  from the RMS survey is at an angular offset of 1.86\asec\ \citep{Mottram07}. This YSO and the MMB \meth\ maser G333.068$-$0.447 are co-spatial within the errors of the MMB ATCA data. The ATLASGAL source AGAL 333.068-00.447 is at a similar separation to the \exOH\ maser of 1.9\asec.
\\
\\
\textit{333.228$-$0.055} ND, position from ATCA-2009. 1665-MHz detection in C98. No 6030-MHz source detected in MMB MX data. No nearby radio or mm- continuum source reported within 5\asec, with the nearest AGAL source over 30\asec\ away. At an offset of 28.1\asec\ the nearest MMB \meth\ maser, G333.234$-$0.060, is unlikely to be associated with the same exciting source as the \exOH\ maser.
\\
\\
\textit{337.097$-$0.929} ND, with position from ATCA-2013. This source was found in the Piggyback MX rather than the main survey MX as such there is no 6030-MHz MX data. No 6030-MHz maser was found in the ATCA-2013 data. No 1665-MHz source reported in C98.  There is the sub-mm source AGAL 337.098-00.929 nearby at an angular offset of 2.4\asec. MMB \meth\ maser G337.097$-$0.929 is at an offset of 5.3\asec. 
\\
\\
\textit{337.844$-$0.374} ND, position from ATCA-2009. No 1665-MHz maser in FC89, C98 or A00. No 6030-MHz source in MMB MX data. Offset by $\sim$1.0\asec\ from IRAS 16367-4701 a source listed as an outflow candidate by \citet{GuzmanGarayBrooksVoronkov12}. At a similar offset, 1.24\asec, lies the MMB \meth\ maser G337.844$-$0.375, at a greater offset but still potentially associated is the ATLASGAL source AGAL 337.844-00.376, which has a position offset from the \exOH\ maser peak of 5.5\asec.
\\
\\
\textit{339.980$-$0.539} ND, from ATCA-2009. Non detection in 6030-MHz in MMB MX data. No 1665-MHz in C98 or FC89. Radio source GPSR 339.980-0.538 at 1.4GHz is offset by 1.77\asec \citep{Zoonematkermani90}. At a larger offset of 2.6\asec\ is the MMB \meth\ maser G339.980$-$0.538 and at greater still offset (5.3\asec) is compact sub-mm source AGAL 339.979-00.539.
\\
\\
\textit{341.974+0.225} ND, from ATCA-2009. Not detected in 6030-MHz in MMB MX data, nor is there a 1665-MHz maser in C98 or FC89. This maser is at an angular offset of 5\asec\ from sub-mm source AGAL 341.974+00.226. The nearest \meth\ maser reported by the MMB is G341.973+0.233 at a separation of 29.8\asec so unlikely to exist around the same ionising source.
\\
\\
\textit{343.354$-$0.067} ND,  ATCA-2009 position. No 1665-MHz detection in FC89, C98 or A00. No 6030-MHz emission seen in MMB MX. This source offset from the MMB \meth\ maser G343.354$-$0.067 by 2.5\asec. There are no other nearby radio or  mm- continuum source reported within 5\asec, with the nearest sub-mmm source (AGAL G343.352$-$00.067) at an offset of 8.2\asec.
\\
\\
\textit{344.419+0.044:} In this known source the -72.5\kms\ feature (0.28Jy) from CV95 is missing in the MMB spectrum. No spectrum in C03 for comparison. The emission at v=-65 to -62.8\kms\ seen in the MMB MX spectrum is also seen in the CV95 spectrum but were then weak $\sim$0.1Jy compared to the now missing -72.5\kms\ feature.  
\\
\\
\textit{345.495+1.469} ND, ATCA-2009 position. Two nearby 1665-MHz sources listed in C98, \textit{G345.494+1.469} and \textit{G345.498+1.467}. The 6035-MHz peak is closest to \textit{G345.494+1.469} in velocity (peak velocity difference 0.4\kms). The \exOH\ detection is positioned to the northeast of radio sources `C' and  `I-E' detected toward IRAS 16562$-$3959 by \citep{GuzmanGarayBrooks10}. Object `C' being the proposed central object and `I-E' the inner eastern lobe of the collimated jet the authors detect toward this massive young stellar object. The nearest MMB \meth\ maser, G345.498+1.467, to the \exOH\ detection (offset by 13.12\asec) is similarly positioned to the northeast of the `C' and  `I-E' radio sources of \citet{GuzmanGarayBrooks10}. In the same region is the sub-mm source AGAL 345.493+01.469 which is offset from the \exOH\ maser by an angular separation of 6.3\asec.
\\
\\
\textit{354.725+0.299} Detected in C03, however incorrectly listed with negative Galactic latitude.
\\
\\
\textit{357.924$-$0.338} ATCA-2009 position. No detection listed in C98 or A00 of 1665-MHz maser, no 6030-MHz detection in MMB MX. \meth\ maser G357.924$-$0.337 from the MMB is offset by 3.9\asec, beyond this no radio or  mm- continuum sources are reported within 5\asec\ with the nearest sub-mm detection at a slightly larger separation of 5.5\asec\ (AGAL 357.923-00.337).

\subsection{Tentative or non-detections}
\label{TND:sec}
Table \ref{nondetect:tab} lists previously detected \exOH\ sources for which the MMB survey MX shows no maser emission or a tentative detection that is below the MX observation noise. A tentative detection is considered such if there is structured emission at approximately the correct velocity of a previous detection (e.g. C03) but it falls below the 3$\sigma$ RMS noise limit for the MX (listed in column 5 of Table \ref{nondetect:tab}). Non-detections show no such structure. Each of these sources is considered to have varied in flux density since its earlier detetcion to be below our detection threshold. The variability of masers in our sample is discussed in \S  \ref{Varbs:sec}.

\begin{table*}
\renewcommand\thetable{2}
\begin{center}
\begin{tabular}{@{}c c c r c c}
\hline
Source & RA & Dec&  $V_{peak}$ & Upper limit  & Status  \\
MMBxOH- & [ h m s] & [d $^{\prime}$ \asec] & [\kms] & [Jy] & N/T$^{a}$  \\
\hline
G003.910+0.001 &17:54:38.74 & $-$25:04:03.1 & 17.8 & 0.33 & T \\
G022.435$-$0.169 & 18:32:43.83 & $-$09:24:32.8 & 29.5 & 0.32 & N \\
G040.623$-$0.138 & 19:06:01.64 & +06:46:36.5 & 32.0 & 0.22 & T \\
G285.263$-$0.050 & 10:31:29.88 & $-$58:02:18.5 & 9.3 & 0.74 & T \\
G306.322$-$0.334 & 13:21:23.02 & $-$63:00:29.3 & $-$23.5 & 0.21& T \\
G316.762$-$0.012 & 14:44:56.17 & $-$59:48:00.7 & $-$37.5 & 0.25 & N \\
G319.398$-$0.012 & 15:30:17.41 & $-$58:36:13.3 & $-$12.7 & 0.25 & N \\
G329.066$-$0.308 & 16:01:09.96 & $-$53:16:02.3 & $-$42.7 & 0.27 & T \\
G336.358$-$0.137 & 16:33:29.19 & $-$48:03:43.7 & $-$75.1 & 0.38 & N \\
G338.075+0.012 & 16:39:39.05 & $-$46:41:28.3 & $-$48.7 & 0.34 & N \\
G338.280+0.542 & 16:38:09.07 & $-$46:11:03.1 & $-$56.8 & 0.33 & N \\
G348.550$-$0.979 & 17:19:20.42 & $-$39:03:51.6 & $-$13.1 & 0.33 & T \\
\hline
\end{tabular}
\end{center}
\caption{List of previous ex-OH maser detections which were either non-detections or tentative (see text) in the MMB survey MX observations. Positions and velocities listed are from \citet{Caswell03}. The upper limit listed in column 5 is 3$\times$ the RMS noise in the MX for that source. $^{a}$ The status is either `N' for non-detection or `T' for tentative.}
\label{nondetect:tab}
\end{table*} 

\section{Discussion}

\subsection{Galactic Distribution}
\begin{figure*}
\begin{center}
\includegraphics[scale=0.5, trim= 00 00 00 10]{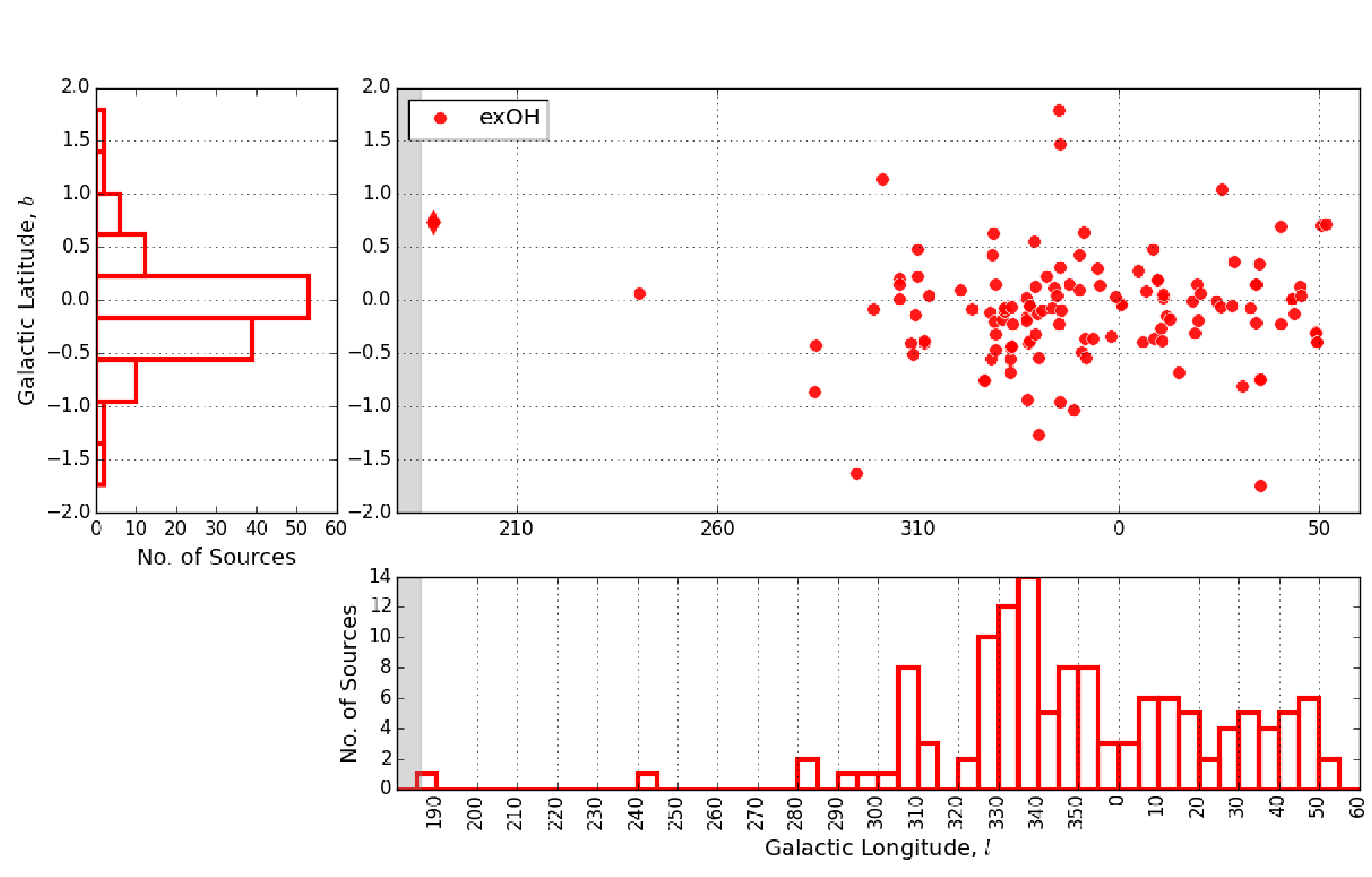}
\caption{Galactic distribution of detected \exOH\ masers within the MMB survey region. \textit{Top right:} Positions of \exOH\ detected during the MMB survey, the grey region between 180$^{\circ}$ and 186$^{\circ}$ is not covered by the survey. The source \textit{189.030+0.783}, is marked as a diamond to highlight that this source's position comes from the Parkes data only. \textit{Top left:} Histogram of \exOH\ distribution in Galactic latitude, $b$. \textit{Bottom} Histogram of \exOH\ distribution in Galactic longitude, $l$, grey region as per \textit{top right plot}.}
\label{GalacticDist:fig}
\end{center}
\end{figure*}

The Galactic 6035-MHz maser population distribution as seen in our data can be seen graphically in Figure \ref{GalacticDist:fig} (main panel). The leftmost and lower images in Figure \ref{GalacticDist:fig} provide histograms of the distributions in Galactic latitude and longitude respectively.

\subsubsection{Longitude Distribution}
We have binned our \exOH\ sample into 5 degree bins to look at the distribution of masers within the Galaxy, as seen in the lower image of Figure \ref{GalacticDist:fig}. Across the whole MMB survey longitude range there is an average maser count per 5$^{\circ}$ bin of 1.8, with a large standard deviation of 3.1. Including only bins with detections there are an average 4.7 masers per bin (with standard deviation of 3.4) implying a significant degree of clustering. 

The majority of the survey range, between the survey boundary at 186$^{\circ}$ and 280$^{\circ}$, is free from maser detections, with only two masers observed in this region (at $l$=189.030$^{\circ}$ and $l$=240.316$^{\circ}$) (1.6\% of the total \exOH\ sample). Given our smaller sample size this would not seem to contradict the detection rate in the \meth\ survey which detected 22 \meth\ masers (2.3\% of the \meth\ sample) in this same longitude range \citep{MMB186to330}.

The clearest peak in the distribution is seen between 325$^{\circ}$  and 340$^{\circ}$, the same longitude range as the Norma and Perseus (inner) tangent points \citep{ValleeTangentPoints}. This peak in the distribution was also seen in the \meth\ maser population of \citet{MMB330to345}. This region contains 28.3\% of the \exOH\ masers detected by the MMB survey. 

Another notable peak, when compared to those surrounding it, appears within the 305$^{\circ}$  and 310$^{\circ}$ bin. This coincides with direction of the Crux tangent point ($\sim 308^{\circ}$ \citealt{ValleeTangentPoints}). 

From 0$^{\circ}$ to the other survey boundary of 60$^{\circ}$ the distribution of masers is fairly uniform, with only a notable dip between 20 and 25$^{\circ}$ and a tail-off after 50$^{\circ}$. The former is likely due to observations being made through the inter-arm region of the Perseus and Sagittarius arms toward the distant Perseus arm. 

\subsubsection{Latitude Distribution}


Of the 127 \exOH\ masers 94\% are within 1$^{\circ}$ of the Galactic plane. Of those outside this latitude range 2 are new detections. In the MMB main survey catalogues \citep{MMB345to006,MMB330to345,MMB006to020,MMB186to330,MMB020to060} (hereafter the MMB catalogue papers), of the 972 6668-MHz masers 911 (94\%) were found to be within 1$^{\circ}$ of the Galactic plane, the same percentage seen in the 6035-MHz sources, indicative that the population of sources being traced by both species is comparable. The region with most 6668-MHz masers at latitudes more extreme than $\lvert b \rvert \geq 1^{\circ}$ was in the longitude range 186$^{\circ}$ to 330$^{\circ}$ \citep{MMB186to330} with 15\% of detections at high/low latitudes. Our small number of 6035-MHz masers (30) in this region does not allow for a robust comparison however 7\% of detections are found at $\lvert b \rvert \geq 1^{\circ}$, which is more consistent with the general trend of 6035-MHz and 6668-MHz within the whole Galaxy.

\subsection{Velocity distribution and Galactic structure.}
\begin{figure*}
\begin{center}
\includegraphics[scale=0.7, trim= 00 00 00 10]{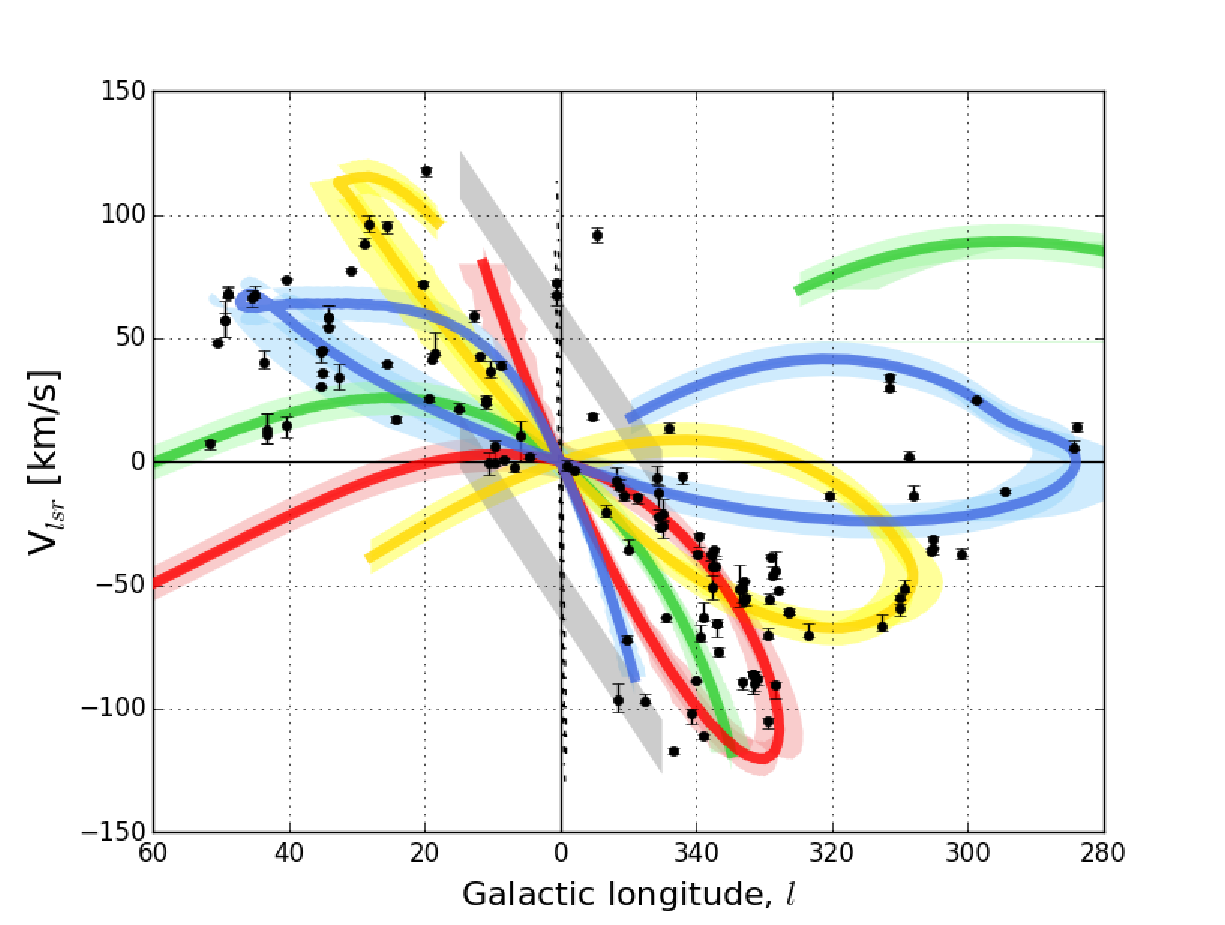}
\caption{\textit{l-V$_{lsr}$} distribution of \exOH\ masers within the galaxy. \exOH\ sources are plotted as black circles at the source peak velocity, the errorbars give the velocity range of the maser source. The spiral arms (coloured as Norma -red, Carina-Sagittarius - blue, Perseus - green, Crux-Scutum - yellow, Local arm - broken pink) follow \citet{GeorgelinGeorgelin76} modified by \citet{TaylorCordes93}, using the \citet{Reid09} rotation curve. The dark central bands in each colour show the loci of each arm with the paler shaded regions showing the parameter space covered by arms of width 1kpc with velocity $\pm$7\kms\ that of the loci. The near and far 3kpc arms are represented as grey regions, following \citet{DameThaddeus08}. The thin black dot-dash line around $l=0^{\circ}$ shows the best-fit open orbit model for molecular clouds in the CMZ from \citet{Kruijssen15}. \textit{Note: For clarity of the majority of sources the x-axis has been truncated and that two sources exist at $l<280^{\circ}$}. }
\label{SpiralArms:fig}
\end{center}
\end{figure*}


Figure \ref{SpiralArms:fig} shows the logitude-velocity ($l-v$) positions of the observed \exOH\ masers overlaid on the spiral arm structure of the Milky Way. The spiral arm positions follow the logarithmic spirals of \citet{GeorgelinGeorgelin76} with the \citet{TaylorCordes93} adjustments. We use the \citet{Reid09} rotation curve to calculate the $V_{lsr}$ of the spiral arms. The 3-kpc arms in Figure \ref{SpiralArms:fig} (grey regions) are as per \citet{DameThaddeus08}. We use these models in order for the plot to be directly comparable with the latest of the MMB catalogue papers. We discuss the velocity range and number count of \exOH\ masers in comparison to \meth\ using the  same longitude ranges as the MMB catalogue papers.

\subsubsection{Longitude 345$^{\circ}$ - 006$^{\circ}$, MMB Paper \textsc{I}}
In the region covering the Galactic Centre we find 23 ex-OH masers, 18.1\% of the total, covering a velocity range of -97 to 92 \kms. There is a significant overlap of all the Galactic spiral arms within this longitude range making comment on maser association with specific spiral arms difficult, however those masers near the extrema of our velocity range do have $l-v$ properties which indicate they are within the near (high negative $v$) or far (high positive $v$) 3-kpc arms. These comprise four sources; \textit{347.628+0.149} and \textit{351.581$-$0.353}, at negative velocity thus in the near 3-kpc arm plus \textit{0.666$-$0.029} and \textit{0.666$-$0.035} at low $l$ but high $v$ making them candidates for inhabiting the far 3-kpc arm. However \citet{MMB345to006} point out that for the latter pair star formation tracing masers in this region are more likely to be part of the Sgr B2 star forming complex. The most extreme positive velocity source in this $l$ range is \textit{354.725+0.299} at a velocity of +91.76\kms, which, as can be seen in Figure \ref{SpiralArms:fig}, makes it an outlier with respect to the spiral arm structure of the Milky Way.  

\citet{MMB345to006} found 183 \meth\ masers in this region, 18.9\% of the \meth\ total, comparable to the proportion of \exOH\ in our sample. The velocity range covered by these masers is -127 to 104 \kms\ exceeding the \exOH\ velocity range of -97 to 92 \kms\ seen in our sample. Two methanol masers G354.701+0.299 and G354.724+0.300 from this region display peculiarly high positive velocities for the longitude, which are ascribed to these sources being in the Galactic bar. Our unusual \exOH\ source \textit{354.725+0.299} is separated spatially by 3.66\asec\ and 2.14\kms\ from G354.724+0.300 placing it also in the Galactic bar.

\paragraph{The Central Molecular Zone:} In recent years (since the publication of \citealt{MMB345to006}) there has been significant advancement in the understanding of the orbital behaviour of molecular gas in the central molecular zone (CMZ) of the Milky Way.

The orbital properties of molecular gas in the inner 250pc ($l=\sim\pm1.7^{\circ}$) of the Milky Way have been found to display an eccentric (about Sgr A*), open (rather than closed) and warped (having vertical motion) orbit \citep{Longmore13,Kruijssen15,Henshaw16}. 

In this region we find four \exOH\ masers \textit{0.666$-$0.029} and \textit{0.666$-$0.035} which \citet{MMB345to006} previously associated with Sgr B2 and \textit{357.924$-$0.338} and \textit{359.137+0.031}. Using the best-fit orbital solution of \citet{Kruijssen15} (plotted as black dot-dash in Figure \ref{SpiralArms:fig}) we find that indeed \textit{0.666$-$0.029} and \textit{0.666$-$0.035} seem to indeed fall very well on this orbital path toward the position of Sgr B2. \textit{357.924$-$0.338} is at too low a latitude to be expected to follow the CMZ orbit and \textit{359.137+0.031} lies outside of this orbital path and at velocities more consistent with association with one of the spiral arms.

\subsubsection{Longitude 006$^{\circ}$ - 020$^{\circ}$, MMB Paper \textsc{II}}
Within this longitude range we detect 16 ex-OH sources, 12.6\% of total, which display a velocity range of -2.3 to 117.8 \kms.  In $l-v$ space there is a significant portion of all the major Galactic spiral arms within this $l$ range, though it does appear that most of our negative velocity \exOH\ masers are likely within the Norma arm rather than the Crux-Scutum arm (see Figure \ref{SpiralArms:fig}).  

\citet{MMB006to020} catalogued 119 \meth\ masers in this longitude range , 12.24\% of the \meth\ total. Again the \exOH\ and \meth\ proportion with respect to the source totals are comparable.
The velocity range, -30.2 to 157\kms, of the \meth\ masers exceeds that of the \exOH\ sample, particularly for those few sources with negative velocity $<-$10\kms\ which are thought to be part of the near 3-kpc arm. Also unlike the \meth\ survey we do not appear to have detected any \exOH\ masers in the far 3-kpc arm, which contains the \meth\ sample's high positive velocity sources in this region.

\subsubsection{Longitude 330$^{\circ}$ - 345$^{\circ}$, MMB Paper \textsc{III}}
A total of 30 ex-OH masers are found within this longitude range, 23.6\% of total, covering a velocity range of -116.9 to 13.7 \kms. Figure \ref{SpiralArms:fig} shows this region contains a significant part of the Norma, Crux-Scutum and Perseus Arms, with the majority of our sources appearing in the region of $l-v$ space where Norma and Crux-Scutum overlap. 

The 198 \meth\ masers found within this region make up 20.4\% of total \meth\ population. This value differs by $\sim$3.2\% from the \exOH\ result, with a greater fraction of \exOH\ found within this longitude range, though given the smaller \exOH\ population this value is within counting statistics and may not be significant. 

The \meth\ maser velocity range covers -127 to 19 \kms\ with just two sources contributing positive velocities. The \exOH\ sample contains only a single source with positive velocity \textit{343.929+0.125} which is positionally coincident with \meth\ maser G343.929+0.125, one of the two contributing the positive velocity from the \meth\ survey.

\subsubsection{Longitude 186$^{\circ}$ - 330$^{\circ}$, MMB Paper \textsc{IV}}
\label{PaperIV:sec}
We find 30 ex-OH masers within this area of sky (23.6\% of the total). \exOH\ masers in this region exhibit velocities in the range of -104.7 to 63.6\kms. This region contains a large section of the Crux-Scutum and the Carina-Sagittarius arms as well as a section of the Perseus arm (unconfused at high velocity $>$63\kms) and the Norma tangent. We find two \exOH\ masers within the Perseus arm which exists in this region, \textit{189.030+0.738} and \textit{240.316+0.071} (not seen in Figure \ref{SpiralArms:fig} due to the truncated x-axis). The remainder of our masers in this region trace the remaining spiral arms well.

\citet{MMB186to330} report 207 \meth\ masers comprising 21.3\% of the total MMB \meth\ sample a comparable percentage to our \exOH\ statistics.

\subsubsection{Longitude 020$^{\circ}$ - 060$^{\circ}$, MMB Paper \textsc{V}}
In this region we find 28 ex-OH masers inhabit this $l$ range, (22.1\% of total) over a velocity range 7.1 to 96.1 \kms. Notably in Figure \ref{SpiralArms:fig}, no \exOH\ masers in our sample appear associated with the negative velocity components of the Norma or Crux-Scutum arms present in this $l$ range.

\citet{MMB020to060} catalogued 265 \meth\ masers in this region, 27.3\% of the total. The proportion of \meth\ masers detected with respect to the total source counts is 5.2\% greater than \exOH\ masers, suggesting an underabundance of the latter species in this region. 

The velocity range of \meth\ sources is -42.8 to 125.5 \kms\ appearing far greater than that of \exOH, though in fact the majority (79\%) of \meth\ masers fall within the same range of \exOH\ with only 12 masers having $V<0$\kms\ and 43 appearing above 100\kms.

\subsection{6030-MHz \exOH\ masers associated with 6035-MHz masers}
\label{sixohthreeoh:sec}
We detect 32 6030-MHz \exOH\ masers within the same sources as 6035-MHz masers in our catalogue, however there are 11 6035-MHz masers from our catalogue for which there is no 6030-MHz MX data and one for which there exist only RHCP data. Disregarding these 12 sources, 28\% of our 6035-MHz sample has a 6030-MHz counter part. Previous studies of these two \exOH\ transitions found percentages of 36\% \citep{Caswell03}. We account for the difference between these two values in three ways. Firstly, by the fact that the MMB data have a typical MX RMS $\sim$ 0.1Jy \citep{GreenTech} which is higher than the typical spectrum RMS (0.03-0.05 Jy) reported by \citet{Caswell03} for their data. This may allow for the detection by \citet{Caswell03} of more lower flux density 6030-MHz counterparts than we were capable of, thus boosting their percentage. Secondly, the MMB is a complete untargeted survey meaning we may be seeing a number of sources at different evolutionary stages to that of the sample in \citet{Caswell03}, which collated targeted surveys toward strong OH emission sites \citep{Caswell97} and ground-state OH masers \citep{Caswell98,Caswell01}.  Finally, given the variability of the \exOH\ transition (see section \ref{Varbs:sec} for a discussion of this in the 6035-MHz line) a number of weak 6030-MHz masers previously detected maybe too weak for detection in our survey.

\subsubsection{Flux density comparison}
\begin{figure}
\begin{center}
\includegraphics[scale=0.45, trim= 00 00 00 00]{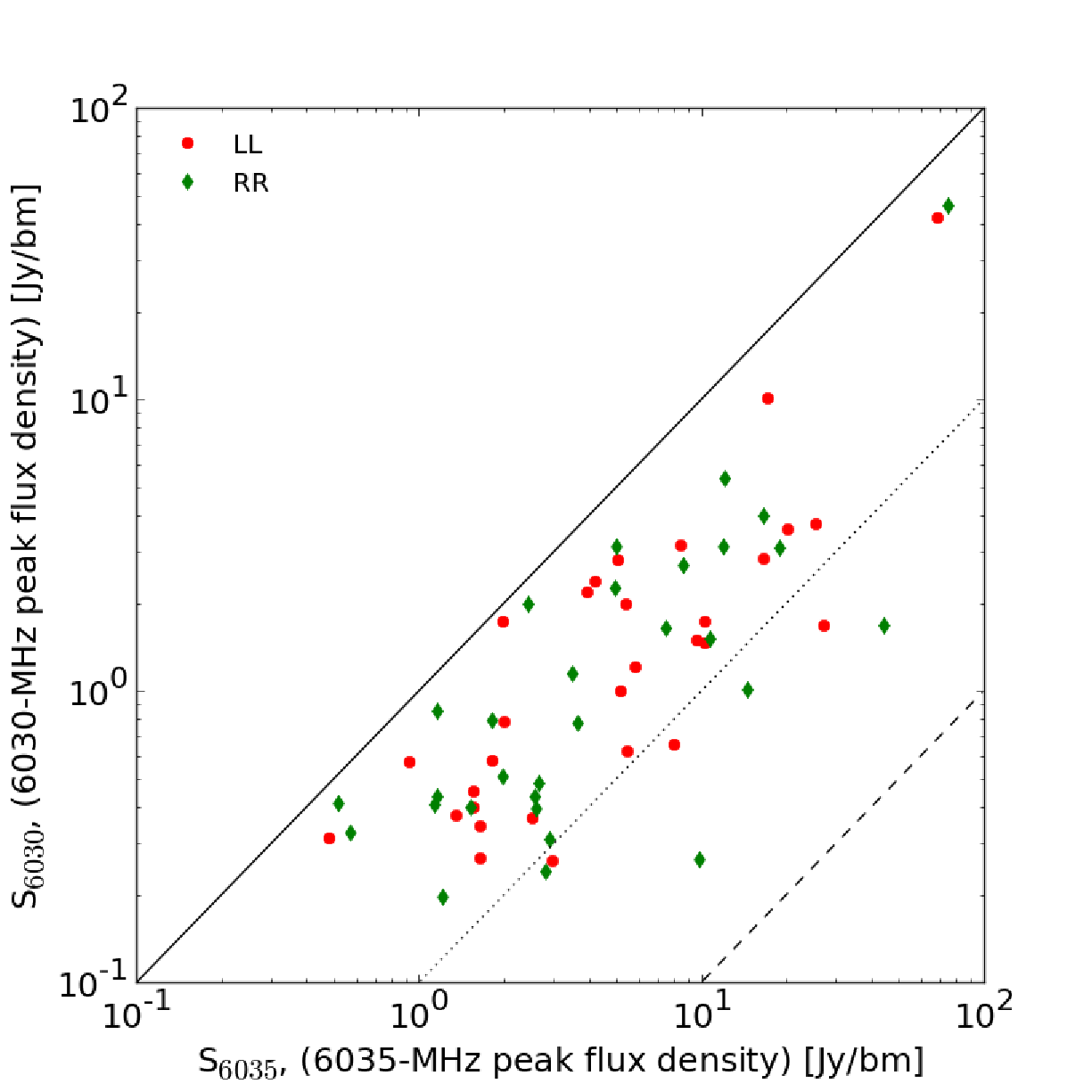}
\caption{Peak 6035-MHz \exOH\ flux density, S$_{6035}$, compared to peak 6030-MHz \exOH\ flux density, S$_{6030}$.  Solid line S$_{6035}$=S$_{6030}$, dotted line S$_{6030}$=0.1$\times$S$_{6035}$, dashed line S$_{6030}$=0.01$\times$S$_{6035}$.}
\label{SixOthreO:fig}
\end{center}
\end{figure}

Figure \ref{SixOthreO:fig} compares the peak flux density of the 6035-MHz maser emission with that of the counterpart source at 6030-MHz at its peak velocity. The largest velocity offset between the two maser species peak emission seen in our data is 5.0\kms\ in source \textit{49.490-0.388} (a feature also seen by C03), as such this is not included in Figure \ref{SixOthreO:fig}. 

For the remainder of sources the velocity displacement between peaks at the different frequencies is always below 2.6 \kms. Where there is a 6030-MHz \exOH\ maser associated with a 6035-MHz maser, the 6030-MHz peak flux density never exceeds the counterpart 6035-MHz maser value. This result is consistent with the \exOH\ maser model results of \citet{Gray92}, \citet{Cragg02} and the majority of observational results \citep[e.g,][]{Baudry97, DesmursBaudry98, Caswell03}. \citet{Baudry97} report 6030/6035-MHz typical flux density ratios of 0.14 to 0.5, and similarly \citet{Caswell03} reports a typical values of 0.125, up to extremes such as 0.01. Such values are consistent with our findings (Figure \ref{SixOthreO:fig}).

\subsection{Methanol maser association}
The works of \citet{Ellingsen07a} and \citet{Breen2010} suggest that masers of different species provide an evolutionary time line of regions of high-mass star formation. The simultaneous occurrence of class II \meth\ and ground-state OH masers occurs later on in the relative lifetime of a high-mass star formation region \citep[][see their Figure 6.]{Breen2010}, whilst it remains unclear about the association epoch of class II \meth\ and \exOH\ species. Given our sample of \exOH\ masers are drawn from the same Galactic plane survey as the class II \meth\  masers we are in a position to give robust information on the association rates of the two species.

\subsubsection{Angular Separation}
\begin{figure}
\begin{center}
\includegraphics[scale=0.45, trim= 00 00 00 00]{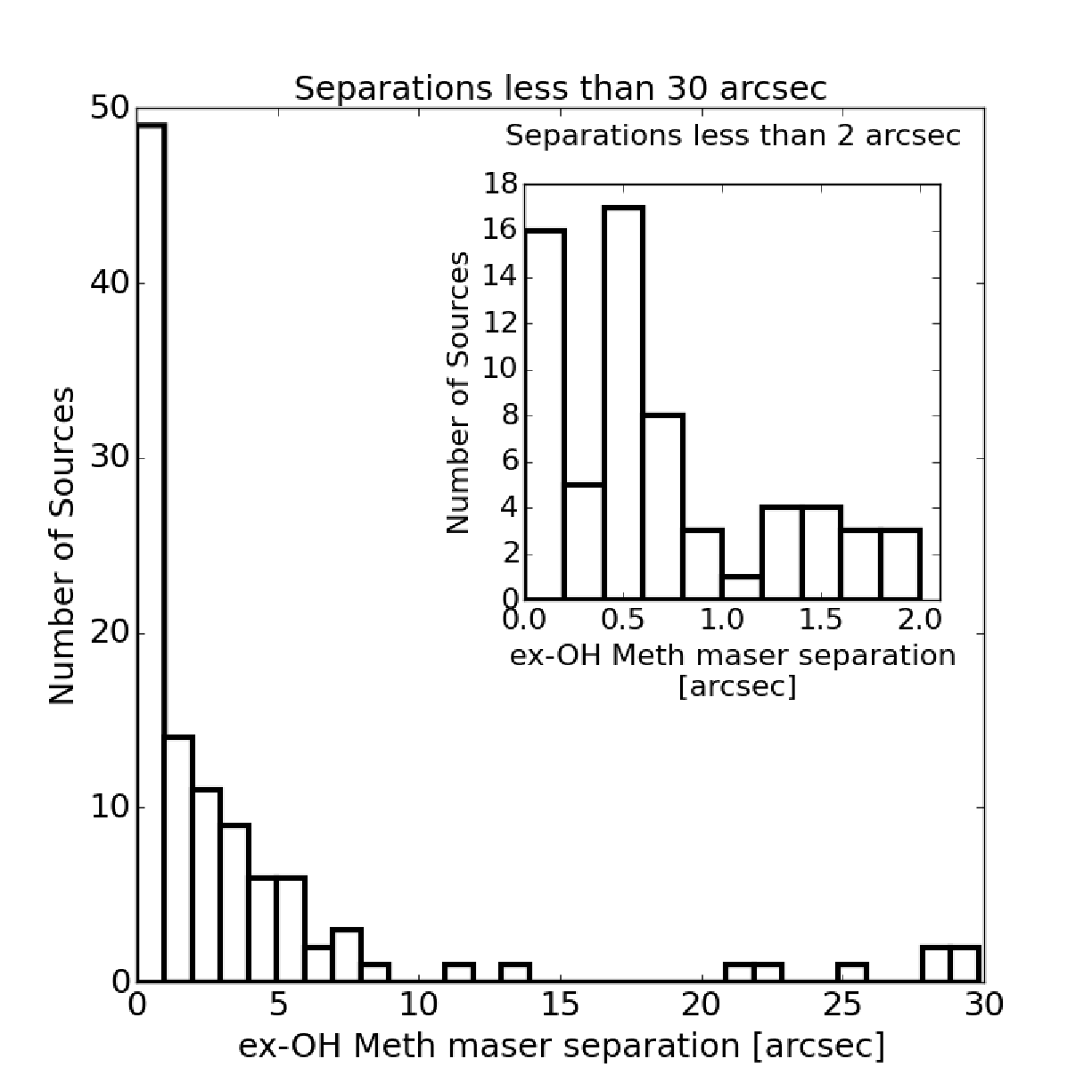}
\caption{Histogram of \exOH\ to nearest class II \meth\ maser from the MMB survey for separations $\leq 40$\asec. The inset figure shows the distribution of separations  $\leq 2$\asec\ equivalent to $\sim$0.05pc at 5kpc. }
\label{Separation:fig}
\end{center}
\end{figure}

For each of the observed \exOH\ masers the closest class II \meth\ maser (in terms of angular separation) was determined; the \meth\ maser positions were taken from the MMB catalogues. Figure \ref{Separation:fig} shows the distribution of angular separation between our \exOH\ masers and the nearest class II \meth\ masers. Within this figure we show only sources with angular separations, $\theta$ $\leq$30\asec\ (equivalent to $\sim$1pc at a distance of 5kpc). Sources at larger separations, referred to hereafter as `isolated' \exOH\ masers, are discussed in $\S$\ref{isolated:sec}. 

As can be seen in Figure \ref{Separation:fig} the majority of \exOH\ masers, $\sim$80\%, are within $\theta \leq$ 10\asec\ ($\sim$0.25pc at 5kpc) of a \meth\ maser, where one would assume that the majority of these sources are associated with the same star forming core (nominal size $\sim$0.1pc), forming one or more stellar sources. Of this 80\% the largest proportion (64 sources), are within 2\asec\ ($\sim$0.05pc at 5kpc) a separation at which one may assume both are associated with the same protostellar object (c.f. the fiducial size of a hyper compact \hii\ region ($\lesssim$0.03pc), e.g. \citealt[][]{Kurtz05}). 

In their pilot study of ground-state OH masers in Carina-Sagittarius \citet{Green12} use an association limit of `coincident within the positional errors'. Applying this same limit to our ex-OH data (0.4\asec, 9.7mpc at 5kpc) we find 26 sources meeting this criteria, for these sources one would confidently assume each maser species are part of the same protostellar system.

\begin{figure}
\begin{center}
\includegraphics[scale=0.45, trim= 00 00 00 00]{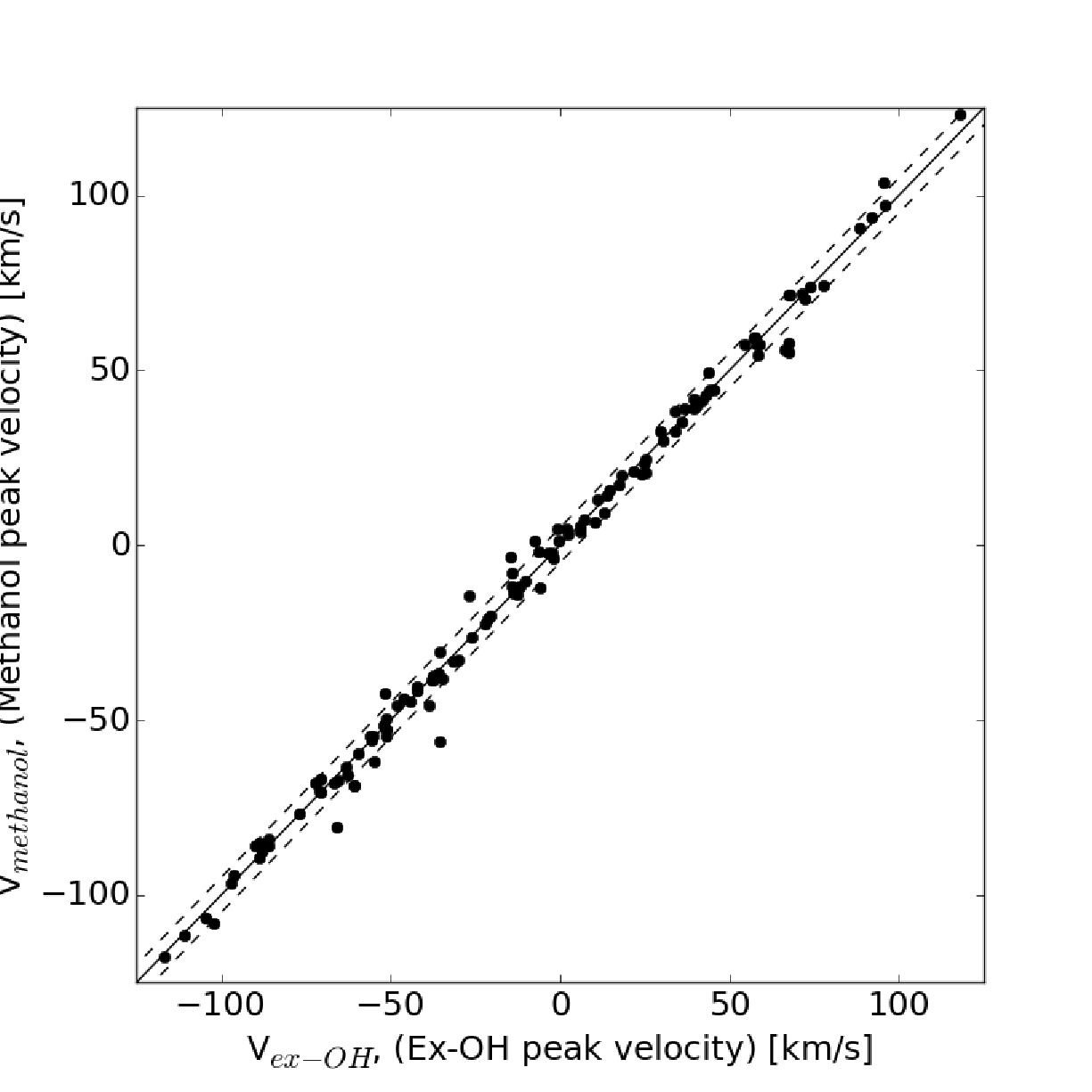}
\caption{Peak $V_{methanol}$ plotted against peak $V_{ex-OH}$ for nearest maser pairings for sources separated by less than the ATCA primary beam at 6035-MHz (569\asec). The solid line gives $V_{methanol}=V_{ex-OH}$ and the dashed lines $V_{methanol}=V_{ex-OH} \pm 5.0$\kms .}
\label{VeloVVelo:fig}
\end{center}
\end{figure}

Figure \ref{VeloVVelo:fig} plots the peak velocity of each maser species against its nearest counterpart for all sources with an ATCA position and clearly shows that there is a tight correlation between the two masing species across all velocities. Figure \ref{SepVelo:fig} gives the velocity difference, $\Delta V$, of potential \exOH\ - \meth\ maser pairs plotted against the angular separation of the sources. In this figure it can be seen that the majority of \exOH\ masers are within $\pm$ 5.0\kms\ (denoted by the two green horizontal dashed lines) of their positional counterpart \meth\ maser. This is particularly true for those sources with angular separations of  $\theta \leq$2\asec\ (left of the vertical dotted line), with only 5 of 64 sources outside the $\pm$5.0\kms\ range, and 2\asec$ < \theta \leq$10\asec\ (left of the vertical dot-dash line), with 7 of 37 sources outside this range. 

Also indicative of true associations for the majority of low angular separation maser pairs is the fact that the mean $\lvert \Delta V \rvert$ increasingly deviates from $\lvert \Delta V \rvert=$ 0 from the $\leq$2\asec\ grouping through  2\asec$ < \theta \leq$10\asec\ up to the $\theta >10$\asec\ group (see column 3 in the upper portion of Table \ref{massep:tab}). There is also an increase in the standard deviation (column 4 Table \ref{massep:tab}) between the each group.

\subsubsection{Physical Separation}
\label{physsep:sec}
A large number of MMB methanol sources were assigned a distance in \citet{GreenMcClure11}, which used H\textsc{i} self-absorption data to resolve the near/far kinematic distance ambiguity for these masers. Of our sample of 126\footnote{We exclude \textit{189.030+0.738} for this analysis due to its low accuracy position (see $\S$ 3)} \exOH\ masers 118 have likely associated \meth\ counterparts in the \citet{GreenMcClure11} study. 

Additionally, \citet{Reid14} have released parallax distance measurements of a large number of \meth\ (both 6.7 and 12.2 GHz) masers, eight of which are likely counterparts to our \exOH\  and the MMB sources. For these sources we use the parallax distance in place of the H\textsc{i} self-absorption distance. Sources for which the near/far distance ambiguity remains, i.e. they have no H\textsc{i} or parallax distance, are not considered further in this association analysis, nor are sources where the \exOH\ and \meth\ maser are separated by $\geq$30\asec. Leaving 91 potential maser pairs.

Assuming that these 91 \exOH\ and \meth\ potential pairs both exist at the same distance we can repeat the analysis from the previous section using physical, $\Theta$, rather than angular separation. 

The lower panel of Table \ref{massep:tab} presents this quantitatively. We divide our sample into maser pairs of separations $\leq$ 0.03pc (the HC\hii\ size scale), between 0.03pc  $<\Theta \leq$0.1pc (between HC\hii\ and protostellar core size scale) and greater that 0.1pc. Again we find an increase in mean $\lvert \Delta V \rvert$ with increasing separation, indicative of looser association between species. 
Figure \ref{SepVeloPhys:fig} follows Figure \ref{SepVelo:fig}, but using the physical separation between maser species for those with assigned distances from \citet{GreenMcClure11} or \citet{Reid14}. We find that only four sources of forty-one in our smallest physical separation group have have $\lvert \Delta V \rvert> 5$ \kms\ (left of dotted line).

From this, one could take a \textit{very} conservative lower limit of true associations with the 972 class II \meth\ masers from the MMB as being 37 \exOH\ sources; at separations $\leq$ 0.03pc with $\lvert \Delta V \rvert\leq5$\kms. 

\subsubsection{Ex-OH maser lifetimes}
\citet{vanderWalt05} used a statistical analysis of the then known number of 6668-MHz \meth\ masers in the Milky Way to estimate the lifetime of an individual \meth\ maser source, finding a value of between 2.5 and 4.5$\times10^{4}$ years. Taking this value and the occurrence of associated pairs between the MMB \meth\ and \exOH\ masers (29.4\%) we can calculate a lifetime for an \exOH\ maser source.

Firstly, we assume that \exOH\ and 6668-MHz \meth\ masers are associated for a single period during the formation of a high-mass star and that all our associated pairs are forming in high-mass star forming regions\footnote{6668-MHz \meth\ maser have never been detected outside of high-mass star forming regions \citep{Minier03,Xu08,Breen13}.}. Given these assumptions the timescale for coexistence for these maser species can be calculated: 
\begin{equation}
\tau_{coexist}=\frac{N_{meth+exOH}}{N_{meth}}\tau_{meth}
\label{coexist:eqn}
\end{equation}

\noindent where $\tau_{coexist}$ is the time of species coexistence, $N_{meth+exOH}$ the number of associated \exOH\ and \meth\ pairs in our survey, $N_{meth}$ the total number of \meth\ maser in the same survey region and $\tau_{meth}$ the \meth\ maser lifetime as calculated by \citet{vanderWalt05}. Using our conservative lower limit of 37 maser pairs and the total of 972 \meth\ maser in the MMB catalogues this gives a timescale of coexistence for these maser species of 950 to 1700 years.

This coexistence timescale allows us to next estimate the lifetime of the \exOH\ masers themselves. We find 29.4\% of our sample are coexisting with \meth\ masers, which is equivalent to the \exOH\ masers spending 29.4\% of their total lifetime in associated with \meth\ meaning that the total lifetime, $\tau_{exOH}$, can be calculated using the equivalent formula to \ref{coexist:eqn}:
\begin{equation}
\tau_{coexist}=\frac{N_{meth+exOH}}{N_{exOH}}\tau_{exOH}
\label{exOHlife:eqn}
\end{equation}

\noindent where $N_{exOH}$ is the total number of \exOH\ maser in our sample and other symbols have their previous meanings. Following this approach we calculate that the total lifetimes of \exOH\ maser sources are between 3.3$\times10^3$ to 5.8$\times10^3$ years. If we expand our associated sample ($N_{meth+exOH}$) to include all sources with separations less than 0.1pc with $\lvert \Delta V \rvert$ between $\pm$5.0\kms\ this gives 55 associated pairs, a coexistence timescale of 1400 to 2500 years and an estimated life time of 4.6$\times10^3$ to 8.3$\times10^3$ years for \exOH. This value is shorter than that expected of the ground-state OH maser of a $few\times10^4$ years \citep{FishReid06}. 

Finally, we note two caveats relavent to the above calculated values. First we do not take into account the potential existence of \exOH\ masers around post main sequence stars, but given the rarity of such masers and the lack of detection of such sources in our sample (see $\S$\ref{isolated:sec}) the effect of such contamination on the estimated value will be of the order a few percent, much smaller the estimated lifetime ranges we calculate. Second, if the assumption that 6668-MHz \meth\ masers and \exOH\ masers exist at only a single epoch during star formation is not true then our values would act as an upper limit for \exOH\ maser lifetime.

\subsubsection{Isolated \exOH\ sources: Separation $\geq$ 30\asec\ }
\label{isolated:sec}

\begin{figure}
\begin{center}
\includegraphics[scale=0.45, trim= 00 00 00 00]{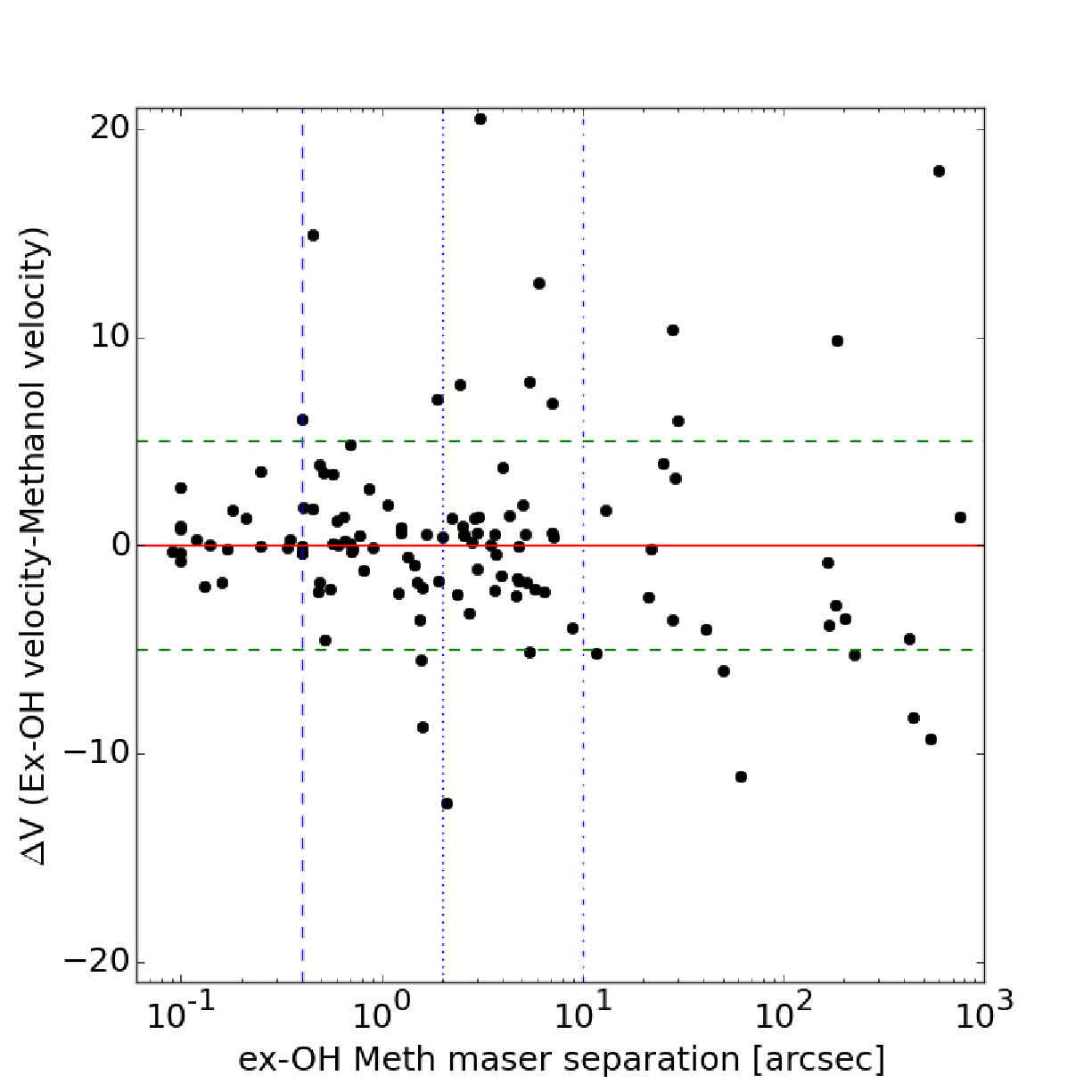}
\caption{The velocity difference between \exOH\ masers and the nearest \meth\ maser as a function of angular separation of the pair. The horizontal red (solid) and green (dashed) lines denoted $\Delta V$ of 0\kms\ and $\pm$5\kms\ respectively. The vertical dotted and dash-dot line mark out angular separations of 2 and 10\asec\ respectively. \textit{Note: the log x-axis excludes those few sources where the maser coordinates are identical thus having a separation of zero.} }
\label{SepVelo:fig}
\end{center}
\end{figure}

\begin{figure}
\begin{center}
\includegraphics[scale=0.45, trim= 00 00 00 00]{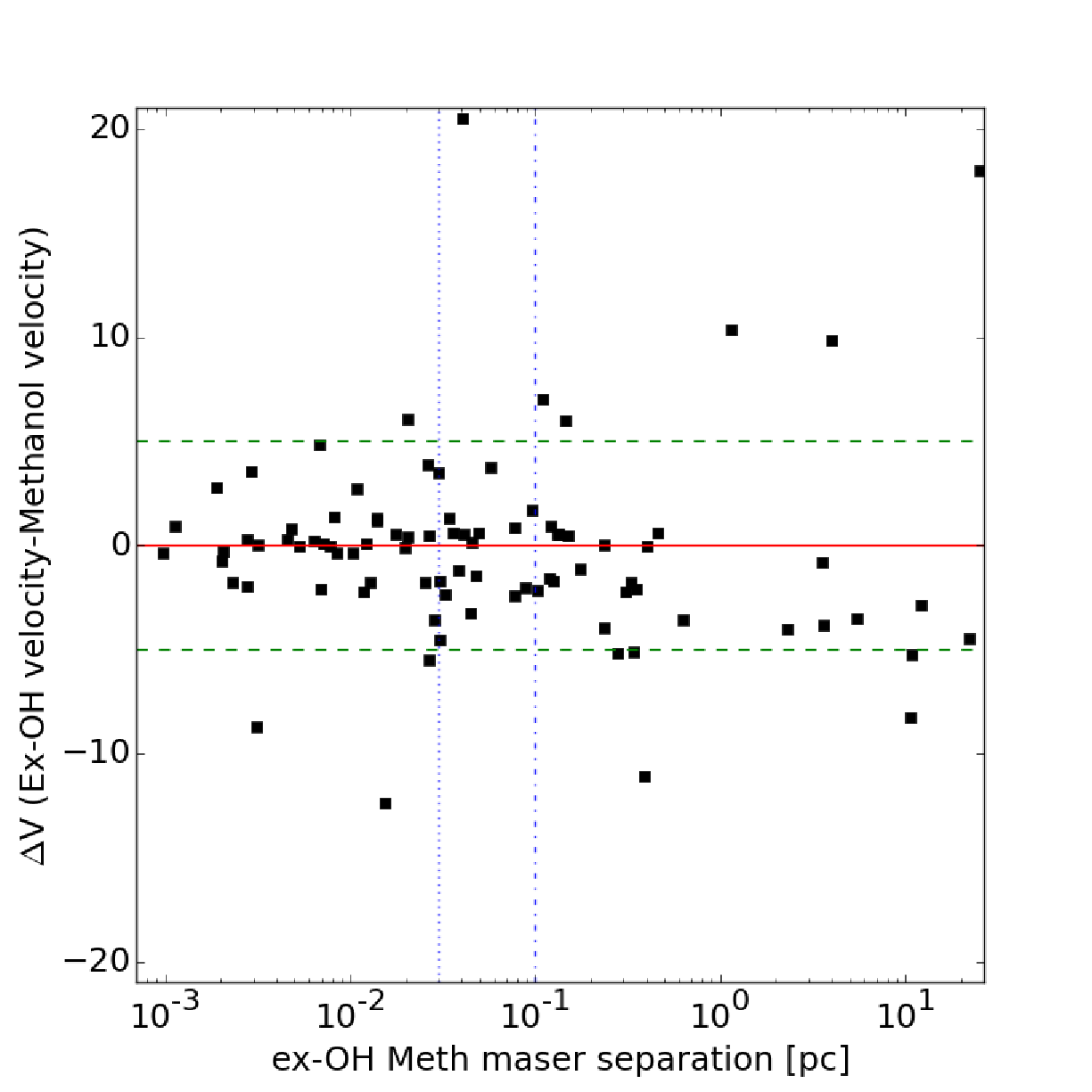}
\caption{The velocity difference between \exOH\ masers and the nearest \meth\ maser as a function of physical separation of the pair, for sources with distances from \citet{GreenMcClure11} only. The horizontal red (solid) and green (dashed) lines denoted $\Delta V$ of 0\kms\ and $\pm$5\kms\ respectively. The vertical dotted and dash-dot line mark out angular separations of 0.03 and 0.1pc respectively. \textit{Note: the log x-axis excludes those few sources where the maser coordinates are identical thus having a separation of zero.}}
\label{SepVeloPhys:fig}
\end{center}
\end{figure}

From Figure \ref{SepVelo:fig} it can also be seen that 11 \exOH\ sources have separations from the nearest \meth\ maser greater than 100\asec\ up to 764\asec\ which at a fiducial 5kpc distance to a high-mass star forming region gives a separation in excess of 2pc up to 18.5pc. Whilst this separation is sufficiently small that the sources could reside in the same giant molecular cloud (typical size 100pc see e.g \citealt{Larson03}), which may account for some of the low $\Delta V$ values for these sources, it is very unlikely that these masers are truly associated and being excited by the same source.  Beyond this, there are a further three sources which appear truly isolated with no counterpart from the MMB catalogues found within tens of arcminutes, these being \textit{8.352$+$0.478}, \textit{240.316$+$0.071} and \textit{284.016$-$0.856}. 

We inspected the literature for counterparts to these three extremely isolated masers, finding an interesting diversity of star formation tracing counterparts between them. Source \textit{8.352$+$0.478} has no counterpart dense core in either BGPS or ATLASGAL within 30\asec\ and no other maser species detections within $1.0$\asec. Source \textit{284.016$-$0.856} similarly has no other maser species present within $1.0$\asec, but does (as noted in $\S$ \ref{NotesObj:sec}) have a likely associated dense core AGAL 284.016$-$00.857 offset by 4.8\asec. The source \textit{240.316+0.071} has a close, $1.0$\asec, water maser counterpart listed in both \citet{Breenwater10} and \citet{Reid14} and a ground-state OH maser at an offset position of 0.74\asec. This source is outside of the survey range for both BGPS or ATLASGAL, but does have nearby, $1.0$\asec\, millimetre and radio wavelength detections associated with the (ultra) compact \hii\ region G240.31+0.07 \citep[see e.g.][]{Chen07,Trinidad11}. Finally, the literature provides no nearby ($\leq$30\asec) evolved/post main sequence star counterparts for any isolated maser sources suggesting that these masers are associated with star formation.

In the modelling of \citet{Cragg02} it was found that the 6035-MHz \exOH\ maser can exist at densities up to $n_H = 10^{8.5}$cm$^{-3}$ which exceeds the density at which the 6668-MHz \meth\ maser is quenched by collisional interaction ($n_H = 10^{8.3}$cm$^{-3}$). The isolated \exOH\ masers we detect may be tracing these extreme high density regions and warrant further investigation to examine their environments.

\begin{table}
\renewcommand\thetable{3}
\begin{center}
\begin{tabular}{@{}c c c c}
\hline
\exOH\ - \meth\ & No. of sources &  $\lvert \Delta V\rvert_{mean}$ &  $\Delta V_{std}$   \\
Separation & & [\kms] & [\kms] \\
\hline
$\leq$ 0.4\asec\ & 26 & 1.06 & 1.64 \\
$\leq$ 2\asec\ & 64 & 1.85 & 3.04 \\
2\asec $< \theta \leq$10\asec\ & 37 & 3.12 & 5.19 \\
$>$ 10\asec\ & 23 & 5.46 & 6.74 \\
\hline
$\leq$ 0.03pc & 41 & 1.93 & 3.15 \\
0.03pc  $< \Theta \leq$0.1pc & 18 & 2.23 & 3.72\\
$>$ 0.1pc & 32 & 4.04 & 5.59 \\
\hline
\end{tabular}
\end{center}
\caption{\exOH\ - \meth\ maser separation distribution. \textit{Upper panel}, angular separation (in arcseconds) for all sources. \textit{Lower panel} physical separation (in parsec) for sources with \citet{GreenMcClure11} or \citet{Reid14} distances only. Column 3 gives the mean velocity difference between positionally associated \exOH\ - \meth\ pairs from the Stokes I peak, and Column 4 the standard deviation of these values.}
\label{massep:tab}
\end{table} 

\subsection{Variability}
\label{Varbs:sec}
To inspect the variability of the \exOH\ sources over time a comparison of the flux densities for sources observed in either CV95 and/or C03 and the MMB MX's was made. The MMB MX data was obtained 2008/2009, CV95 data were mostly taken in 1993/1994, giving a minimum time baseline of 14 years. The C03 data were taken in 2001 giving a minimum time baseline of 7 years. The CV95/C03 reported peak flux densities and spectra were inspected to ensure that for sources which we had an MX, the CV95/C03 peak was within ~1.0km/s of the MX peak for the same polarisation.  Only masers where the peak matched in velocity and polarisation were used in the comparison, thus excluding those with multi-component spectra where a different velocity component now dominates (as discussed in \S \ref{NotesObj:sec}).

\begin{figure}
\begin{center}
\includegraphics[scale=0.45, trim= 00 00 00 00]{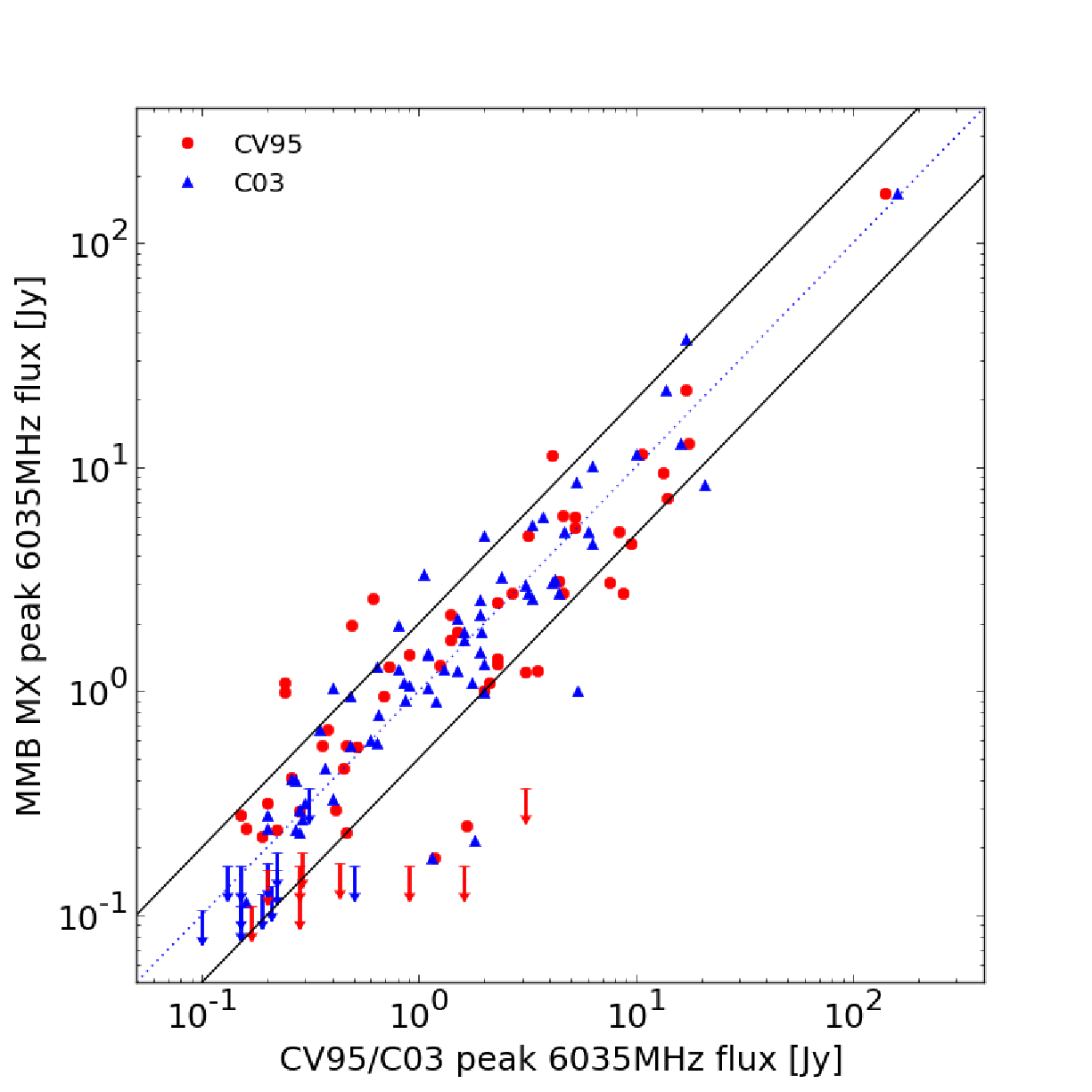}
\caption{Comparison of \exOH\ masers observed as part of the MMB with those previously observed in CV95 (red) and C03 (blue), sources which are tentative or none-detections (see \S \ref{TND:sec}) are marked as downward arrows of the same colour. The dashed line indicated y=x, and the solid lines y=2x and y=x/2.}
\label{Variability:fig}
\end{center}
\end{figure}

In total the comparison was made with 61 sources from CV95 and 77 sources from C03. Figure \ref{Variability:fig} plots the comparison of the MMB peak flux densities, including the sources which were found to be tentative or non-detections from the MMB MX data (see \S \ref{TND:sec}) for which $3\sigma$ upper limits were used. The ratios of flux densities for CV95/MMB and C03/MMB have median values of 0.92 and 0.94 respectively, but as can be seen from Figure \ref{Variability:fig} a number of sources, 19 from CV95 and 11 from C03, have varied by a factor of $\geq$2 over at least one of the two periods. These highly variable sources are listed in Table \ref{VariabilityTab:tab}.  For those sources where the variation is $\geq$2 over only one epoch, the variation over the other epoch is also listed (where possible) to highlight any trend within the variation.

The majority of these highly variable sources show the variation acting in the same direction, either increasing or decreasing, over both epochs to the MMB. Four sources (\textit{19.486+0.151, 49.486-0.389, 345.003-0.224 and 345.487+0.314}) have variation acting in the opposite direction between the CV95-MMB and C03-MMB epochs. An additional 4 sources (\textit{11.904-0.141, 15.035-0.677, 35.200-1.736 and 345.010+1.792}) show a switch in sense of variation between the CV95 to C03 and the C03 to MMB measurements, but had sufficient variation over the CV95 to MMB period to show an overall continuing positive or negative trend when comparing CV95-MMB and C03-MMB. For these 8 sources the variation must have reached a maximum/minimum at some point between either CV95 and C03 or C03 and MMB observations which may be indicative of variation on a timescale of years or less in these sources, making them interesting candidates for single dish monitoring observations.

\begin{table*}
\renewcommand\thetable{4}
\begin{center}
\begin{tabular}{@{}l r c r c}
\hline
Source Name & S$_{MMB}$ & MMB/CV95 & S$_{MMB}$& MMB/C03\\
MMBOH- & Peak [Jy] &  Variation factor & Peak [Jy] & Variation factor \\ 
\hline
\hline
G000.666$-$0.029 & 9.08(R) & 0.48 & 9.08(R) & 0.72\\
G011.034+0.062 & 2.18(L) & 4.54 & 2.18(L) & 1.28\\
G011.904$-$0.141 & 9.87(R) & 1.54 & 9.87(R) & 2.47\\
G015.035$-$0.677 & 14.58(R) & 0.52 & 16.68(L) & 0.40\\
G019.486+0.151 & 1.98(R) & 4.13 & 1.81(L) & 0.75\\
G035.025+0.350 & 6.09(R) & 0.40 & 6.09(R) & 0.74\\
G035.198$-$0.743 & 3.91(L) & 3.99 & 3.91(L) & 2.44\\
G035.200$-$1.736 & 0.50(R) & 0.15 & 0.43(L) & 0.12\\
G049.486$-$0.389 & 2.05(R) & 0.45 & 2.05(R) & 2.56\\
G284.351$-$0.418 & 2.44(R) & 0.39 & 1.97(L) & 0.49\\
G285.263$-$0.050 & $<$0.74 & 0.12 & - & - \\
G316.762$-$0.012 & $<$0.25 & 0.45 & - & -\\
G330.953$-$0.182 & 1.35(R) & 1.78 & 6.67(L) & 3.15\\
G331.542$-$0.066 & 22.58(R) & 2.75 & 11.08(L) & 1.68\\
G337.606$-$0.052 & 1.18(R) & 0.15 & 0.36(L) & 0.16\\
G338.075+0.012 & $<$0.34 & 0.40 & - & -\\
G338.280+0.542 & $<$0.33 & 0.10 & - & -\\
G339.884$-$1.259 & - & - & 74.85(R) & 2.20\\
G343.929+0.125 & 2.45(L) & 0.35 & 2.45(L) & 0.82\\
G345.010+1.792 & 2.00(L) & 0.5 & 2.00(L) & 0.19\\
G345.003$-$0.224 & 5.51(R) & 0.32 & 6.45(L) & 1.34\\
G345.487+0.314 & 5.17(L) & 4.24 & 5.17(L) & 0.78\\
G348.550$-$0.979 & $<$0.33 & 0.18 & $<$0.33 & 0.33\\
\hline
\end{tabular}
\caption{Table of sources which have increased or decreased by a factor of $\geq\pm$2 or more. Negative values show a decrease in flux density between epochs and positive an increase. For each epoch the polarisation handedness is denoted with R or L in parenthesis in the two MMB column, highlighting which those sources where the handedness has changed. For tentative or non-detected MMB sources the upper limit value is given and only variability between epochs where the previous detection was above this upper limit is listed.}
\label{VariabilityTab:tab}
\end{center}
\end{table*} 

\section{Summary}
We present the results from the first complete Galactic Plane survey of \exOH\ masers at 6035-MHz. This survey was carried out as part of the Methanol MultiBeam Survey using the Parkes 64-m radio telescope. The catalogue of sources comprises 127 sources with 47 new detections. All the new detections have high accuracy positions from interferometric observations with the ATCA.  In addition to the measured 6035-MHz sources we also catalogue associated 6030-MHz \exOH\ masers, detecting 32 masers at this frequency. The 6030-MHz emission is typically weaker that at 6035 consistent with previous studies.

The catalogue is compared to the main MMB catalogue of 6668-MHz \meth\ masers (comprising 972 detections), from which we find an unambiguous association (sources separated by $\leq 0.4$\asec, $\leq$9.7mpc at 5kpc) for just 26 sources. Increasing the association limit to 2\asec\ (0.05pc at 5kpc, the approximate scale of a single proto-stellar core) we find 64 \meth$-$\exOH\ pairs. In our sample 91 \exOH\ sources have a robust distance measurements allowing physical separations to be calculated. Of these we find 37 likely \meth$-$\exOH\ maser pairs with physical separations of $\leq$ 0.03pc and 55 pairings separated by $\leq$ 0.1pc.  

Assuming a single epoch of coexistence of \meth\ and \exOH\ masers around a high-mass protostellar source and using the \citet{vanderWalt05} estimated \meth\ maser lifetime, we are able to put constraints on the coexistence period of the two maser species and the \exOH\ maser lifetime, using our associated pairs with distance measurements. We find a period of maser overlap of  $0.95 - 2.5\times10^3$ and a lifetime for \exOH\ masers of $3.3 - 8.3\times10^3$ years.

Finally we compare the observed peak 6035-MHz emission to that observed at previous epochs by \citet{CaswellVaile95} and \citet{Caswell03}. We find over these periods that the majority of \exOH\ masers have varied, mostly by less that a factor of two. However, twenty-three sources have varied by more than a factor of two. Some of these sources have also changed the direction of variability (increasing to decreasing or vice versa) between the three observing times, suggestive of variation on short timescales.

One important feature of \exOH\ masers is their sensitivity to the local magnetic field. A companion paper  will present the results of the polarisation data of the \exOH\ masers and the derived on the magnetic field strengths (Avison et al. \textit{in prep}).

\section*{Acknowledgements}
This work in its initial form was presented as part of LQ's PhD thesis at the University of Manchester \citep{QuinnPhD} and was revised, re-examined, expanded and completed by AA. The Parkes Observatory and the ATCA are part of the Australia Telescope which is funded by the Commonwealth of Australia for operation as a National Facility managed by CSIRO. AA is funded by the STFC at the UK ARC Node. SLB is the recipient of an Australian Research Council DECRA Fellowship (project number DE130101270) and a L'Or\'eal-UNESCO for Women in Science Fellowship.
\bibliographystyle{mnras}
\DeclareRobustCommand{\VAN}[2]{#2}
\bibliography{exOHposVar_accepted}



\onecolumn
\footnotesize{
\begin{landscape}
\renewcommand\thetable{1}
\begin{longtable}{p{2.0cm} l l c p{2.0cm} | c c c | c c c }
\caption{MMB Survey \exOH\ Catalogue.  Columns 1: Source Name, which is defined as the source position in Galactic coordinates (rounded at the third decimal point). Columns 2 \& 3: source position ins Right Ascension and Declination (sources with positions from MMB observations have three decimal places in RA and two in Dec, all previously detected masers have two and one decimal accuracy respectively). Column 4: The \exOH\ transition for the listed maser, counterpart 6030-MHz masers are listed directly below their 6035-MHz partner. Column 5: Reference for the position listed, with the following abbreviations ATCA 09, ATCA 13 and ATCA 14 denote our ATCA observation resutls from 2009, October 2013 and February 2014 respectively observations. C03 = \citet{Caswell03}, C01 = \citet{Caswell01}. Columns 6 to 11: The \exOH\ maser peak flux density, velocity of peak and a minimum-maximum velocity range over which emission is seen in the spectrum, first for left hand circular polarisation (LHCP) and then right hand circular polarisation (RHCP). Sources in \textit{italics} are values for Stokes I (total intensity) flux density taken from the respective ATCA observation as these sources are not present in the nearest Parkes MX. As such only columns 7 to 9 are filled in. $\dagger$  Sources \textit{333.136$-$0.432} and \textit{333.135$-$0.431} have a blended spectrum preventing a complete velocity range to be estimated from the MX data.}\\ 
\hline
 Source Name& Source & Position &  &  & & MX \textit{/(ATCA)} & & & MX &\\
 &  &  &  & &  & LHCP \textit{/(Stokes I)}  &  &  & RHCP & \\
 MMBOH- & RA & Dec & Transition & Position Ref & S$_{peak}$ & V$_{peak}$ & V$_{range}$ & S$_{peak}$& V$_{peak}$ & V$_{range}$ \\ 
 G\textit{lll.lll$\pm$bb.bbb} & hh:mm:ss & dd:mm:ss.s(s) & [MHz] &  & [Jy] & [km/s] & [km/s] & [Jy] & [km/s] & [km/s] \\
\hline
\hline
\endfirsthead
\hline
 Source Name & Source & Position & &  & & MX\textit{/(ATCA)} & & & MX &\\
  &  &  &  & &  & LHCP \textit{/(Stokes I)} &  &  & RHCP & \\
 MMBOH- & RA & Dec & Transition & Position Ref & S$_{peak}$ & V$_{peak}$ & V$_{range}$ & S$_{peak}$& V$_{peak}$ & V$_{range}$ \\ 
 G\textit{lll.lll$\pm$bb.bbb} & hh:mm:ss & dd:mm:ss.s(s) & [MHz] &  & [Jy] & [km/s] & [km/s] & [Jy] & [km/s] & [km/s] \\
\hline
\hline
\endhead
\hline
\multicolumn{3}{r@{}}{continued \ldots}\\
\endfoot
\hline
\endlastfoot
0.666$-$0.029 & 17:47:18.64 & $-$28:22:54.6 & 6035 & C03 & 20.07 & 72.23 & 68.5, 73.6 & 9.08 & 72.06 & 68.4, 73.5\\
 &  &  & 6030 & C03 & 1.50 & 72.28 & 70.1, 73.2 & 2.26 & 70.24 & 69.7, 72.8\\
0.666$-$0.035 & 17:47:20.14 & $-$28:23:06.2 & 6035 & C03 & 2.06 & 67.58 & 60.0, 68.0 & 2.10 & 67.12 & 66.6, 67.6\\
4.682+0.278 & 17:55:18.808 & $-$24:46:24.97 & 6035 & ATCA 09 & 0.71 & 2.01 & 1.3, 3.1 & 0.22 & 2.11 & 1.3, 3.1\\
5.885$-$0.392 & 18:00:30.36 & $-$24:04:03.1 & 6035 & C03 & 0.52 & 10.45 & 0.1, 16.4 & 0.63 & 9.51 & $-$1.0, 16.4\\
6.882+0.094 & 18:00:49.375 & $-$22:57:37.80 & 6035 & ATCA 09 & 5.81 & $-$2.34 & $-$4.1, 0.0 & 1.16 & $-$2.05 & $-$4.1, $-$0.5\\
  &  &  & 6030 & ATCA 09 & 1.22 & $-$2.32 & $-$4.3, $-$1.6 & 0.43 & $-1$2.62 & $-$3.3, $-$1.4\\
8.352+0.478 & 18:02:31.248 & $-$21:29:33.67 & 6035 & ATCA 13 & 2.18 & 0.48 & 0.1, 1.6 & 1.07 & 0.38 & $-$0.1, 1.0\\
8.669$-$0.356 & 18:06:19.01 & $-$21:37:32.7 & 6035 & C03 & 0.66 & 39.31 & 37.8, 41.1 & 0.58 & 39.31 & 38.1, 41.1\\
9.620+0.194 & 18:06:14.916 & $-$20:31:39.07 & 6035 & ATCA 09 & 0.89 & 6.00 & $-$1.7, 8.7 & 0.59 & 6.09 & $-$1.2, 7.9\\
9.622+0.196 & 18:06:14.639 & $-$20:31:27.67 & 6035 & ATCA 09 & 0.24 & $-$0.60 & $-$1.0, 0.6 & 0.25 & 0.08 & $-$0.8, 0.6\\
10.322$-$0.258 & 18:09:23.307 & $-$20:08:02.20 & 6035 & ATCA 09 & 0.75 & 36.59 & 34.2, 41.0 & 1.15 & 35.62 & 34.2, 38.0\\
10.623$-$0.384 & 18:10:28.65 & $-$19:55:49.6 & 6035 & C03 & 0.48 & $-$0.60 & $-$5.6, 3.7 & 0.57 & $-$0.69 & $-$4.8, 3.2\\
  &  &  & 6030 & C03 & 0.32 & 0.80 & $-$2.0, 3.4 & 0.33 & 0.21 & $-$2.1, 2.4\\
10.960+0.022 & 18:09:39.673 & $-$19:26:23.00 & 6035 & ATCA 13 & 1.50 & 25.12 & 23.5, 26.9 & 3.02 & 24.73 & 23.6, 25.5\\
11.034+0.062 & 18:09:39.85 & $-$19:21:20.1 & 6035 & C03 & 2.18 & 24.05 & 21.3, 25.7 & 1.31 & 23.27 & 21, 25.3\\
11.904$-$0.141 & 18:12:11.44 & $-$18:41:29.0 & 6035 & C03 & 1.64 & 42.88 & 41.1, 43.3 & 9.87 & 42.97 & 41.1, 43.9\\
  &  &  & 6030 & C03 & 0.27 & 42.01 & 41.4, 44.1 & 0.26 & 42.99 & 42.1, 43.8\\
12.681$-$0.182 & 18:13:54.771 & $-$18:01:43.70 & 6035 & ATCA 09 & 0.83 & 58.82 & 56.4, 61.2 & 0.52 & 59.21 & 56.9, 61.0\\
15.035$-$0.677 & 18:20:24.81 & $-$16:11:34.1 & 6035 & C03 & 16.68 & 21.52 & 20.9, 24.1 & 14.58 & 21.52 & 21.0, 23.7\\
  &  &  & 6030 & C03 & 2.87 & 21.64 & 21.3, 22.2 & 1.01 & 21.64 & 21.1, 22.4\\
18.460$-$0.005 & 18:24:36.400 & $-$12:51:09.89 & 6035 & ATCA 09 & 0.37 & 43.88 & 43.0, 52.2 & 0.52 & 43.49 & 42.9, 52.5\\
18.836$-$0.299 & 18:26:23.681 & $-$12:39:30.80 & 6035 & ATCA 09 & 0.35 & 41.69 & 41.1, 43.8 & 0.60 & 41.50 & 40.5, 43.3\\
19.486+0.151 & 18:26:00.39 & $-$11:52:21.9 & 6035 & C03 & 1.81 & 25.41 & 24.4, 26.1 & 1.98 & 25.41 & 24.0, 26.7\\
  &  &  & 6030 & C03 & 0.58 & 25.45 & 24.7, 26.0 & 0.51 & 25.45 & 25.0, 26.0\\
19.752$-$0.191 & 18:27:44.910 & $-$11:47:49.94 & 6035 & PIGGY & 0.37 & 117.84 & 115.7, 119.1 & 0.29 & 116.39 & 114.6, 119.1\\
20.237+0.065 & 18:27:44.56 & $-$11:14:54.6 & 6035 & C03 & 2.51 & 71.51 & 71.0, 73.1 & 2.59 & 71.51 & 70.1, 74.5\\
  &  &  & 6030 & C03 & 0.37 & 71.44 & 71.2, 71.8 & 0.39 & 71.54 & 71.2, 71.8\\
24.147$-$0.010 & 18:35:21.020 & $-$07:48:58.99 & 6035 & ATCA 09 & 0.27 & 17.42 & 16.1, 18.2 & 6.67 & 17.42 & 15.8, 18.3\\
25.648+1.050 & 18:34:20.575 & $-$05:59:45.40 & 6035 & ATCA 13 & 2.50 & 39.58 & 38.7, 40.3 & 0.72 & 38.71 & 38.2, 40.0\\
25.509$-$0.060 & 18:38:03.249 & $-$06:37:46.90 & 6035 & ATCA 09 & 1.04 & 95.50 & 92.3, 97.1 & 1.20 & 95.40 & 91.8, 97.8\\
28.200$-$0.049 & 18:42:58.07 & $-$04:13:57.0 & 6035 & C03 & 5.48 & 96.08 & 92.9, 100.0 & 3.50 & 96.27 & 92, 101.0\\
  &  &  & 6030 & C03 & 0.63 & 96.29 & 90.1, 99.4 & 1.15 & 95.02 & 92.8, 98.9\\
28.819+0.366 & 18:42:37.335 & $-$03:29:34.57 & 6035 & ATCA 14 & 0.60 & 88.38 & 87.3, 90.8 & 0.27 & 88.00 & 86.9, 91.8\\
30.778$-$0.801 & 18:50:21.583 & $-$02:16:55.00 & 6035 & ATCA 09 & 0.52 & 77.24 & 77.3, 78.4 & 1.22 & 77.62 & 77.0, 78.2\\
32.744$-$0.076 & 18:51:21.88 & $-$00:12:05.5 & 6035 & C03 & 0.47 & 33.95 & 29.3, 39.7 & 0.91 & 33.66 & 29.3, 41.7\\
34.258+0.153 & 18:53:18.68 & +01:15:00.3 & 6035 & C03 & 0.46 & 58.19 & 54.7, 63.2 & 0.71 & 58.48 & 51.7, 63.8\\
34.258+0.153b & 18:53:18.68 & +01:15:00.0 & 6035 & C03, ATCA 09 & 3.95 & 54.31 & 55.1, 63.1 & 2.67 & 54.50 & 54.0, 56.4\\
  &  &  & 6030 & C03 & 2.20 & 54.34 & 53.7, 54.9 & 0.48 & 54.64 & 54.2, 56.3\\
34.261$-$0.213 & 18:54:37.360 & +01:05:08.60 & 6035 & ATCA 09 & 0.62 & 58.45 & 53.9, 54.9 & 0.71 & 58.35 & 54.3, 55.4\\
35.025+0.350 & 18:54:00.66 & +02:01:19.3 & 6035 & C03 & 3.25 & 45.32 & 44.8, 45.8 & 6.09 & 45.61 & 45.1, 46.1\\
35.133$-$0.744 & 18:58:06.146 & +01:37:09.50 & 6035 & ATCA 13 & 2.24 & 35.80 & 34.9, 36.6 & 0.98 & 36.48 & 35.2, 36.8\\
35.198$-$0.743 & 18:58:13.06 & +01:40:37.7 & 6035 & C03 & 3.91 & 30.46 & 29.9, 30.7 & $-$ & $-$ & - ,-\\
35.200$-$1.736 & 19:01:45.55 & +01:13:33.3 & 6035 & C01, C03 & 0.43 & 44.34 & 40.2, 45.7 & 0.50 & 42.98 & 39.1, 45.3\\
40.282$-$0.220 & 19:05:41.209 & +06:26:11.80 & 6035 & ATCA 14 & 0.86 & 73.88 & 73.5, 74.3 & 0.23 & 74.56 & 74, 74.7\\
40.426+0.701 & 19:02:39.62 & +06:59:12.0 & 6035 & C01, C03 & 0.29 & 14.48 & 10.0, 18.5 & 0.49 & 16.04 & 9.6, 18.5\\
43.149+0.013 & 19:10:11.05 & +09:05:22.1 & 6035 & C01, C03 & 3.65 & 11.16 & 9.0, 14.4 & 3.38 & 10.86 & 7.9, 13.2\\
43.165+0.013 & 19:10:12.85 & +09:06:12.0 & 6035 & C01, C03 & 1.56 & 13.19 & 16.6, 18.3 & 1.21 & 13.09 & 16.0, 18.3\\
  &  &  & 6030 & C03 & 0.40 & 13.49 & 16.7, 18.1 & 0.20 & 13.39 & 16.7, 18.1\\
43.796$-$0.127 & 19:11:53.97 & +09:35:51.8 & 6035 & C01, C03 & 0.54 & 40.29 & 37.7, 44.9 & 0.59 & 40.29 & 38.2, 45.4\\
45.123+0.133 & 19:13:27.80 & +10:53:39.1 & 6035 & C01, C03 & 1.79 & 67.51 & 66.4, 70.9 & 4.06 & 67.60 & 65.3, 71.3\\
45.466+0.045 & 19:14:25.66 & +11:09:26.5 & 6035 & C01, C03 & 10.25 & 66.38 & 62.4, 69.5 & 7.48 & 64.93 & 62.1, 69.3\\
  &  &  & 6030 & C03 & 1.73 & 65.23 & 64.1, 68.2 & 1.64 & 64.84 & 64.2, 68.7\\
48.988$-$0.300 & 19:22:26.035 & +14:06:30.99 & 6035 & ATCA 09 & 1.65 & 67.53 & 66.3, 71.2 & 3.58 & 67.63 & 65.4, 71.0\\
\textit{49.046$-$0.290} & \textit{19:22:30.778} & \textit{+14:09:52.59} & \textit{6035} & \textit{ATCA 09} & \textit{1.21} & \textit{67.97} & \textit{66.5, 72.1} &  &  &\\
49.490$-$0.388 & 19:23:43.93 & +14:30:34.9 & 6035 & C01, C03 & 6.29 & 57.40 & 50.8, 60.5 & 6.98 & 57.70 & 51.1, 65.5\\
  &  &  & 6030 & C01, C03 & $-$ & $-$ & -,- & 1.59 & 52.69 & 51.4, 55.1\\
49.486$-$0.389 & 19:23:43.85 & +14:30:29.9 & 6035 & C01, C03 & 1.99 & 55.10 & 54.3, 56.1 & 2.05 & 55.80 & 54.5, 56.5\\
50.478+0.705 & 19:21:40.286 & +15:53:46.39 & 6035 & ATCA 09 & 4.80 & 48.12 & 47.4, 48.9 & 0.82 & 48.22 & 47.1, 48.8\\
51.681+0.714 & 19:24:00.392 & +16:57:39.400 & 6035 & ATCA 14 & 0.31 & 7.14 & 5.1, 8.6 & 0.45 & 7.04 & 5.7, 8.0\\
189.030+0.783 & - & - & 6035 & MX only, see $\S$\ref{SurRes:ref}  &0.6 & 3.36 & 2.2,4.5 & 0.98 & 3.27 & 2.3, 4.3\\
240.316+0.071 & 07:44:51.97 & $-$24:07:42.3 & 6035 & CV95, C03 & 2.64 & 63.62 & 62.2, 64.3 & 0.95 & 63.62 & 61.7, 64.3\\
284.016$-$0.856 & 10:20:16.299 & $-$58:03:50.20 & 6035 & ATCA 09 & 0.60 & 14.25 & 12.5, 15.9 & 1.62 & 14.05 & 13.3, 15.5\\
284.351$-$0.418 & 10:24:10.68 & $-$57:52:34.0 & 6035 & C03 & 1.97 & 5.86 & 5.1, 8.9 & 2.44 & 5.76 & 1.1, 12.5\\
  &  &  & 6030 & C03 & 1.73 & 5.87 & 4.3, 9.5 & 2.00 & 5.78 & 3.6, 8.8\\
294.511$-$1.621 & 11:35:32.21 & $-$63:14:43.0 & 6035 & C03 & 5.44 & $-$11.99 & $-$12.3, $-$11.6 & 4.71 & $-$11.99 & $-$12.3, $-$11.6\\
298.723$-$0.086 & 12:14:39.674 & $-$62:39:23.10 & 6035 & ATCA 09 & 0.28 & 24.84 & 24.6, 25.2 & 0.20 & 24.06 & 23.9, 24.5\\
300.969+1.148 & 12:34:53.27 & $-$61:39:39.9 & 6035 & C03 & 17.10 & $-$37.39 & $-$39.2, $-$35.3 & 12.08 & $-$37.58 & $-$39.4, $-$35.4\\
  &  &  & 6030 & C03 & 10.09 & $-$37.17 & $-$37.7, $-$36.8 & 5.39 & $-$37.66 & $-$38.1, $-$37.2\\
305.200+0.019 & 13:11:16.90 & $-$62:45:54.7 & 6035 & C03 & 1.70 & $-$31.37 & $-$37.7, $-$30.4 & 3.31 & $-$31.56 & $-$38.6, $-$30.0\\
305.208+0.206 & 13:11:13.800 & $-$62:34:41.1 & 6035 & ATCA 13 & 0.98 & $-$34.90 & $-$35.1, $-$34.2 & 1.25 & $-$35.39 & $-$35.6, $-$35.1\\
305.362+0.150 & 13:12:35.933 & $-$62:37:17.30 & 6035 & ATCA 09 & 0.41 & $-$36.05 & $-$36.5, $-$35.3 & 0.36 & $-$35.76 & $-$36.8, $-$35.0\\
308.056$-$0.396 & 13:36:32.327 & $-$62:49:05.20 & 6035 & ATCA 09 & 15.93 & $-$14.11 & $-$15.3, $-$9.8 & 25.46 & $-$13.62 & $-$16.0, $-$10.5\\
308.651$-$0.507 & 13:41:50.399 & $-$62:49:04.90 & 6035 & ATCA 09 & 0.35 & 2.27 & 1.2, 4.4 & 0.42 & 2.17 & 1.4, 4.2\\
309.384$-$0.135 & 13:47:24.192 & $-$62:18:11.80 & 6035 & ATCA 09 & 0.51 & $-$51.37 & $-$52.9, $-$47.9 & 2.25 & $-$52.05 & $-$53.4, $-$49.7\\
309.901+0.231 & 13:51:00.966 & $-$61:49:54.79 & 6035 & ATCA 09 & 0.56 & $-$55.16 & $-$55.8, $-$54.1 & 0.73 & $-$55.16 & $-$56.0, $-$54.0\\
309.921+0.479 & 13:50:41.77 & $-$61:35:10.1 & 6035 & C01, C03 & 20.20 & $-$59.43 & $-$62.4, $-$57.3 & 18.84 & $-$61.56 & $-$62.8, $-$57.2\\
  &  &  & 6030 & C03 & 3.62 & $-$58.05 & $-$62.6, $-$56.9 & 3.11 & $-$59.32 & $-$62.7, $-$57.2\\
311.596$-$0.398 & 14:06:18.35 & $-$62:00:15.3 & 6035 & C03 & 1.82 & 29.74 & 28.8, 32 & 1.61 & 30.61 & 27.9, 32.4\\
311.643$-$0.380 & 14:06:38.74 & $-$61:58:23.1 & 6035 & C03 & 1.15 & 33.91 & 28.6, 35.2 & 0.71 & 30.61 & 27.7, 35.7\\
312.598+0.045 & 14:13:14.958 & $-$61:16:53.90 & 6035 & ATCA 09 & 0.51 & $-$66.72 & $-$68.7, $-$62.1 & 0.66 & $-$66.33 & $-$67.9, $-$65.8\\
320.427+0.103 & 15:09:39.787 & $-$57:59:40.68 & 6035 & ATCA 09 & 0.60 & $-$14.05 & $-$15.0, $-$12.5 & 0.25 & $-$13.66 & $-$15.3, $-$12.4\\
323.459$-$0.079 & 15:29:19.332 & $-$56:31:21.26 & 6035 & C97 & 27.14 & $-$70.50 & $-$70.9, $-$65.4 & 44.18 & $-$70.21 & $-$70.7, $-$65.4\\
  &  &  & 6030 & C03 & 1.69 & $-$70.39 & $-$64.0, $-$71.5 & 1.69 & $-$70.19 & $-$65.0, $-$70.7\\
\textit{326.447$-$0.749} & \textit{15:49:18.525} & \textit{$-$55:16:56.98} & \textit{6035} & \textit{ATCA 09} & 0.55 & \textit{$-$60.92} & \textit{$-$61.7, $-$60.2} &  &  &\\
326.448$-$0.749 & 15:49:18.701 & $-$55:16:53.98 & 6035 & ATCA 09 & 0.92 & $-$60.88 & $-$61.8, $-$59.7 & 1.82 & $-$60.78 & $-$62.2, $-$59.7\\
  &  &  & 6030 & ATCA 09 & 0.57 & $-$60.76 & $-$59.9, $-$61.4 & 0.80 & $-$60.57 & $-$60.0, $-$61.3\\
327.944$-$0.116 & 15:54:34.002 & $-$53:50:47.90 & 6035 & ATCA 09 & 0.57 & $-$52.05 & $-$52.6, $-$51.3 & 0.65 & $-$51.95 & $-$52.4, $-$51.6\\
328.236$-$0.548 & 15:57:58.348 & $-$53:59:25.20 & 6035 & C03, ATCA 09 & 1.35 & $-$44.21 & $-$47.4, $-$36.4 & 0.52 & $-$46.05 & $-$48.1, $-$35.7\\
  &  &  & 6030 & C03, ATCA 09 & 0.38 & $-$45.73 & $-$46.2, $-$45.4 & 0.41 & $-$46.04 & $-$46.4, $-$45.7\\
328.307+0.430 & 15:54:06.44 & $-$53:11:41.1 & 6035 & C98, C03 & 3.39 & $-$90.52 & $-$95.5, $-$88.9 & 2.96 & $-$90.33 & $-$94.2, $-$89.2\\
328.808+0.633 & 15:55:48.39 & $-$52:43:06.7 & 6035 & C03 & 22.76 & $-$46.08 & $-$47.1, $-$42.3 & 13.70 & $-$45.79 & $-$47.3, $-$42.1\\
329.031$-$0.197 & 16:00:30.38 & $-$53:12:25.5 & 6035 & C98 & 0.21 & $-$38.77 & $-$39.0, $-$37.5 & 0.17 & $-$32.46 & $-$32.8, $-$32\\
329.184$-$0.314 & 16:01:47.103 & $-$53:11:40.60 & 6035 & ATCA 09 & 0.24 & $-$55.48 & $-$57.7, $-$53.3 & 0.19 & $-$55.68 & $-$56.3, $-$54.6\\
329.339+0.148 & 16:00:33.15 & $-$52:44:39.8 & 6035 & C03 & 0.63 & $-$104.72 & $-$107.8, $-$103.5 & 0.81 & $-$104.04 & $-$107.3, $-$103.2\\
329.405$-$0.459 & 16:03:32.15 & $-$53:09:29.9 & 6035 & C03 & 0.33 & $-$70.47 & $-$71.4, $-$67.2 & 0.17 & $-$70.36 & $-$71.3, $-$69.8\\
330.953$-$0.182 & 16:09:52.38 & $-$51:54:57.6 & 6035 & C03 & 6.67 & $-$87.82 & $-$90.2, $-$86.7 & 1.35 & $-$88.01 & $-$90.5, $-$86.7\\
331.512$-$0.102 & 16:12:09.856 & $-$51:28:35.86 & 6035 & ATCA 09, C03 & 2.97 & $-$89.86 & $-$93.0, $-$88.4 & 2.82 & $-$89.04 & $-$93.0, $-$88.4\\
  &  &  & 6030 & C03 & 0.26 & $-$89.39 & $-$90.7, $-$88.6 & 0.24 & $-$89.10 & $-$90.2, $-$88.6\\
331.512$-$0.102 & 16:12:10.049 & $-$51:28:34.96 & 6035 & ATCA 09, C03 & 2.68 & $-$86.83 & $-$88.0, $-$84.3 & 5.40 & $-$85.98 & $-$88.0, $-$84.3\\
331.543$-$0.066 & 16:12:09.137 & $-$51:25:45.43 & 6035 & ATCA 09, C03 & 11.08 & $-$85.98 & $-$94.0, $-$84.6 & 22.58 & $-$85.89 & $-$87.9, $-$84.7\\
332.824$-$0.548 & 16:20:10.758 & $-$50:53:18.1 & 6035 &  ATCA 13 & 1.14 & $-$54.98 & $-$58.4, $-$53.7 & 2.49 & $-$54.69 & $-$57.7, $-$54.0\\
332.964$-$0.679 & 16:21:23.057 & $-$50:52:56.49 & 6035 & ATCA 09 & 0.21 & $-$48.26 & $-$48.6, $-$47.9 & 0.15 & $-$48.36 & $-$48.6, $-$47.9\\
333.068$-$0.447 & 16:20:49.096 & $-$50:38:38.87 & 6035 & ATCA 09 & 0.74 & $-$56.10 & $-$56.2, $-$55.7 & 0.38 & $-$55.90 & $-$56.2, $-$55.7\\
333.136$-$0.432 & 16:21:03.112 & $-$50:35:08.70 & 6035 & ATCA 09/ C03 & 7.79 & $-$50.24 & $\dagger$, $-$47.8 & 7.32 & $-$50.43 & $\dagger$, $-$47.8\\
333.135$-$0.431 & 16:21:02.92 & $-$50:35:10.0 & 6035 & C03 & 10.31 & $-$51.21 & $-$59.9, $\dagger$ & 10.76 & $-$51.40 & $-$59.9, $\dagger$\\
  &  &  & 6030 & C03 & 1.47 & $-$51.10 & $-$61.2, $-$48.9 & 1.51 & $-$51.10 & $-$61.2, $-$48.9\\
333.228$-$0.055 & 16:19:47.960 & $-$50:15:11.50 & 6035 & ATCA 09 & 0.23 & $-$88.87 & $-$92.1, $-$87.7 & 0.26 & $-$90.13 & $-$91.2, $-$87.7\\
333.608$-$0.215 & 16:22:11.162 & $-$50:05:56.75 & 6035 & C03 & 2.17 & $-$51.61 & $-$59.0, $-$41.9 & 1.55 & 47.63 & $-$52.5, $-$50.8\\
336.822+0.028 & 16:34:38.29 & $-$47:36:32.9 & 6035 & C01 & 1.17 & $-$77.13 & $-$79.1, $-$75.0 & 1.13 & $-$77.52 & $-$78.7, $-$76.8\\
336.941$-$0.156 & 16:35:55.20 & $-$47:38:45.8 & 6035 & C03 & 4.19 & $-$65.39 & $-$71.2, $-$64.5 & 1.52 & $-$64.90 & $-$71.2, $-$63.6\\
  &  &  & 6030 & C03 & 2.38 & $-$65.27 & $-$69.5, $-$68.1 & 0.40 & $-$64.89 & $-$68.1, $-$63.7\\
336.983$-$0.183 & 16:36:12.39 & $-$47:37:57.8 & 6035 & C03 & 0.56 & $-$65.70 & $-$66.6, $-$63.5 & 0.28 & $-$65.12 & $-$66.3, $-$63.9\\
337.098$-$0.928 & 16:39:58.012 & $-$48:02:44.399 & 6035 & ATCA 13 & 0.64 & $-$42.10 & $-$39.1, $-$37.8 & 0.21 & $-$42.01 & $-$42.3, $-$41.7\\
337.404$-$0.402 & 16:38:50.45 & $-$47:28:03.2 & 6035 & C03 & 1.56 & $-$35.51 & $-$42.2, $-$35.0 & 1.14 & $-$35.22 & $-$42.2, $-$34.7\\
  &  &  & 6030 & C03 & 0.46 & $-$35.50 & $-$36.3, $-$35.2 & 0.41 & $-$35.01 & $-$36.3, $-$34.6\\
337.606$-$0.052 & 16:38:09.54 & $-$47:04:59.9 & 6035 & C03 & 0.36 & $-$42.21 & $-$50.3, $-$37.3 & 0.38 & $-$42.41 & $-$43.8, $-$40.1\\
337.705$-$0.053 & 16:38:29.67 & $-$47:00:35.8 & 6035 & C03 & 1.65 & $-$51.13 & $-$55.7, $-$46.3 & 2.90 & $-$50.65 & $-$56.5, $-$46.4\\
  &  &  & 6030 & C03 & 0.35 & $-$49.66 & $-$55.8, $-$44.7 & 0.31 & $-$48.50 & $-$55.5, $-$46.7\\
337.844$-$0.374 & 16:40:26.716 & $-$47:07:12.25 & 6035 & ATCA 09 & 0.93 & $-$37.75 & $-$39.8, $-$36.9 & 2.17 & $-$37.95 & $-$39.7, $-$37.5\\
338.925+0.557 & 16:40:33.53 & $-$45:41:37.2 & 6035 & C03 & 0.47 & $-$62.80 & $-$64.5, $-$57.0 & 0.28 & $-$62.80 & $-$64.5, $-$57.5\\
339.053$-$0.315 & 16:44:48.99 & $-$46:10:13.1 & 6035 & C03 & 0.23 & $-$110.90 & $-$112.2, $-$110.1 & 0.24 & $-$111.28 & $-$112.5, $-$110.6\\
339.282+0.136 & 16:43:43.12 & $-$45:42:08.0 & 6035 & C03 & 0.59 & $-$70.87 & $-$73.0, $-$66.0 & 0.76 & $-$71.07 & $-$73.4, $-$66.2\\
339.622$-$0.121 & 16:46:05.96 & $-$45:36:44.1 & 6035 & C03 & 0.96 & $-$30.17 & $-$34.7, $-$28.9 & 1.90 & $-$30.47 & $-$34.7, $-$29.4\\
339.884$-$1.259 & 16:52:04.61 & $-$46:08:34.0 & 6035 & C01, C03 & 68.48 & $-$37.36 & $-$38.3, $-$36.6 & 74.85 & $-$37.36 & $-$38.5, $-$36.9\\
  &  &  & 6030 & C01, C03 & 42.14 & $-$37.15 & $-$38.4, $-$36.0 & 46.51 & $-$37.44 & $-$38.6, $-$35.7\\
339.980$-$0.539 & 16:49:14.949 & $-$45:36:31.13 & 6035 & ATCA 09 & 1.81 & $-$88.68 & $-$89.3, $-$88.5 & 1.12 & $-$88.97 & $-$89.3, $-$88.7\\
340.785$-$0.096 & 16:50:14.84 & $-$44:42:26.7 & 6035 & C03 & 5.93 & $-$102.07 & $-$106.3, $-$100.1 & 6.20 & $-$101.88 & $-$107.0, $-$100.1\\
341.974+0.225 & 16:53:05.355 & $-$43:35:09.69 & 6035 & ATCA 09 & 0.74 & $-$6.02 & $-$8.7, $-$4.0 & 0.32 & $-$4.66 & $-$112.5, $-$110.6\\
343.354$-$0.067 & 16:59:04.452 & $-$42:41:34.39 & 6035 & ATCA 09 & 0.52 & $-$116.89 & $-$118.5, $-$115.9 & 1.09 & $-$117.09 & $-$118.2, $-$116.4\\
343.929+0.125 & 17:00:10.91 & $-$42:07:19.3 & 6035 & C03 & 2.45 & 13.67 & 11.5, 16.2 & 1.49 & 14.06 & 12.0, 15.6\\
344.419+0.044 & 17:02:08.60 & $-$41:47:10.2 & 6035 & C03 & 0.73 & $-$63.26 & $-$65.0, $-$62.8 & 0.45 & $-$62.97 & $-$65.0, $-$62.5\\
345.010+1.792 & 16:56:47.58 & $-$40:14:25.7 & 6035 & C03 & 2.00 & $-$21.38 & $-$23.6, $-$14.8 & 1.16 & $-$17.69 & $-$24.0, $-$14.8\\
  &  &  & 6030 & C03 & 0.79 & $-$20.58 & $-$22.5, $-$18.9 & 0.85 & $-$20.29 & $-$22.3, $-$18.9\\
345.003$-$0.224 & 17:05:11.20 & $-$41:29:07.0 & 6035 & C03 & 6.45 & $-$25.95 & $-$31.1, $-$24.2 & 5.51 & $-$25.75 & $-$31.7, $-$24.4\\
345.407$-$0.951 & 17:09:35.40 & $-$41:35:55.0 & 6035 & C03 & 0.82 & $-$26.66 & $-$27.0, $-$25.5 & 0.80 & $-$26.27 & $-$26.8, $-$25.8\\
345.487+0.314 & 17:04:28.12 & $-$40:46:25.3 & 6035 & C01, C03 & 5.17 & $-$22.05 & $-$22.6, $-$21.5 & 2.57 & $-$22.15 & $-$22.6, $-$21.5\\
  &  &  & 6030 & C01, C03 & 1.00 & $-$22.03 & $-$23.2, $-$21.1 & 0.44 & $-$21.84 & $-$23.4, $-$21.1\\
345.495+1.469 & 16:59:41.741 & $-$40:03:39.70 & 6035 & ATCA 09 & 0.48 & $-$12.51 & $-$19.5, $-$9.4 & 0.55 & $-$12.31 & $-$19.5, $-$10.5\\
345.698$-$0.090 & 17:06:50.60 & $-$40:50:59.6 & 6035 & C03 & 8.44 & $-$6.45 & $-$9.2, $-$1.9 & 8.58 & $-$6.45 & $-$7.0, $-$3.0\\
  &  &  & 6030 & C03 & 3.17 & $-$4.88 & $-$7.3, $-$2.4 & 2.72 & $-$4.78 & $-$6.6, $-$2.7\\
347.628+0.149 & 17:11:50.89 & $-$39:09:29.0 & 6035 & C03 & 5.41 & $-$96.87 & $-$97.5, $-$94.1 & 11.92 & $-$96.48 & $-$97.8, $-$94.7\\
  &  &  & 6030 & C03 & 2.00 & $-$96.85 & $-$97.6, $-$96.5 & 3.15 & $-$96.57 & $-$97.1, $-$96.1\\
348.698$-$1.027 & 17:19:58.98 & $-$38:58:13.5 & 6035 & C03 & 2.58 & $-$14.61 & $-$17.1, $-$13.1 & 0.56 & $-$15.87 & $-$16.8, $-$13.3\\
350.014+0.434 & 17:17:45 & $-$37:03:12.9 & 6035 & C98 & 0.29 & $-$35.43 & $-$36.0, $-$31.5 & 0.31 & $-$35.62 & $-$37.0, $-$31.4\\
350.113+0.095 & 17:19:25.58 & $-$37:10:04.4 & 6035 & C03 & 0.38 & $-$72.06 & $-$72.6, $-$70.5 & 0.45 & $-$72.25 & $-$72.7, $-$70.7\\
350.686$-$0.491 & 17:23:28.63 & $-$37:01:48.1 & 6035 & C03 & 1.58 & $-$13.64 & $-$15.9, $-$13.3 & 2.11 & $-$13.83 & $-$16.3, $-$13.3\\
351.417+0.645 & 17:20:53.37 & $-$35:47:01.2 & 6035 & C03 & 333.07 & $-$10.25 & $-$11.2, $-$6.9 & 253.32 & $-$10.45 & $-$11.5, $-$7.3\\
  &  &  & 6030 & C03 & 39.21 & $-$10.24 & $-$11.1, $-$6.1 & 26.66 & $-$10.63 & $-$11.5, $-$7.2\\
351.581$-$0.353 & 17:25:25.08 & $-$36:12:46.1 & 6035 & C03 & 7.98 & $-$96.41 & $-$101.3, $-$89.8 & 3.65 & $-$96.31 & $-$101.3, $-$89.5\\
  &  &  & 6030 & C03 & 0.66 & $-$96.00 & $-$103.9, $-$91.6 & 0.78 & $-$96.30 & $-$103.9, $-$91.6\\
351.775$-$0.536 & 17:26:42.56 & $-$36:09:16.0 & 6035 & C03 & 2.31 & $-$7.44 & $-$9.5, $-$2.3 & 4.40 & $-$7.64 & $-$9.0, $-$2.4\\
353.410$-$0.360 & 17:30:26.18 & $-$34:41:46.0 & 6035 & C03 & 25.48 & $-$20.65 & $-$22.4, $-$17.5 & 16.56 & $-$20.75 & $-$22.9, $-$17.5\\
  &  &  & 6030 & C03 & 3.74 & $-$21.61 & $-$22.7, $-$20.3 & 4.00 & $-$22.29 & $-$23.2, $-$21.7\\
354.725+0.299 & 17:31:15.831 & $-$33:14:04.70 & 6035 & C03, ATCA 09 & 0.97 & 91.76 & 88.7, 95 & 0.89 & 90.20 & 88.7, 94.7\\
355.344+0.147 & 17:33:29.05 & $-$32:47:58.8 & 6035 & C03 & 5.08 & 18.11 & 17.4, 19.6 & 5.00 & 17.81 & 17.1, 18.9\\
  &  &  & 6030 & C03 & 2.81 & 18.22 & 16.8, 19.8 & 3.13 & 17.83 & 16.8, 19.2\\
357.924$-$0.338 & 17:41:55.472 & $-$30:52:50.77 & 6035 & ATCA 09 & 0.73 & $-$3.57 & $-$4.0, $-$3.0 & 0.91 & $-$3.57 & $-$4.0, $-$3.2\\
359.137+0.031 & 17:43:25.64 & $-$29:39:18.3 & 6035 & C03 & 2.93 & $-$1.93 & $-$3.3, 0.9 & 2.58 & $-$1.93 & $-$3.8, 0.5\\
\label{Results:tab}
\end{longtable} 
\end{landscape}
}
\twocolumn

\newpage
\onecolumn
\begin{figure}
\centering
	\subfloat[\textit{MMBOH-G000.666-00.029 \& MMBOH-G000.666-00.035}]{\includegraphics[width=0.33\textwidth]{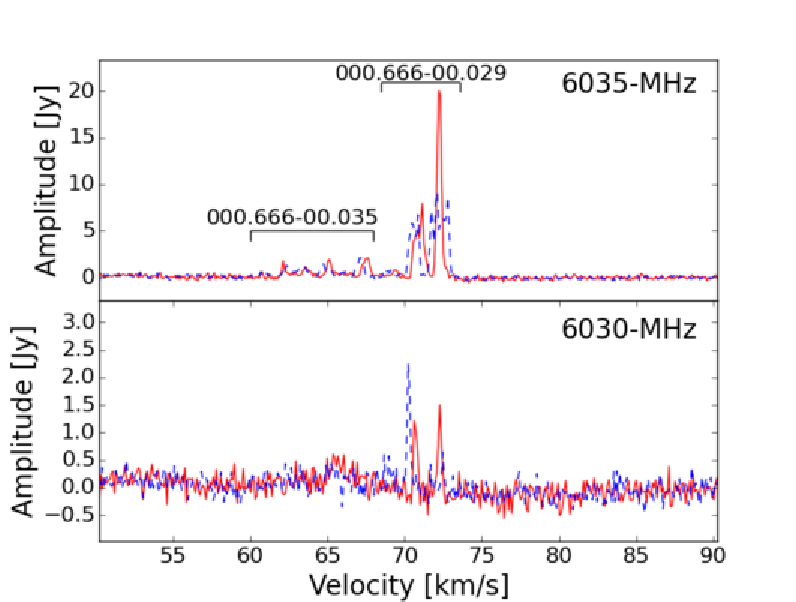}}
	\subfloat[\textit{MMBOH-G004.682+00.278}]{\includegraphics[width=0.33\textwidth]{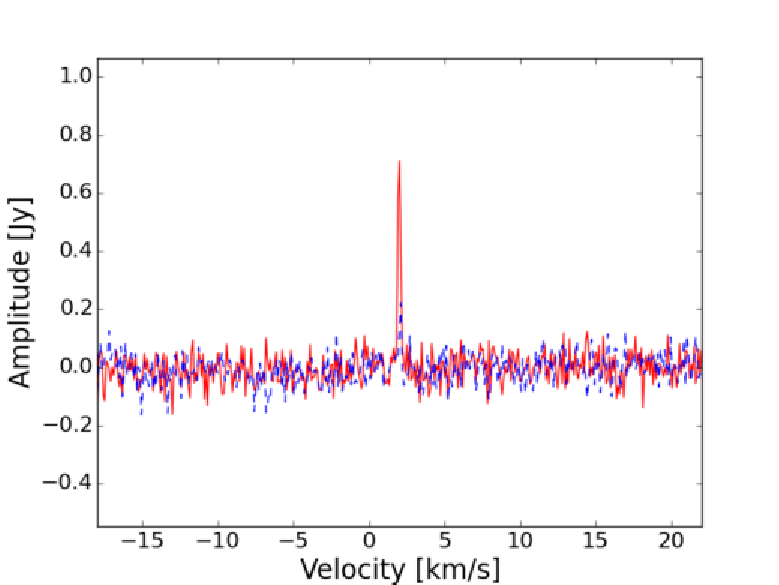}}
	\subfloat[\textit{MMBOH-G005.885-00.392}]{\includegraphics[width=0.33\textwidth]{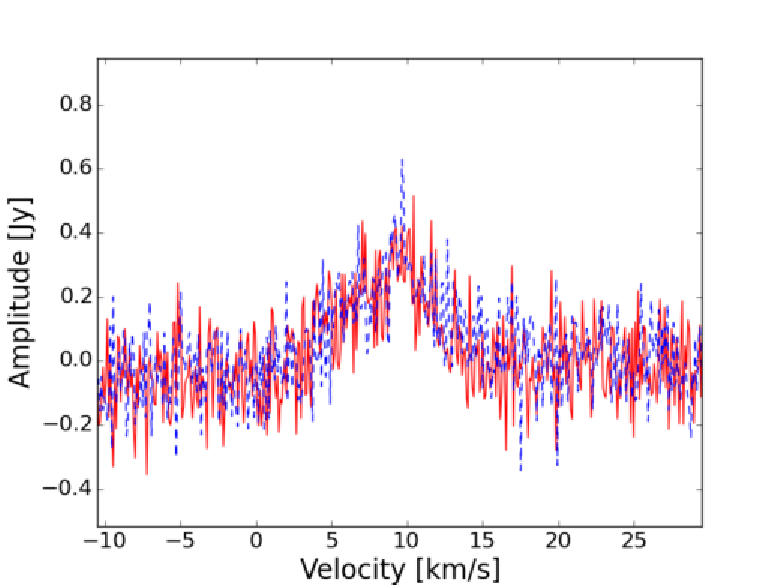}}
\qquad
	\subfloat[\textit{MMBOH-G006.882+00.094}]{\includegraphics[width=0.33\textwidth]{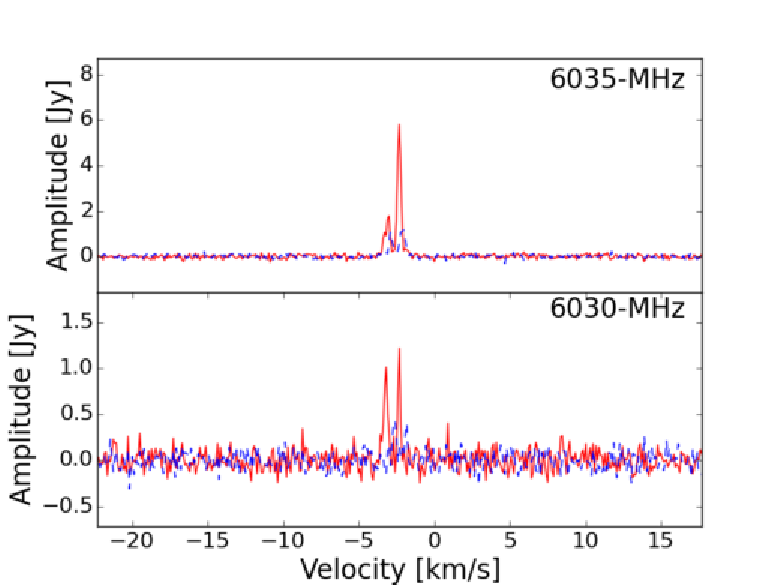}}
	\subfloat[\textit{MMBOH-G008.352+00.478}]{\includegraphics[width=0.33\textwidth]{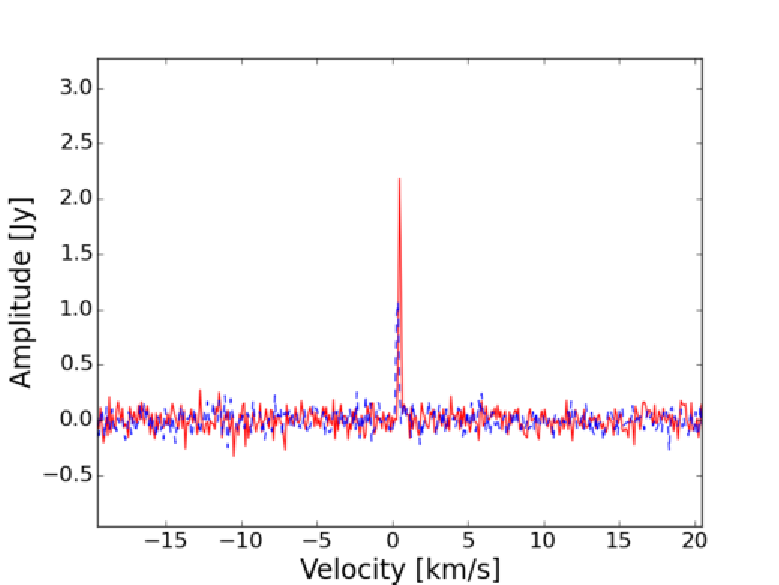}}
	\subfloat[\textit{MMBOH-G008.669-00.356}]{\includegraphics[width=0.33\textwidth]{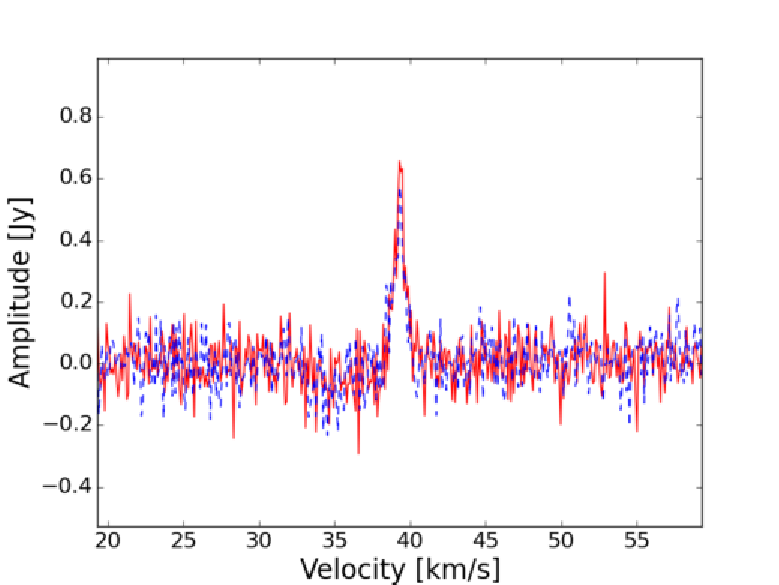}}
\qquad
	\subfloat[\textit{MMBOH-G009.620+00.194 \& MMBOH-G009.622+00.196}]{\includegraphics[width=0.33\textwidth]{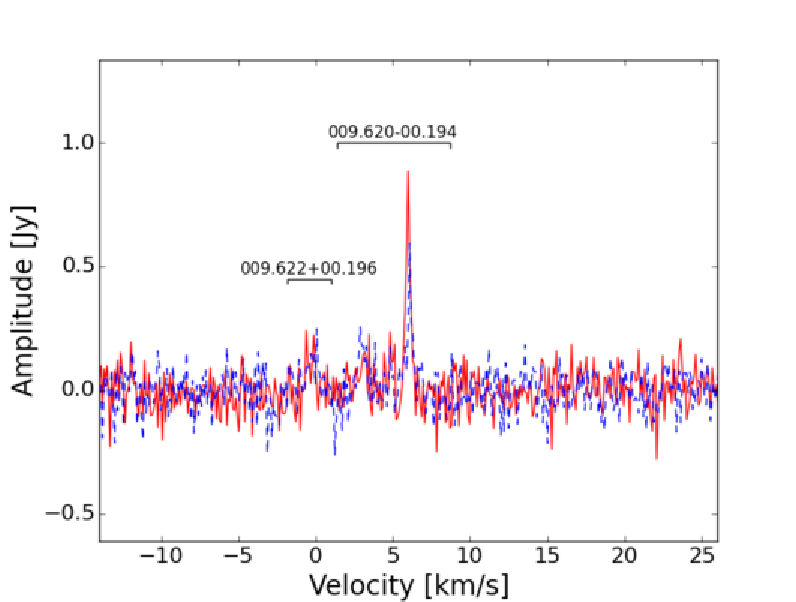}}
	\subfloat[\textit{MMBOH-G010.322-00.258}]{\includegraphics[width=0.33\textwidth]{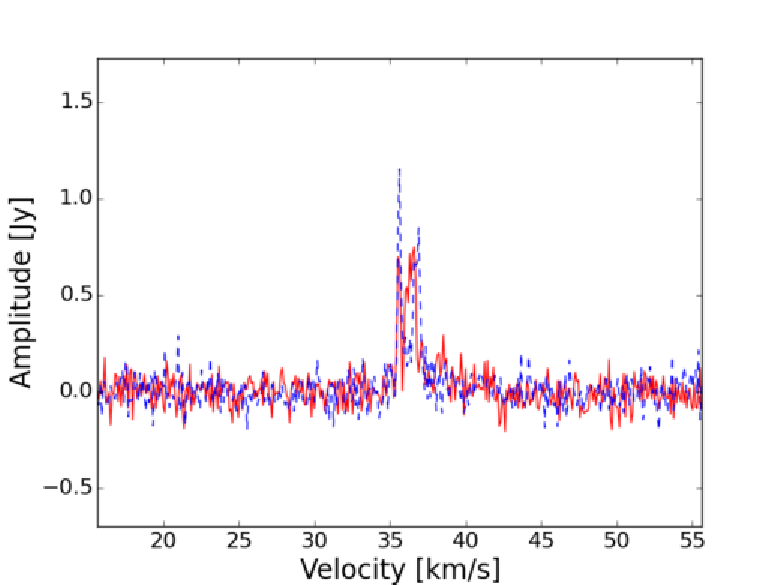}}
	\subfloat[\textit{MMBOH-G010.623-00.384}]{\includegraphics[width=0.33\textwidth]{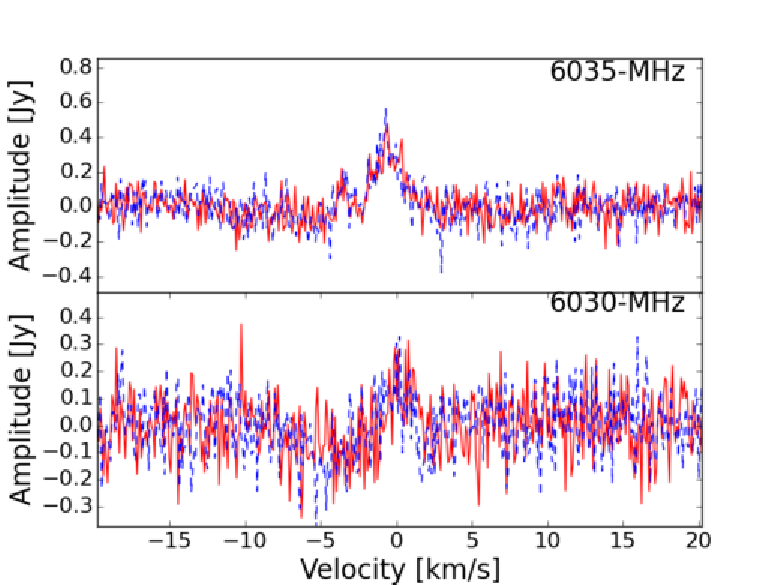}}
\qquad
	\subfloat[\textit{MMBOH-G010.960+00.022}]{\includegraphics[width=0.33\textwidth]{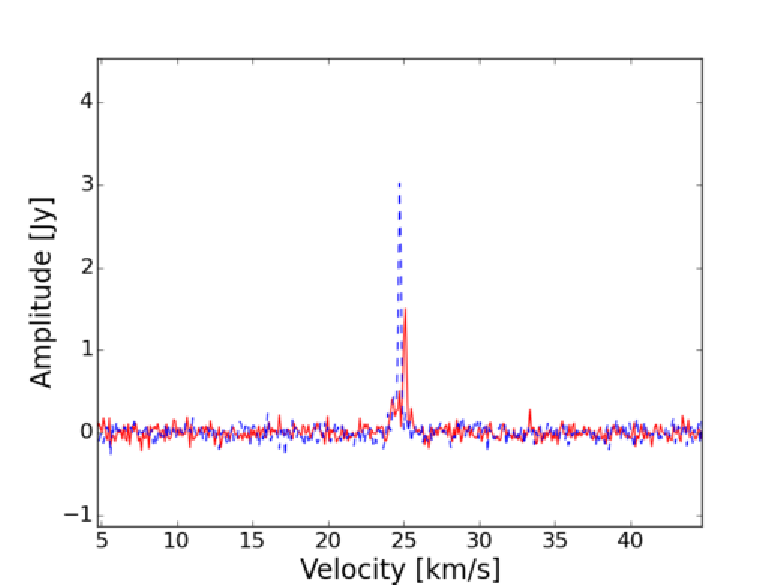}}
	\subfloat[\textit{MMBOH-G011.034+00.062}]{\includegraphics[width=0.33\textwidth]{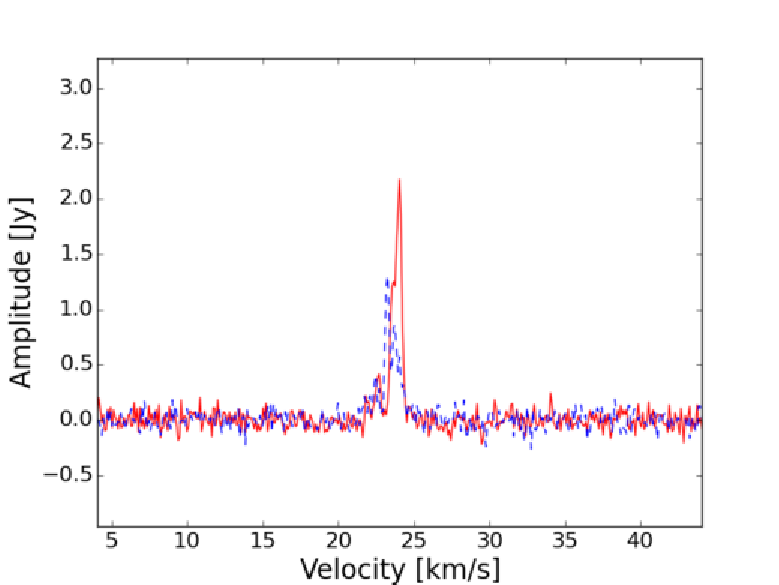}} 
	\subfloat[\textit{MMBOH-G011.904-00.141}]{\includegraphics[width=0.33\textwidth]{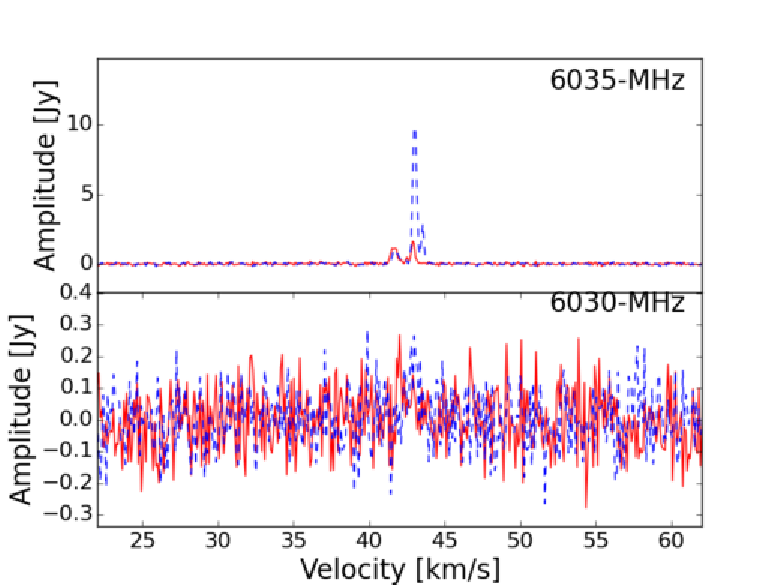}}
	\caption{MMB ex-OH maser spectra from the Parkes `MX' observations. Red (solid) and blue (dashed) lines are left and right hand circular polarisations respectively. Spectra for two sources are presented (\textit{49.046$-$0.290} and \textit{326.447$-$0.749}) in black (dash-dot) lines, these are Stokes I spectra from ATCA as these sources are not present in the nearest Parkes MX. The full version of this figure is available in the online version of this paper.}
	\label{abcd}
\end{figure}
\renewcommand\thefigure{11}\renewcommand\thefigure{11}
\begin{figure}
\ContinuedFloat
\centering
	\subfloat[\textit{MMBOH-G012.681-00.182}]{\includegraphics[width=0.33\textwidth]{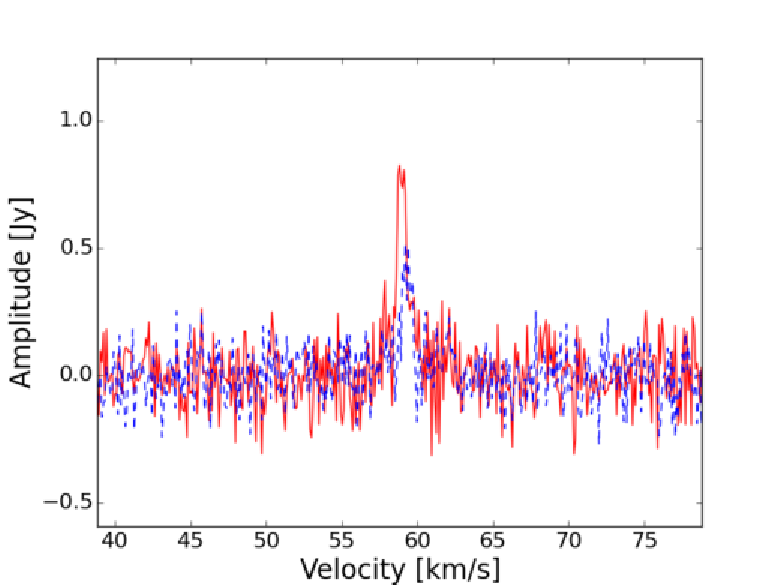}}
	\subfloat[\textit{MMBOH-G015.035-00.677}]{\includegraphics[width=0.33\textwidth]{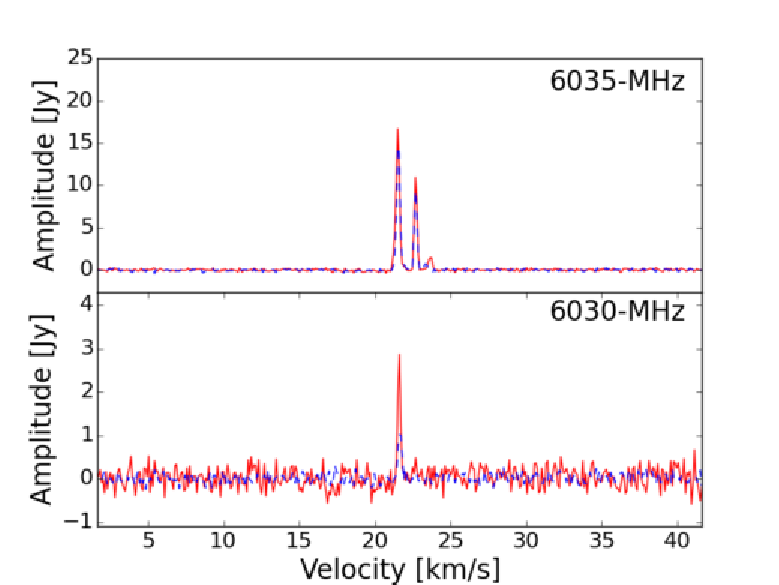}}
	\subfloat[\textit{MMBOH-G018.460-00.005}]{\includegraphics[width=0.33\textwidth]{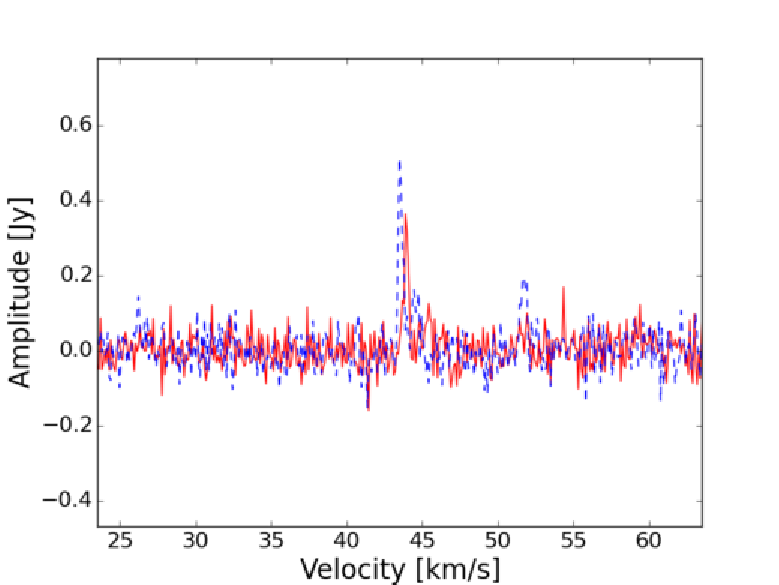}}
\qquad
	\subfloat[\textit{MMBOH-G018.836-00.299}]{\includegraphics[width=0.33\textwidth]{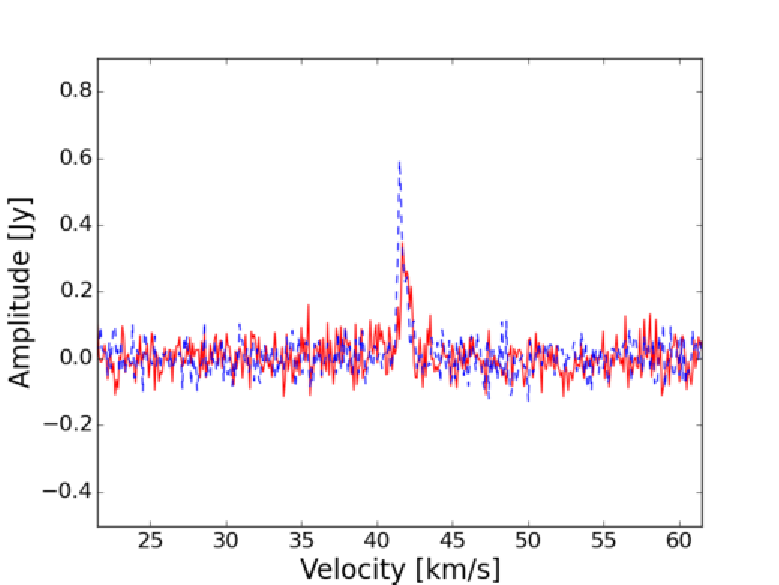}}
	\subfloat[\textit{MMBOH-G019.486+00.151}]{\includegraphics[width=0.33\textwidth]{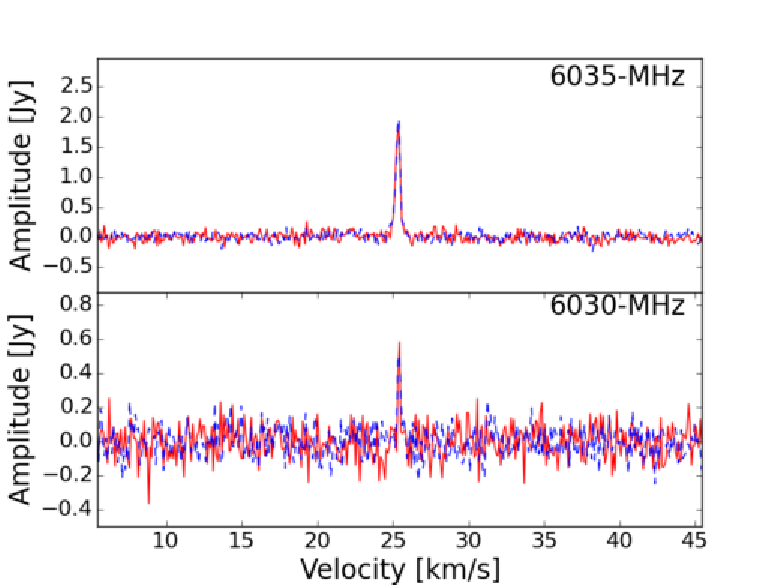}}
	\subfloat[\textit{MMBOH-G019.752-00.191}]{\includegraphics[width=0.33\textwidth]{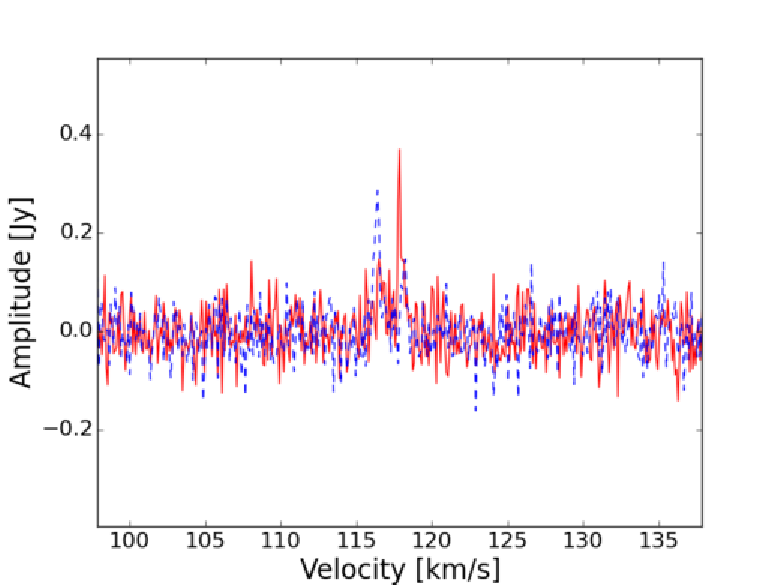}}
\qquad
	\subfloat[\textit{MMBOH-G020.237+00.065}]{\includegraphics[width=0.33\textwidth]{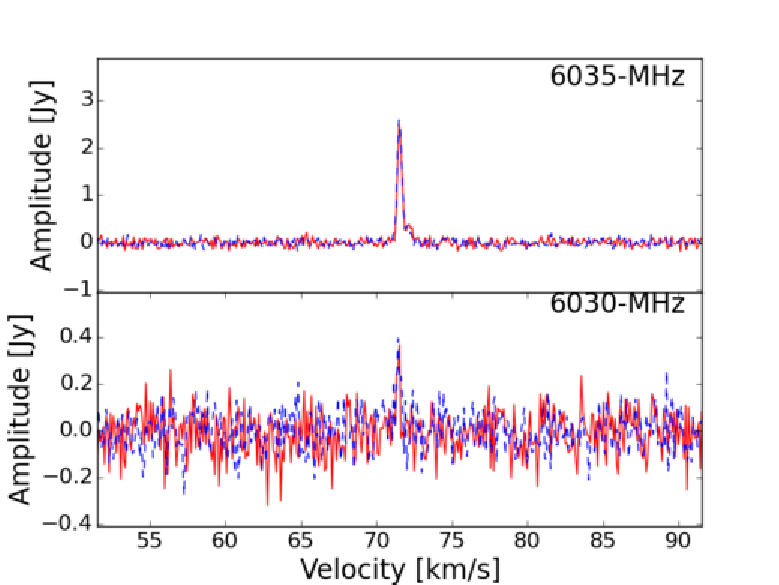}}
	\subfloat[\textit{MMBOH-G024.147-00.010}]{\includegraphics[width=0.33\textwidth]{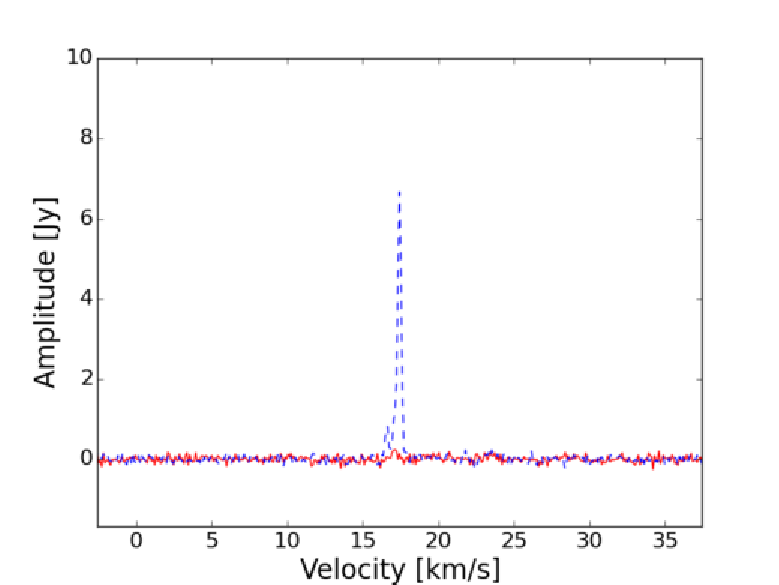}}
	\subfloat[\textit{MMBOH-G025.509-00.060}]{\includegraphics[width=0.33\textwidth]{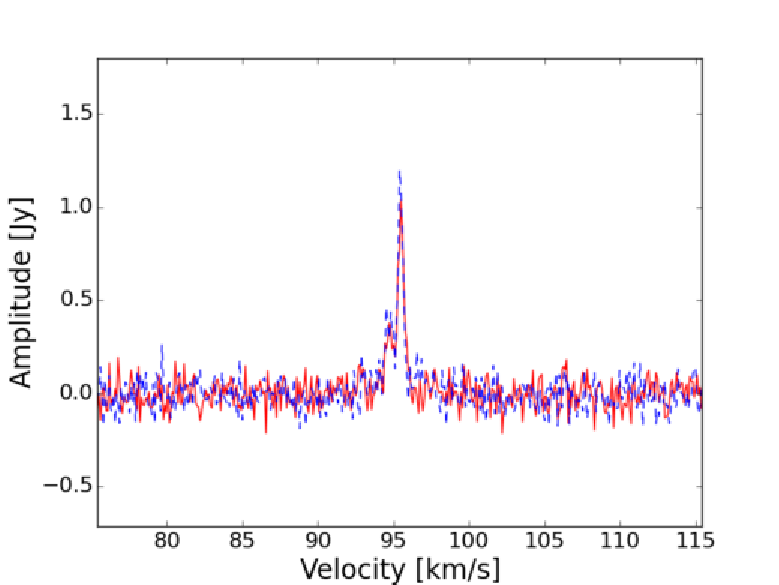}}
\qquad
	\subfloat[\textit{MMBOH-G025.648+01.050}]{\includegraphics[width=0.33\textwidth]{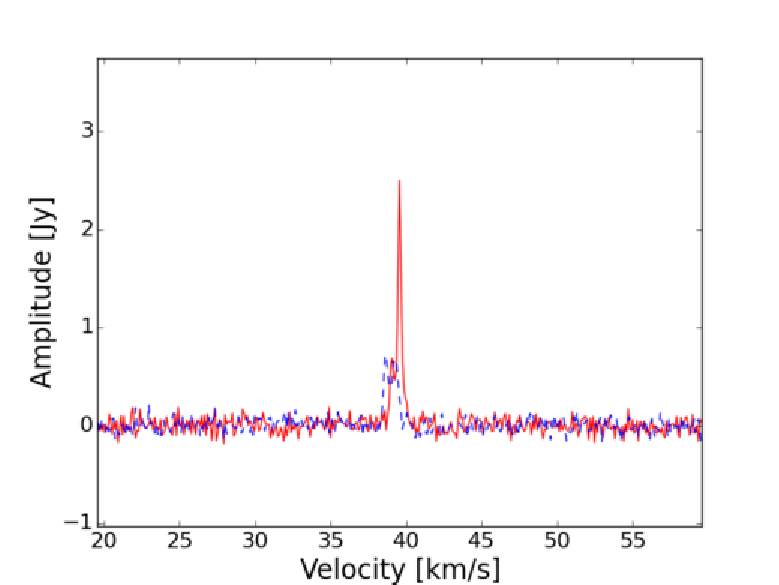}} 
	\subfloat[\textit{MMBOH-G028.200-00.049}]{\includegraphics[width=0.33\textwidth]{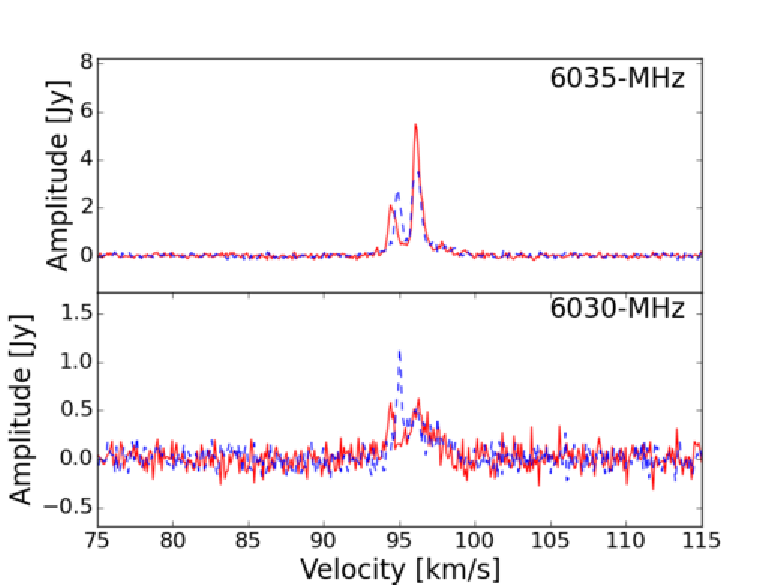}}
	\subfloat[\textit{MMBOH-G028.819+00.366}]{\includegraphics[width=0.33\textwidth]{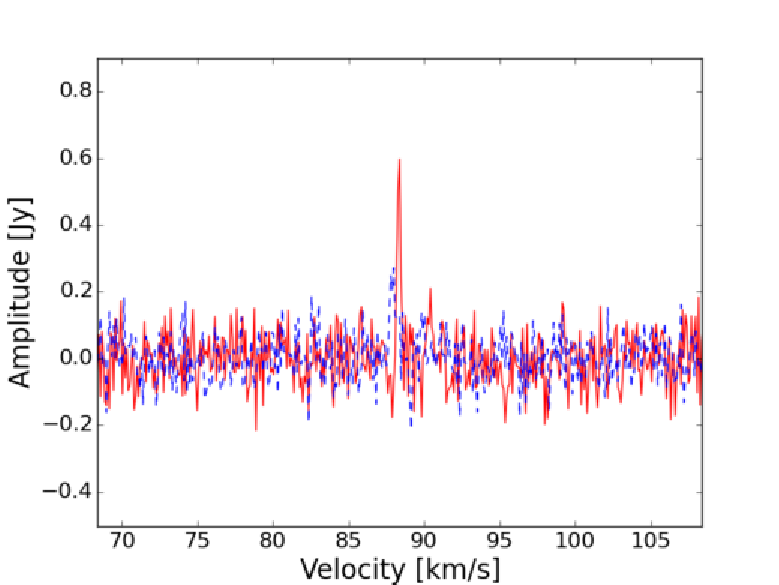}}
	\caption{(continued)}
\end{figure}
\renewcommand\thefigure{11}
\begin{figure}
\ContinuedFloat
\centering
	\subfloat[\textit{MMBOH-G030.778-00.801}]{\includegraphics[width=0.33\textwidth]{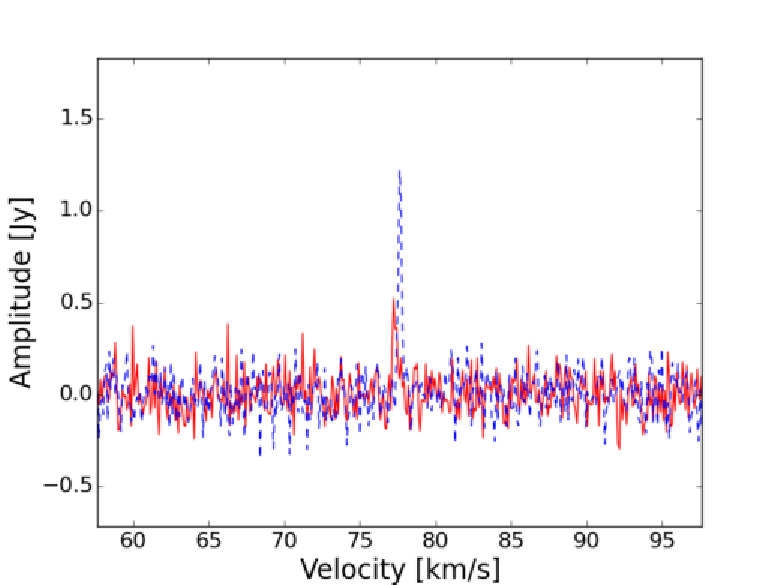}}
	\subfloat[\textit{MMBOH-G032.744-00.076}]{\includegraphics[width=0.33\textwidth]{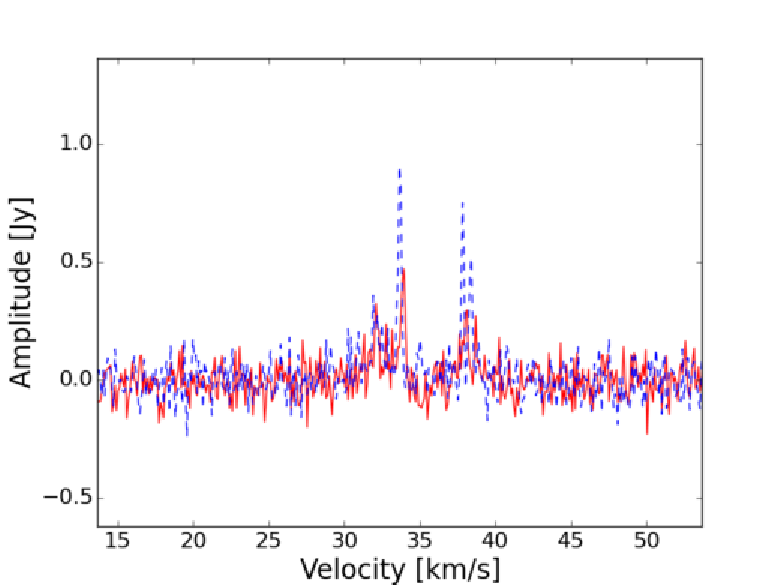}}
	\subfloat[\textit{MMBOH-G034.258+00.153}]{\includegraphics[width=0.33\textwidth]{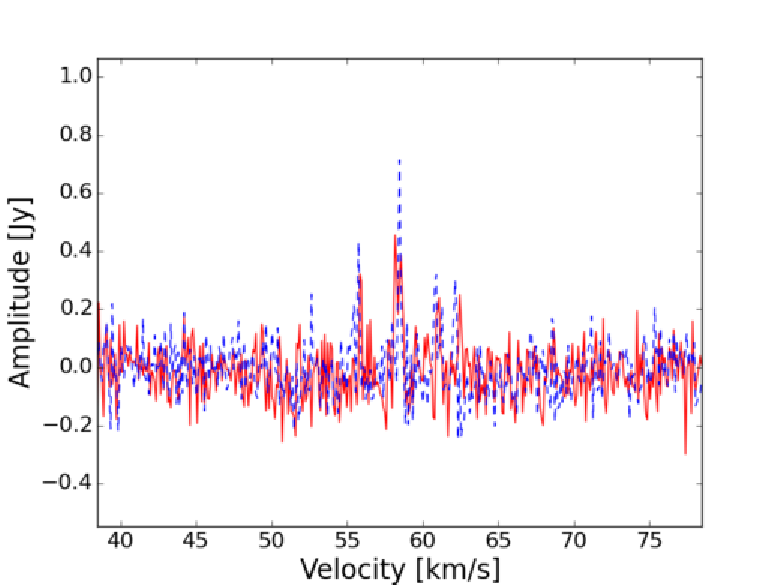}}
\qquad
	\subfloat[\textit{MMBOH-G034.258+00.153b}]{\includegraphics[width=0.33\textwidth]{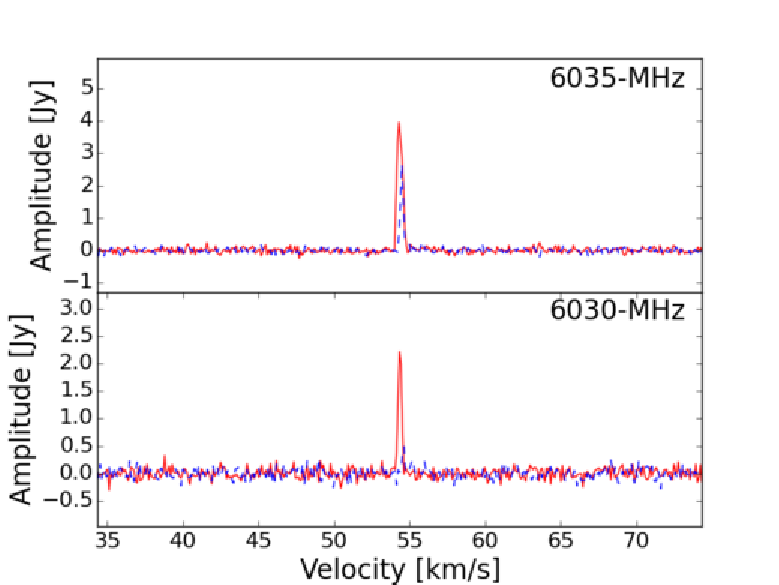}}
	\subfloat[\textit{MMBOH-G034.261-00.213}]{\includegraphics[width=0.33\textwidth]{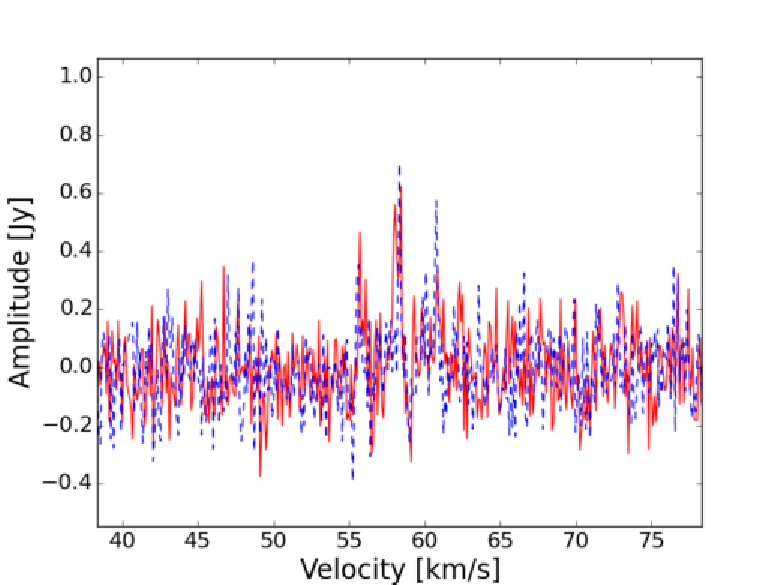}}
	\subfloat[\textit{MMBOH-G035.025+00.350}]{\includegraphics[width=0.33\textwidth]{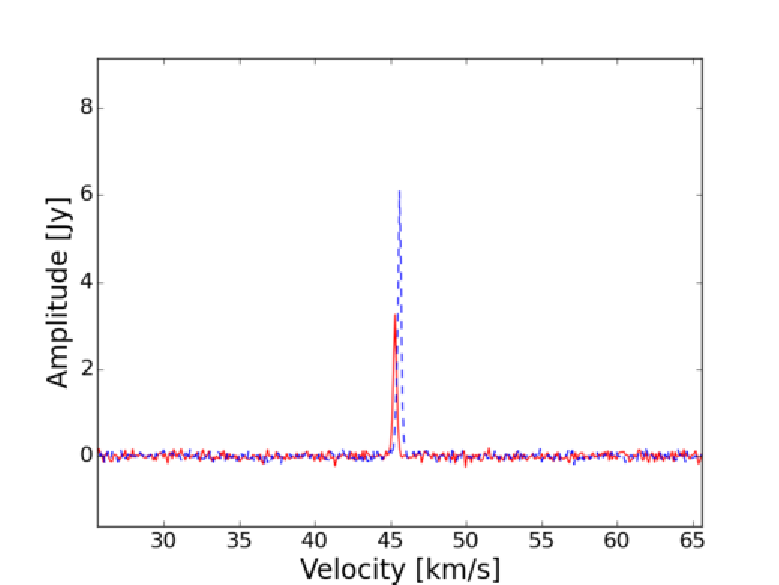}}
\qquad
	\subfloat[\textit{MMBOH-G035.133-00.744}]{\includegraphics[width=0.33\textwidth]{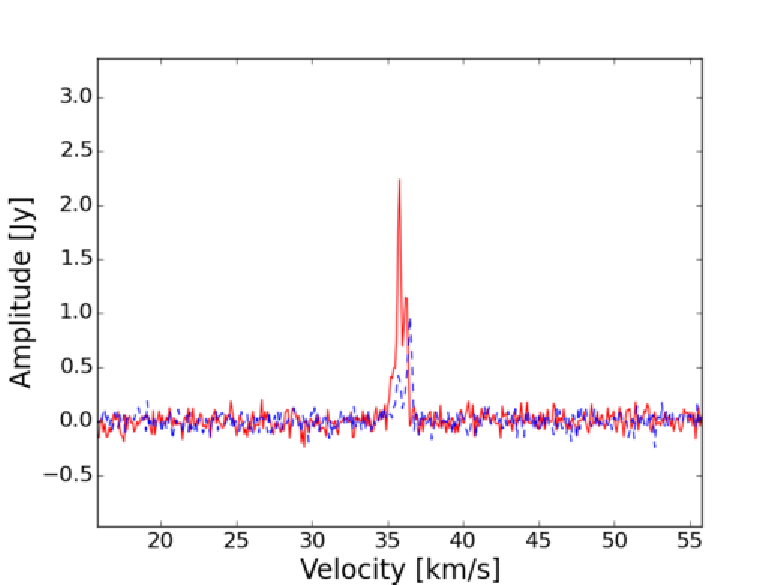}}
	\subfloat[\textit{MMBOH-G035.198-00.743}]{\includegraphics[width=0.33\textwidth]{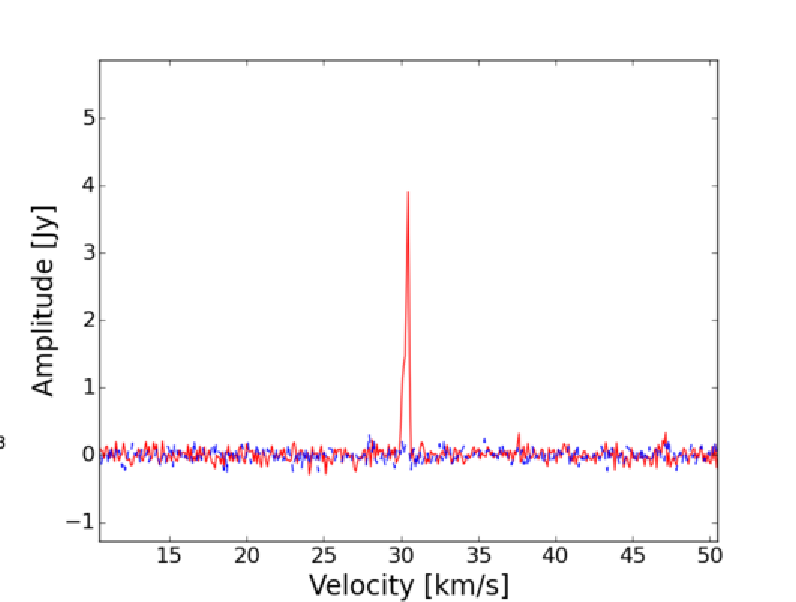}}
	\subfloat[\textit{MMBOH-G035.200-01.736}]{\includegraphics[width=0.33\textwidth]{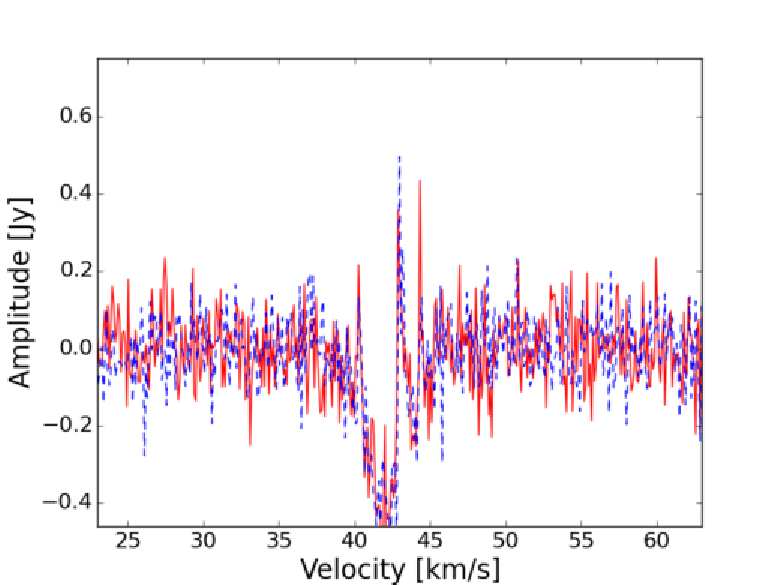}}
\qquad
	\subfloat[\textit{MMBOH-G040.282-00.220}]{\includegraphics[width=0.33\textwidth]{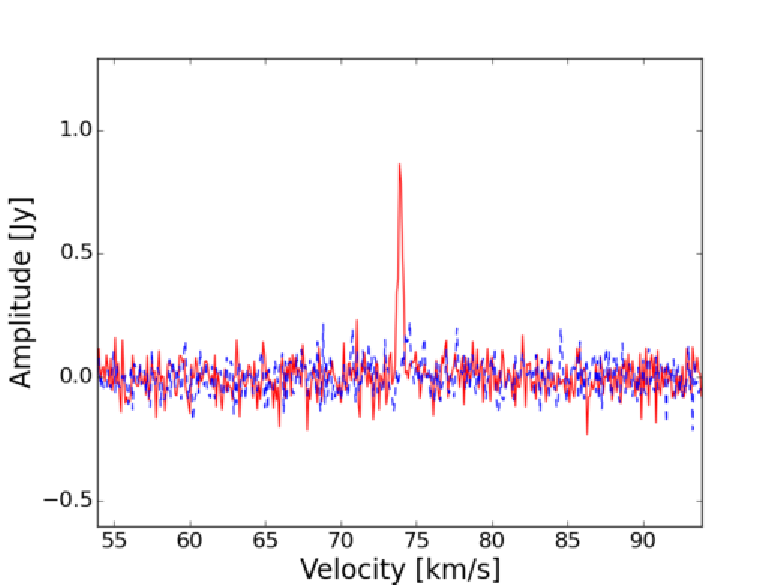}} 
	\subfloat[\textit{MMBOH-G040.426+00.701}]{\includegraphics[width=0.33\textwidth]{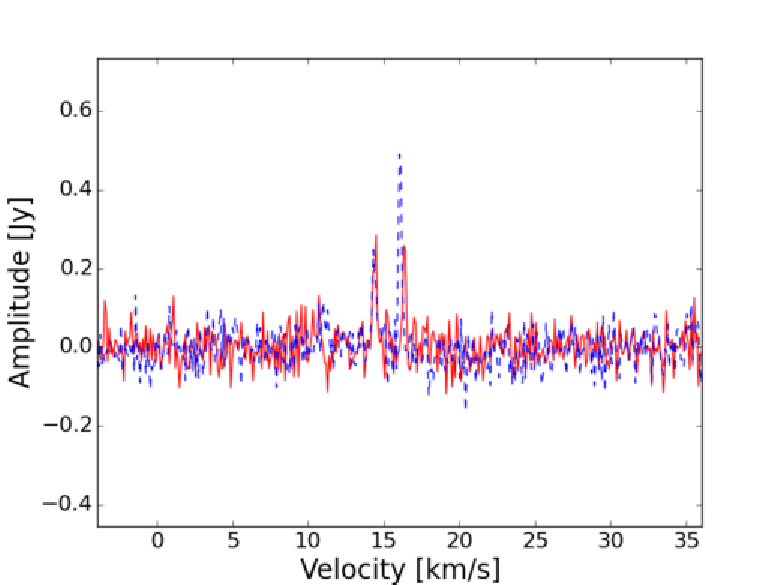}}
	\subfloat[\textit{MMBOH-G043.149+00.013}]{\includegraphics[width=0.33\textwidth]{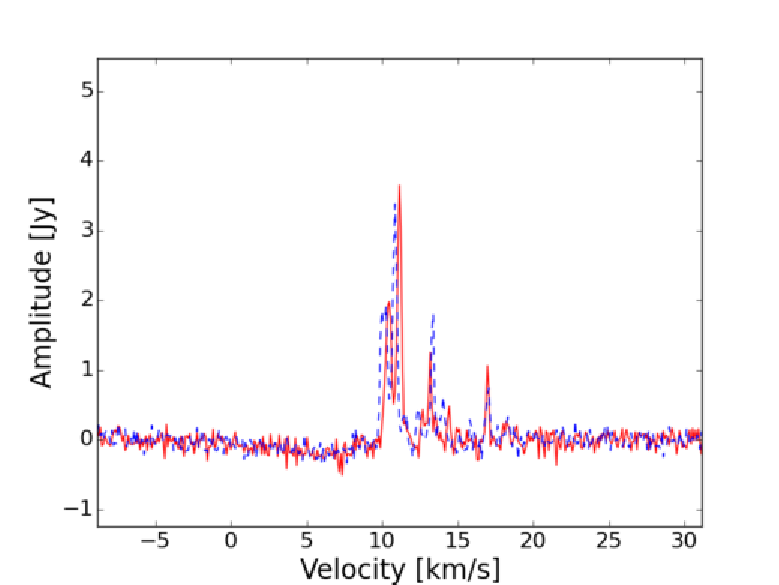}}
	\caption{(continued)}
\end{figure}
\renewcommand\thefigure{11}
\begin{figure}
\ContinuedFloat
\centering
	\subfloat[\textit{MMBOH-G043.165+00.013}]{\includegraphics[width=0.33\textwidth]{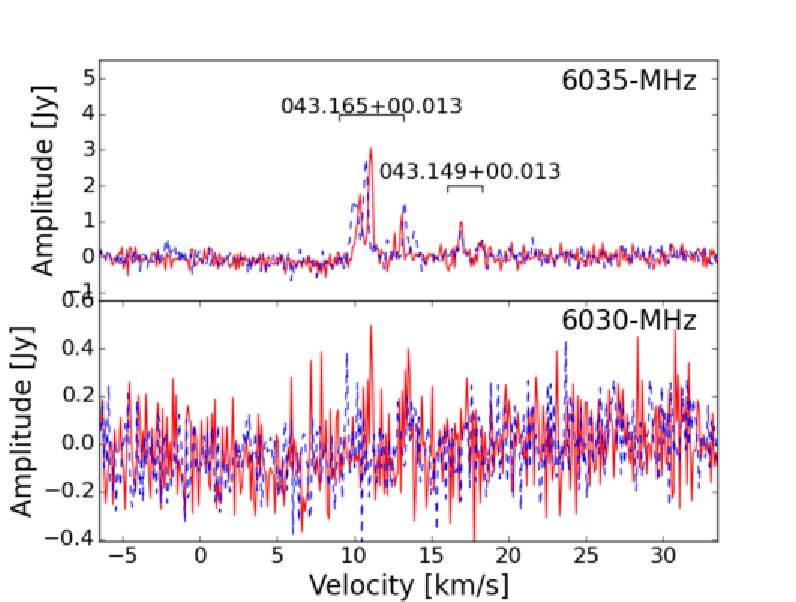}}
	\subfloat[\textit{MMBOH-G043.796-00.127}]{\includegraphics[width=0.33\textwidth]{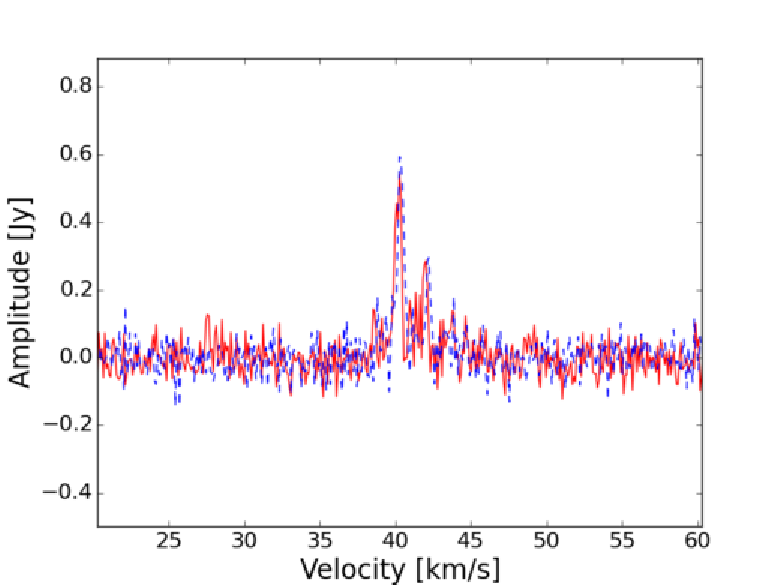}}
	\subfloat[\textit{MMBOH-G045.123+00.133}]{\includegraphics[width=0.33\textwidth]{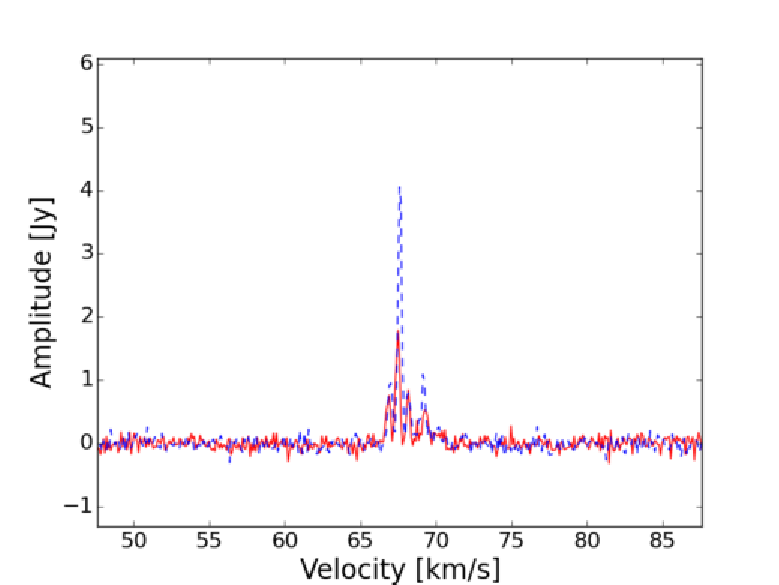}}
\qquad
	\subfloat[\textit{MMBOH-G045.466+00.045}]{\includegraphics[width=0.33\textwidth]{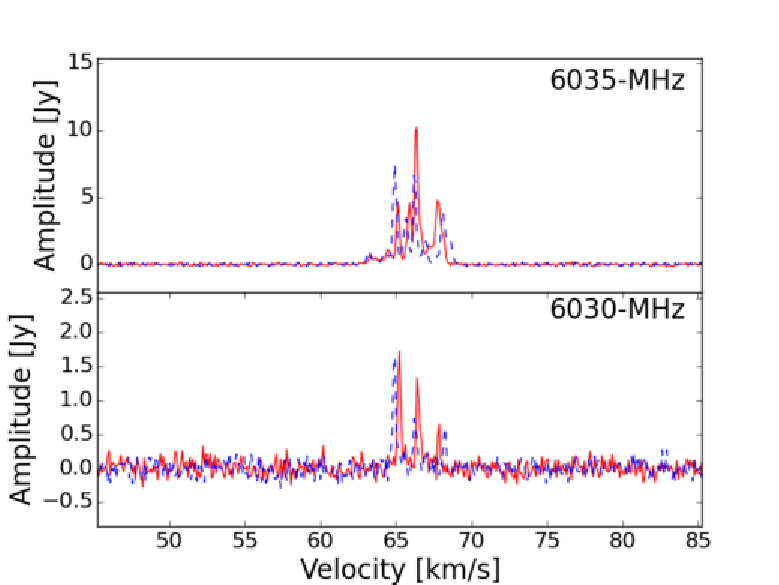}}
	\subfloat[\textit{MMBOH-G048.988-00.300}]{\includegraphics[width=0.33\textwidth]{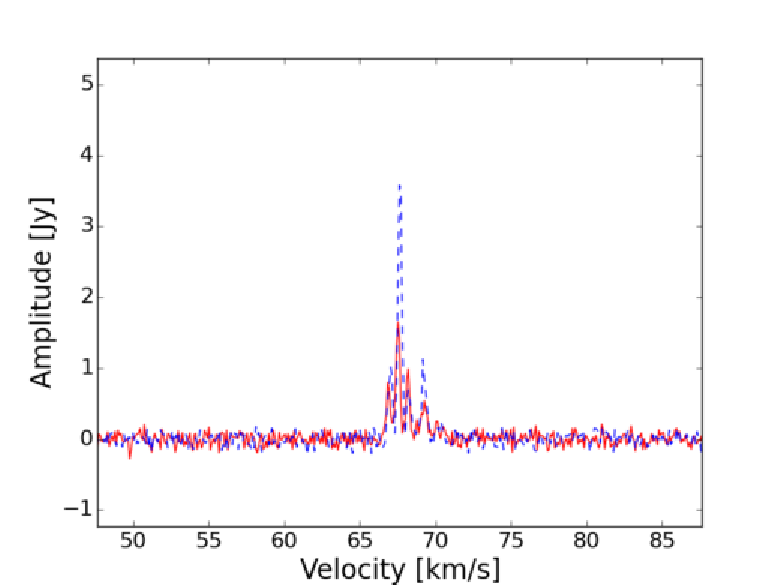}}
	\subfloat[\textit{MMBOH-G049.046-00.290}]{\includegraphics[width=0.33\textwidth]{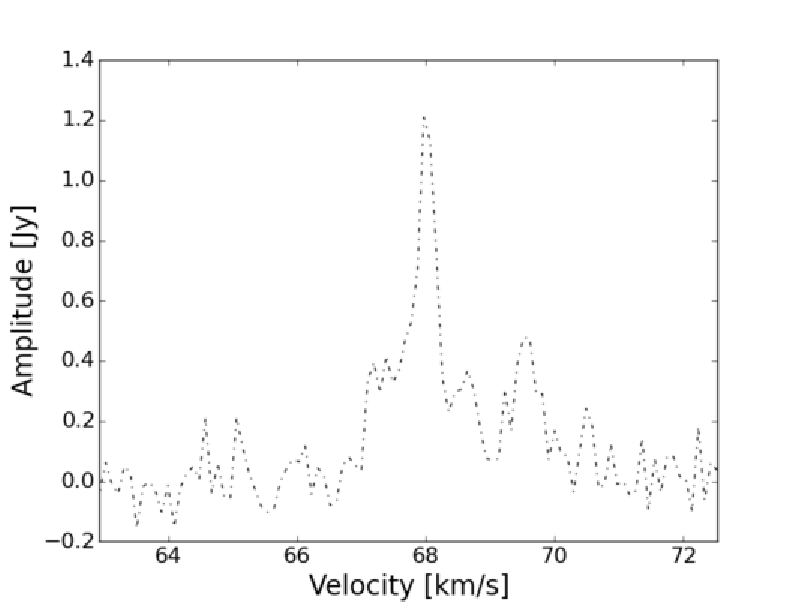}}
\qquad
	\subfloat[\textit{MMBOH-G049.486-00.389}]{\includegraphics[width=0.33\textwidth]{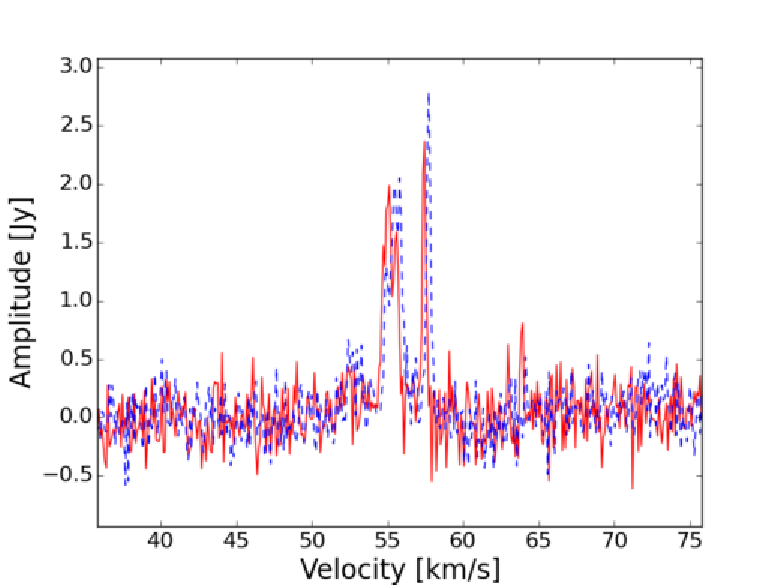}}
	\subfloat[\textit{MMBOH-G049.490-00.388}]{\includegraphics[width=0.33\textwidth]{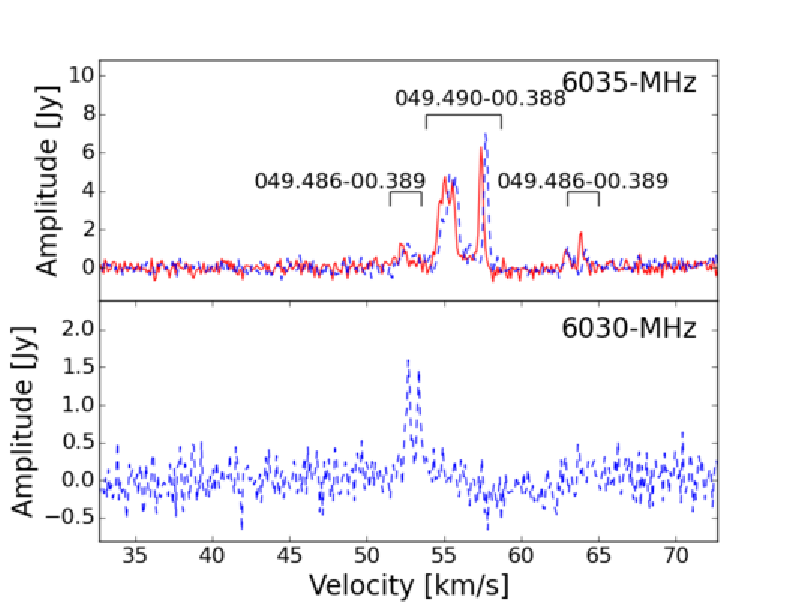}}
	\subfloat[\textit{MMBOH-G050.478+00.705}]{\includegraphics[width=0.33\textwidth]{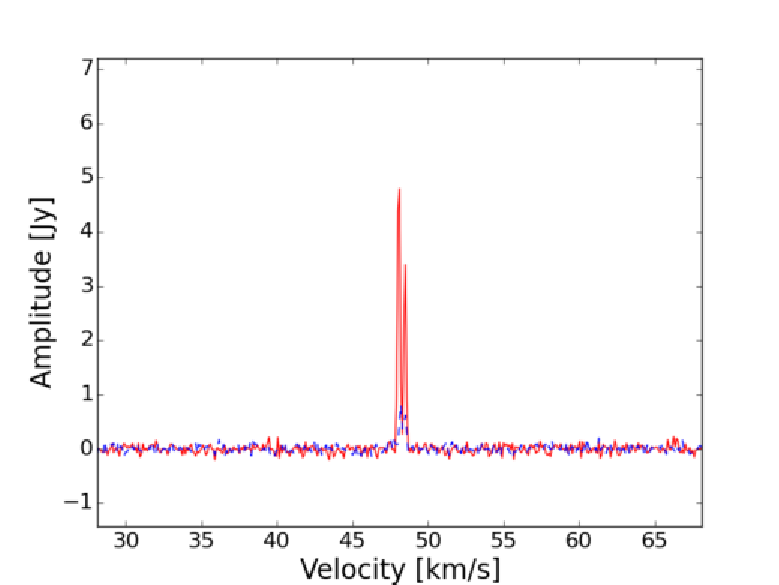}}
\qquad
	\subfloat[\textit{MMBOH-G051.681+00.714}]{\includegraphics[width=0.33\textwidth]{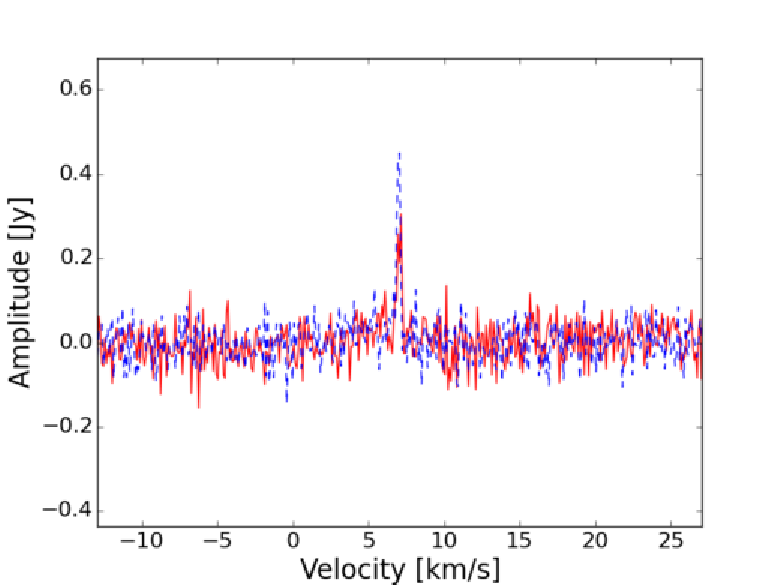}} 
	\subfloat[\textit{MMBOH-G189.030+00.783}]{\includegraphics[width=0.33\textwidth]{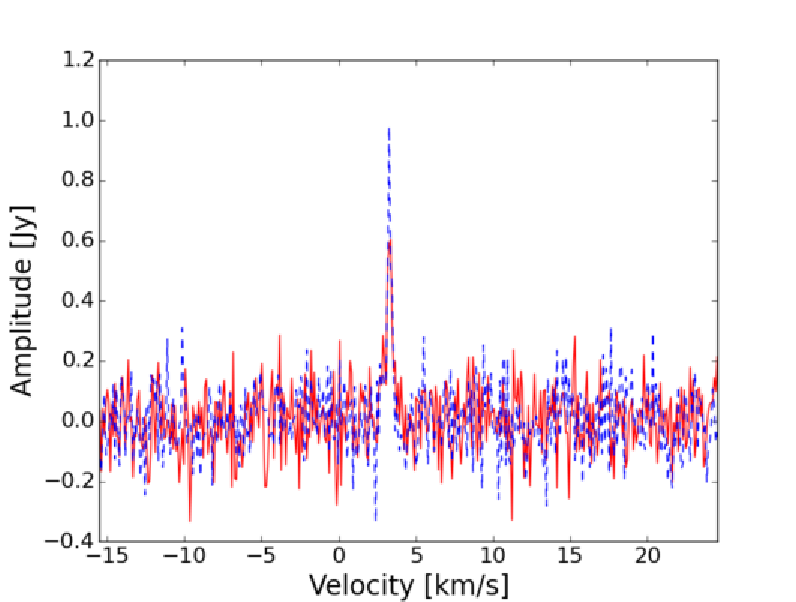}} 
	\subfloat[\textit{MMBOH-G240.316+00.071}]{\includegraphics[width=0.33\textwidth]{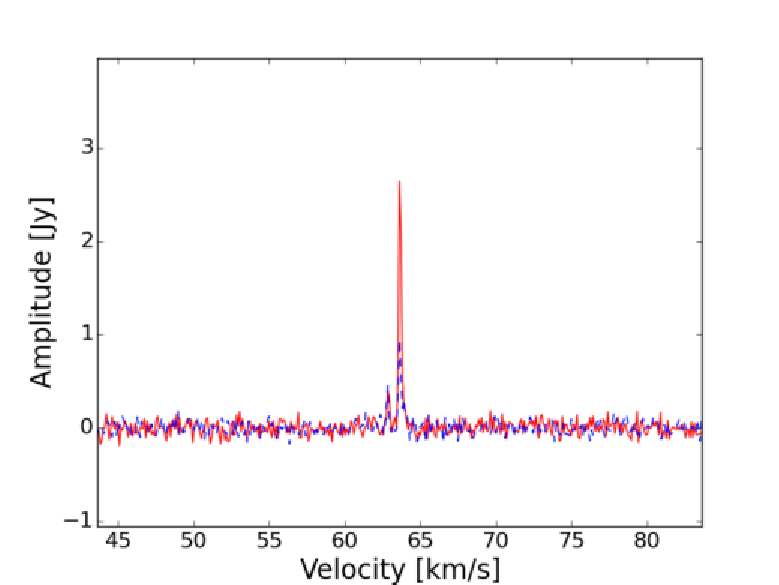}}
	\caption{(continued)}
\end{figure}
\renewcommand\thefigure{11}
\begin{figure}
\ContinuedFloat
\centering
	\subfloat[\textit{MMBOH-G284.016-00.856}]{\includegraphics[width=0.33\textwidth]{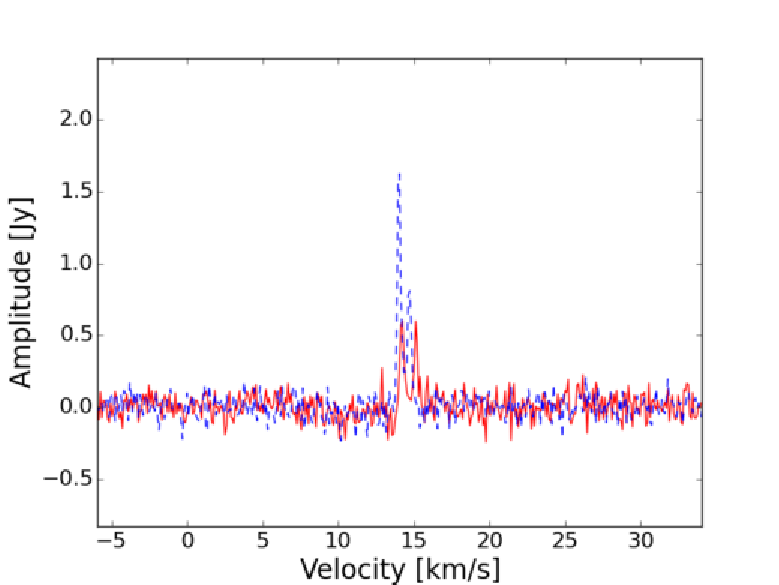}}
	\subfloat[\textit{MMBOH-G284.351-00.418}]{\includegraphics[width=0.33\textwidth]{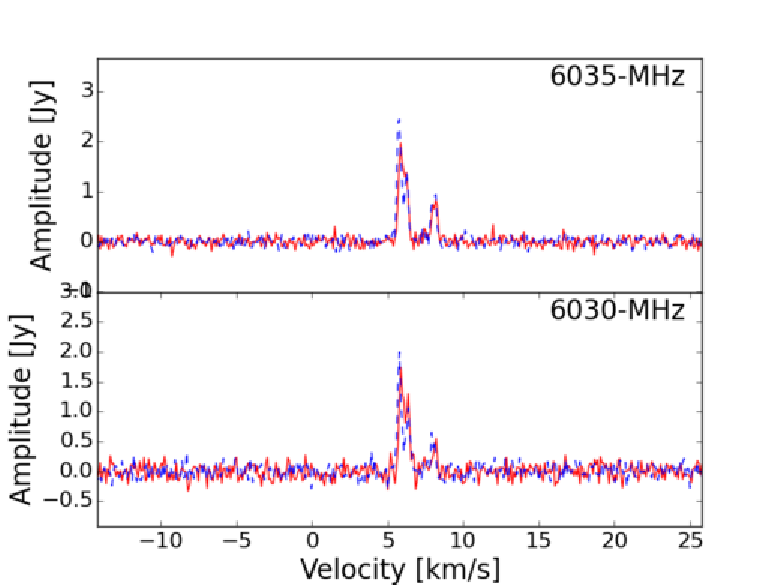}}
	\subfloat[\textit{MMBOH-G294.511-01.621}]{\includegraphics[width=0.33\textwidth]{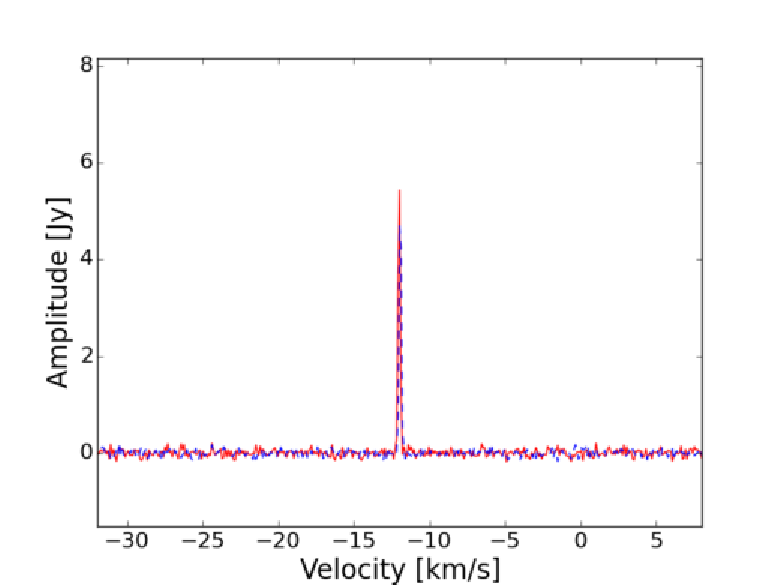}}
\qquad
	\subfloat[\textit{MMBOH-G298.723-00.086}]{\includegraphics[width=0.33\textwidth]{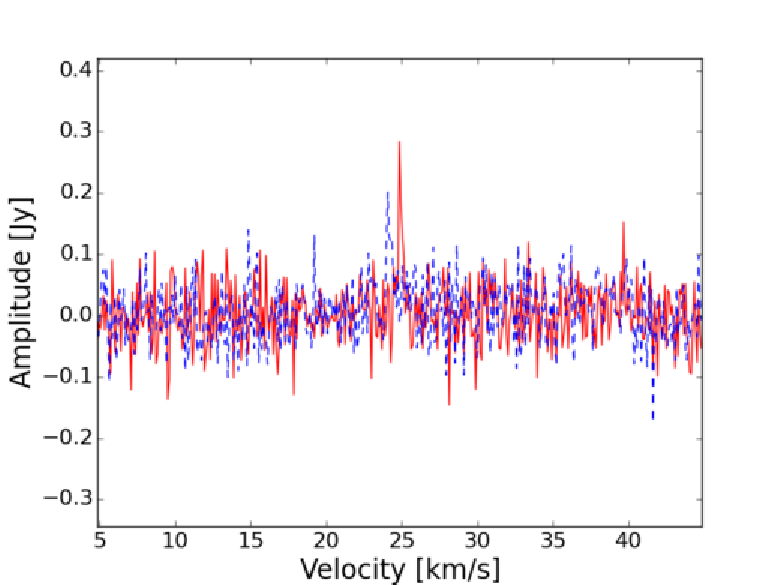}}
	\subfloat[\textit{MMBOH-G300.969+01.148}]{\includegraphics[width=0.33\textwidth]{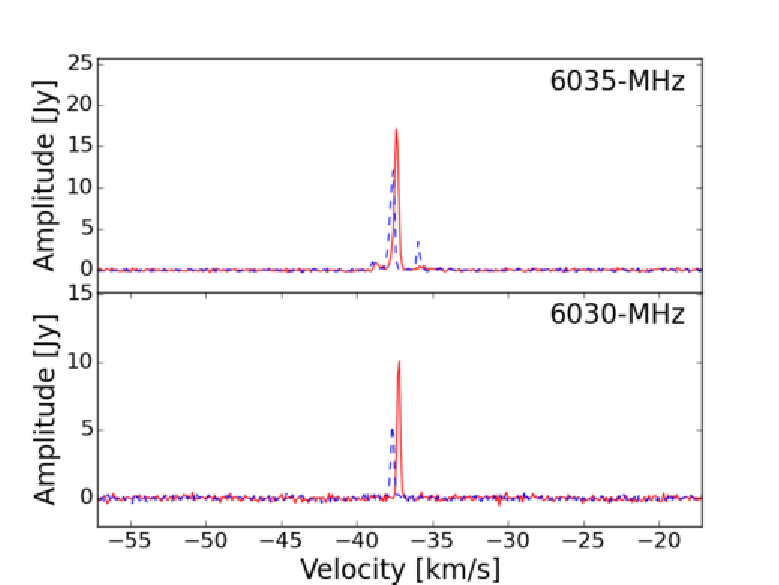}}
	\subfloat[\textit{MMBOH-G305.200+00.019}]{\includegraphics[width=0.33\textwidth]{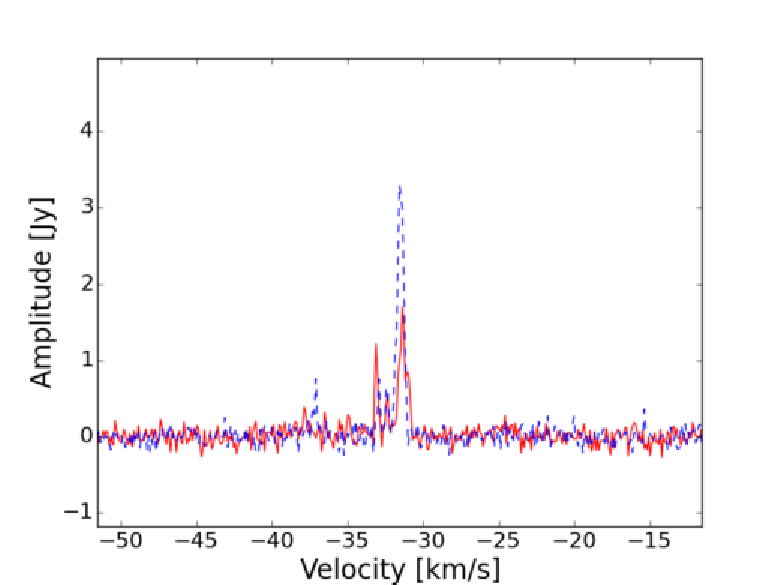}}
\qquad
	\subfloat[\textit{MMBOH-G305.208+00.206}]{\includegraphics[width=0.33\textwidth]{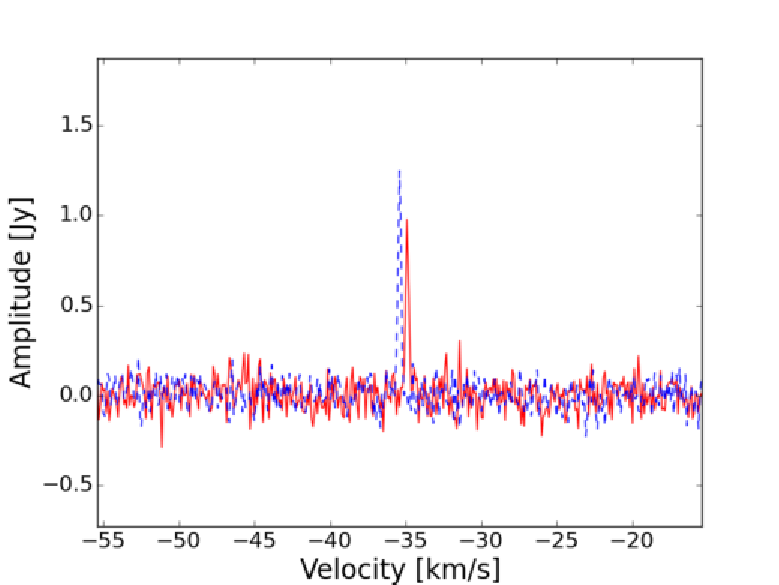}}
	\subfloat[\textit{MMBOH-G305.362+00.150}]{\includegraphics[width=0.33\textwidth]{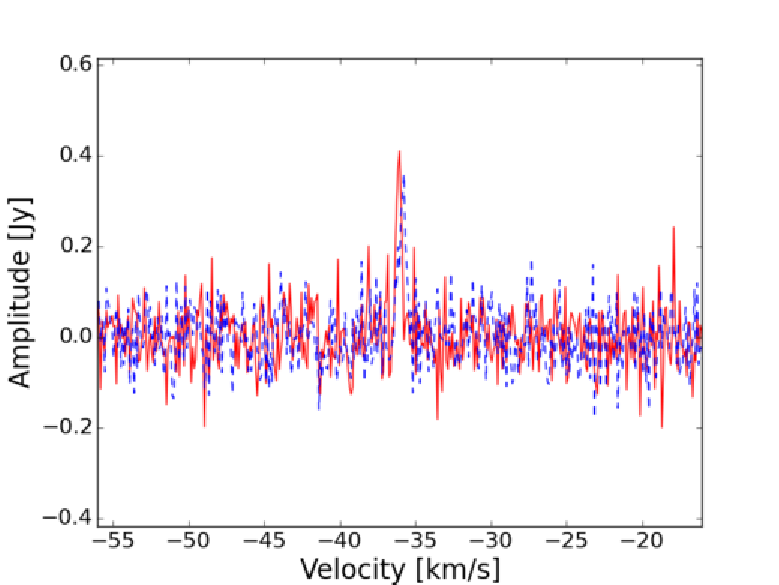}}
	\subfloat[\textit{MMBOH-G308.056-00.396}]{\includegraphics[width=0.33\textwidth]{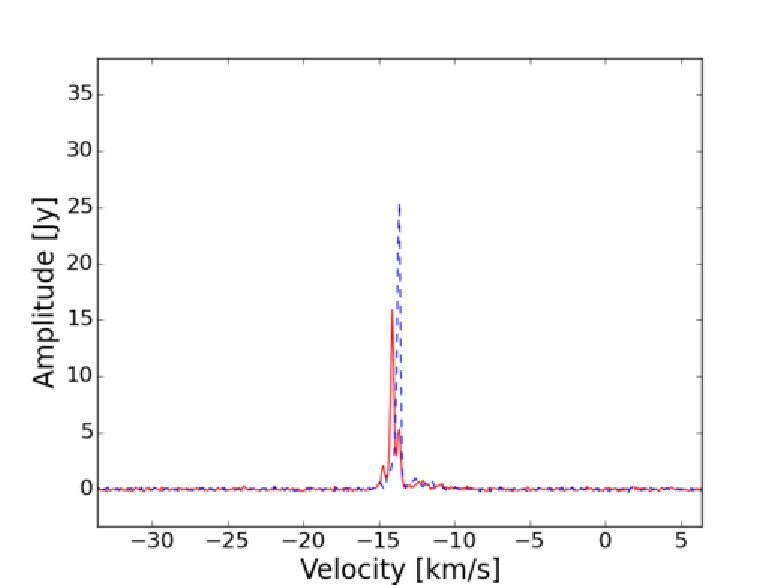}}
\qquad
	\subfloat[\textit{MMBOH-G308.651-00.507}]{\includegraphics[width=0.33\textwidth]{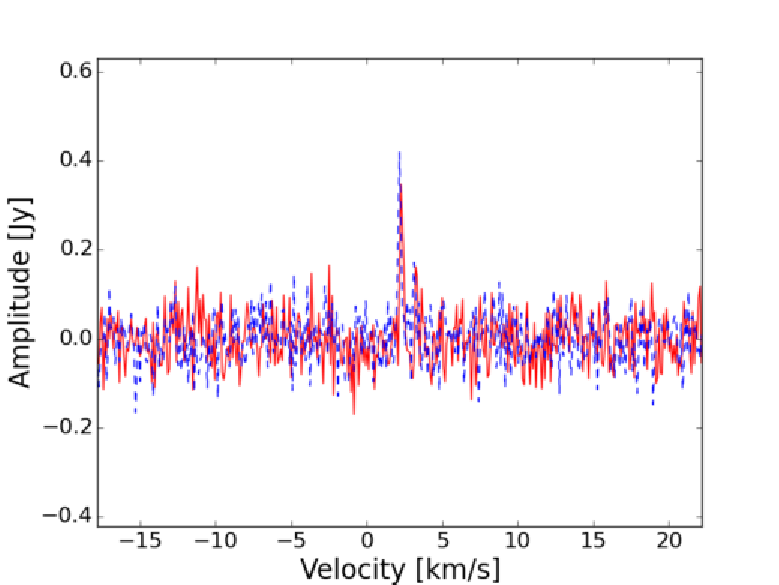}}
	\subfloat[\textit{MMBOH-G309.384-00.135}]{\includegraphics[width=0.33\textwidth]{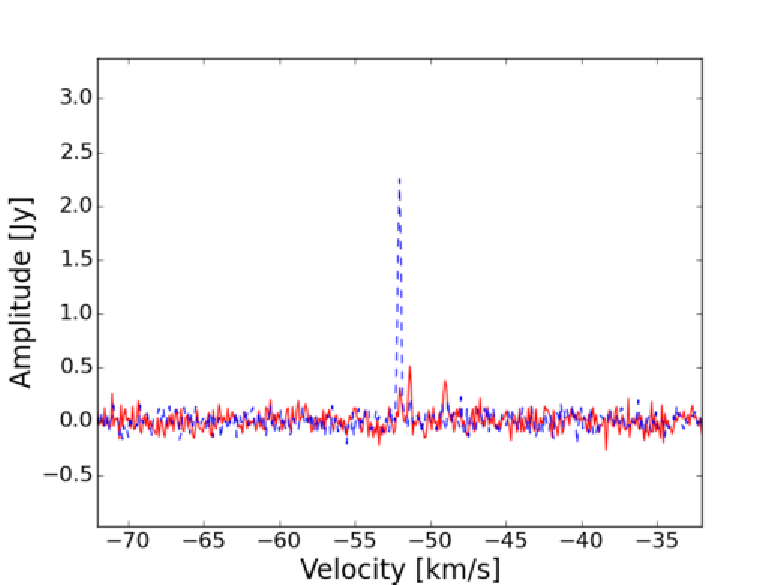}} 
	\subfloat[\textit{MMBOH-G309.901+00.231}]{\includegraphics[width=0.33\textwidth]{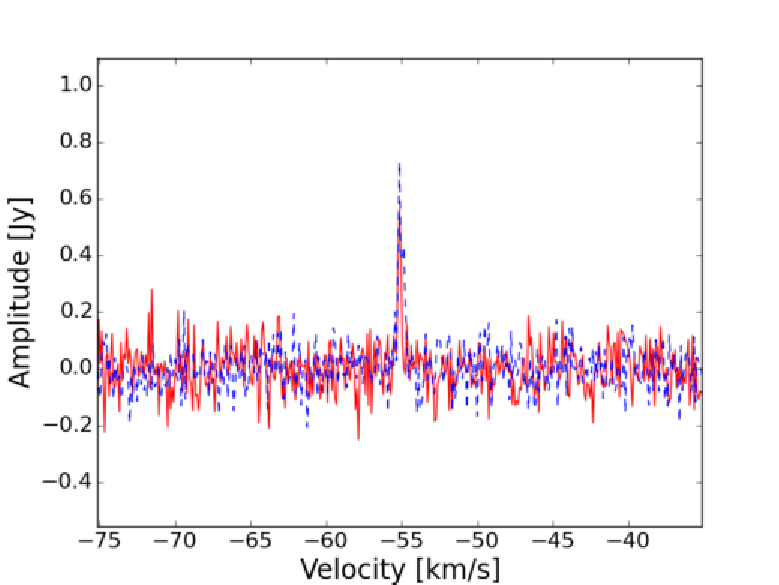}}
	\caption{(continued)}
\end{figure}
\renewcommand\thefigure{11}
\begin{figure}
\ContinuedFloat
\centering
	\subfloat[\textit{MMBOH-G309.921+00.479}]{\includegraphics[width=0.33\textwidth]{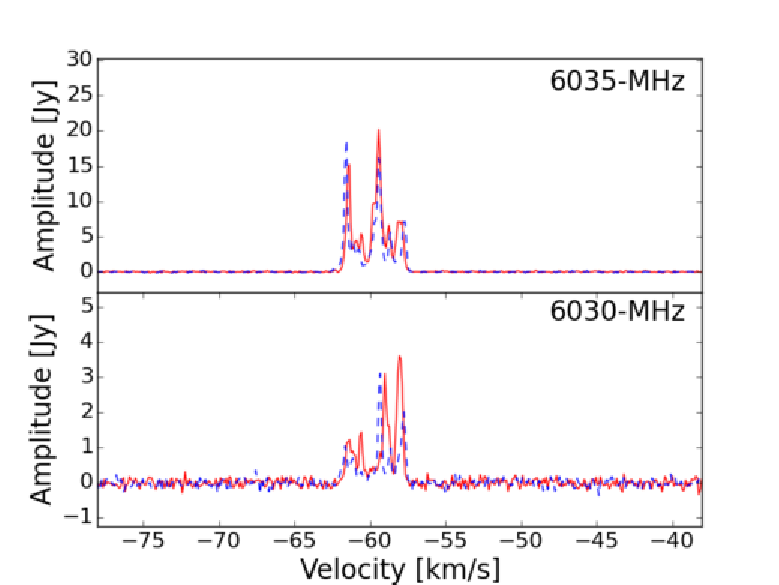}}
	\subfloat[\textit{MMBOH-G311.596-00.398}]{\includegraphics[width=0.33\textwidth]{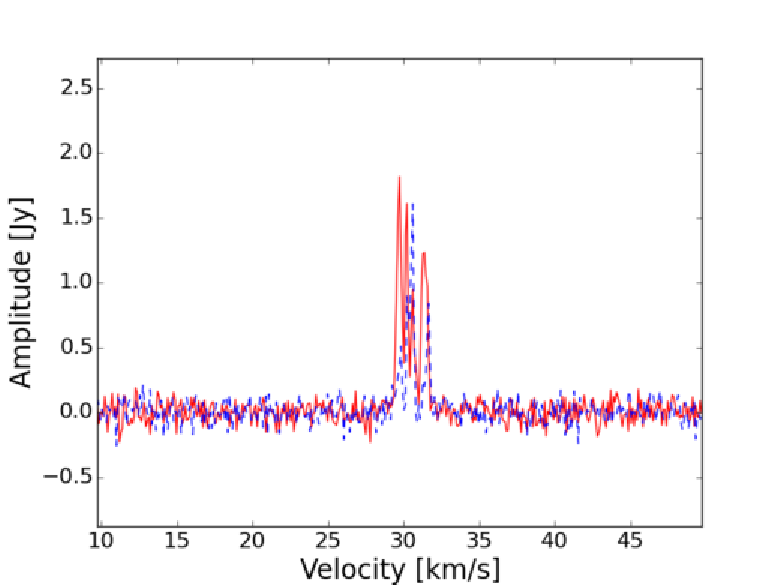}}
	\subfloat[\textit{MMBOH-G311.643-00.380}]{\includegraphics[width=0.33\textwidth]{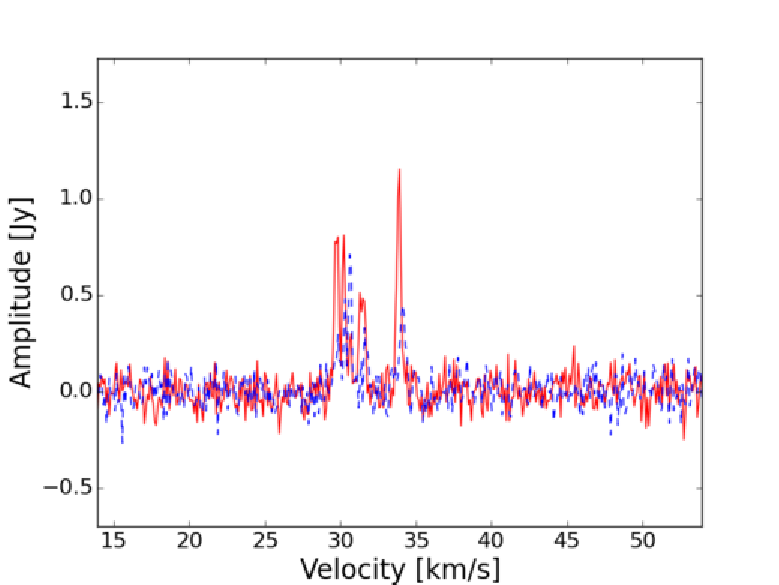}}
\qquad
	\subfloat[\textit{MMBOH-G312.598+00.045}]{\includegraphics[width=0.33\textwidth]{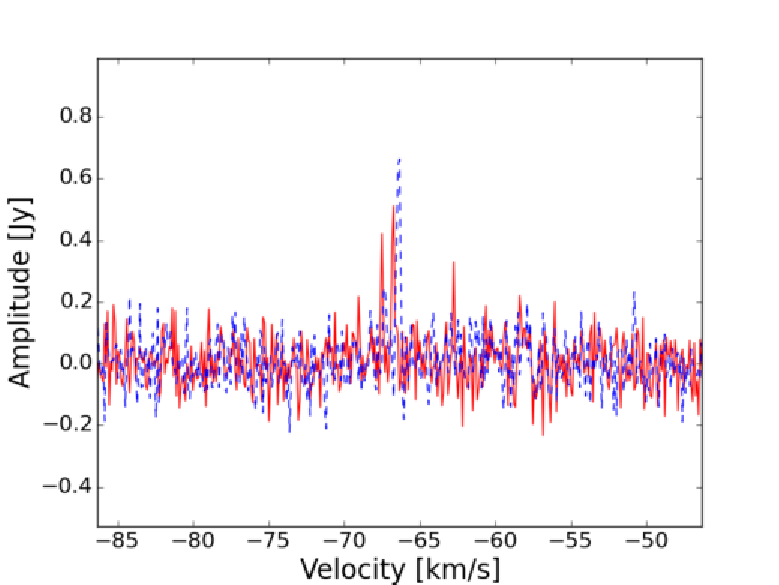}}
	\subfloat[\textit{MMBOH-G320.427+00.103}]{\includegraphics[width=0.33\textwidth]{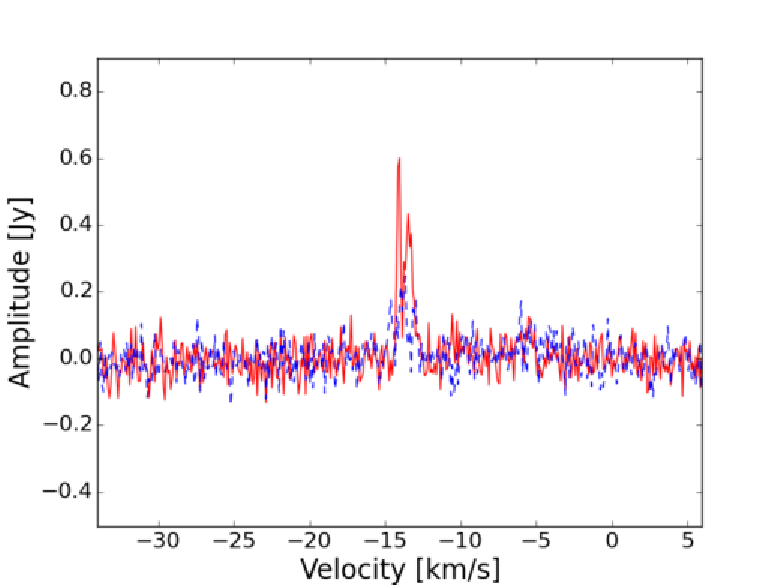}}
	\subfloat[\textit{MMBOH-G323.459-00.079}]{\includegraphics[width=0.33\textwidth]{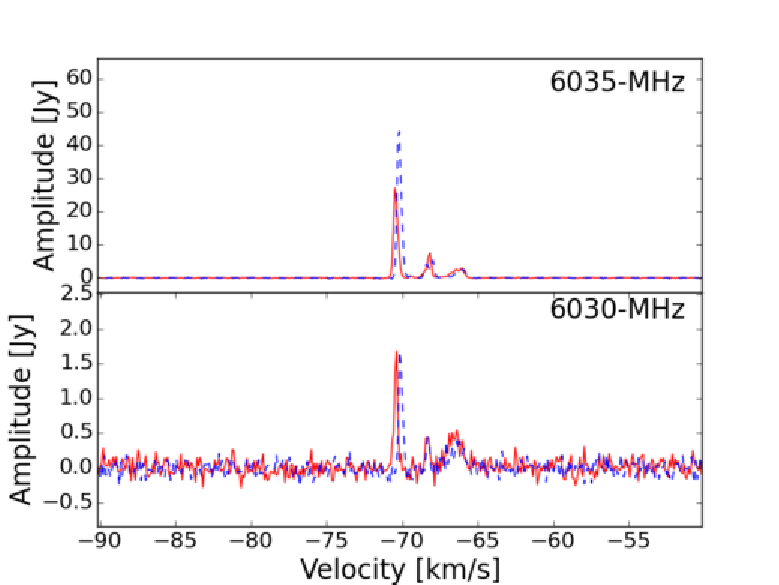}}
\qquad
	\subfloat[\textit{MMBOH-G326.447-00.749}]{\includegraphics[width=0.33\textwidth]{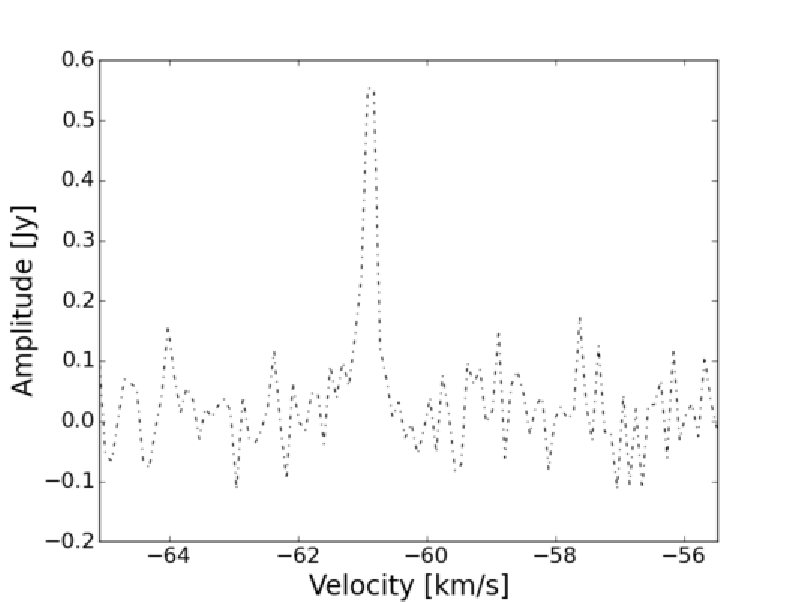}}
	\subfloat[\textit{MMBOH-G326.448-00.749}]{\includegraphics[width=0.33\textwidth]{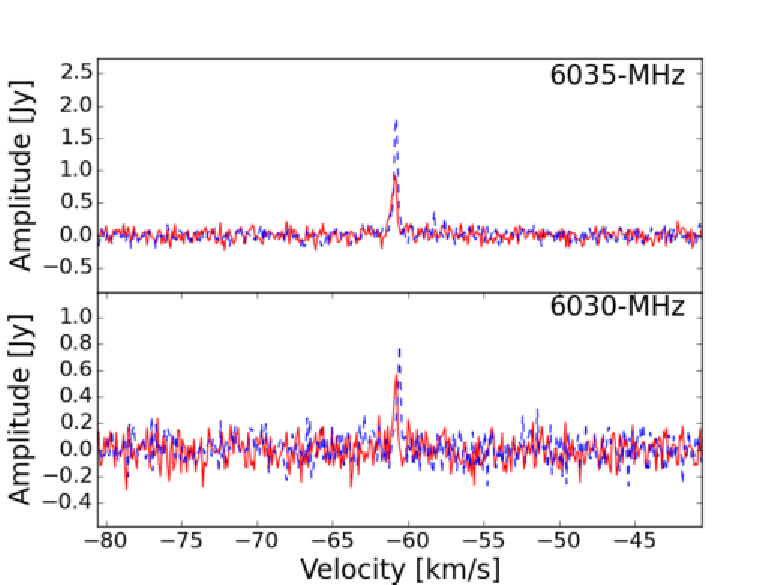}}
	\subfloat[\textit{MMBOH-G327.944-00.116}]{\includegraphics[width=0.33\textwidth]{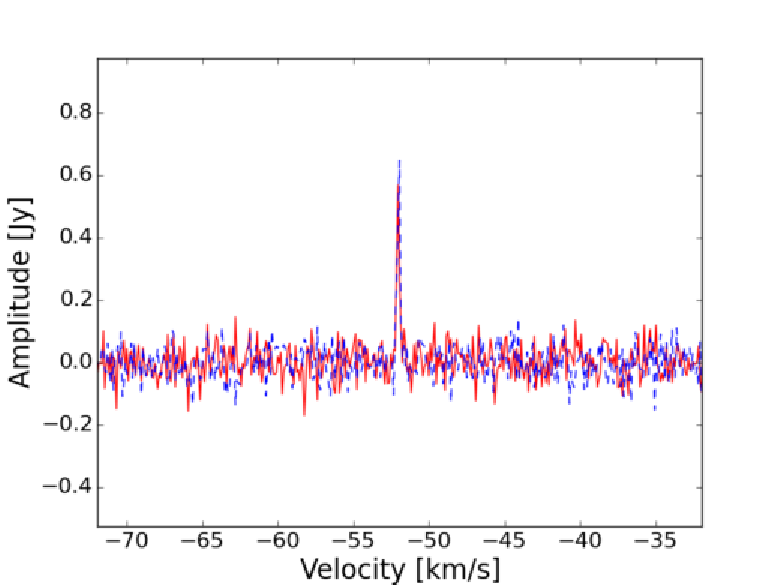}}
\qquad
	\subfloat[\textit{MMBOH-G328.236-00.548}]{\includegraphics[width=0.33\textwidth]{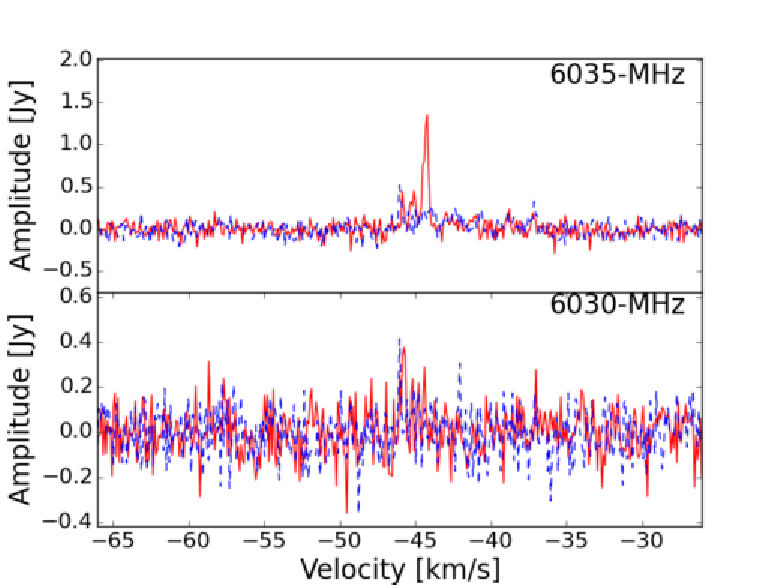}}
	\subfloat[\textit{MMBOH-G328.307+00.430}]{\includegraphics[width=0.33\textwidth]{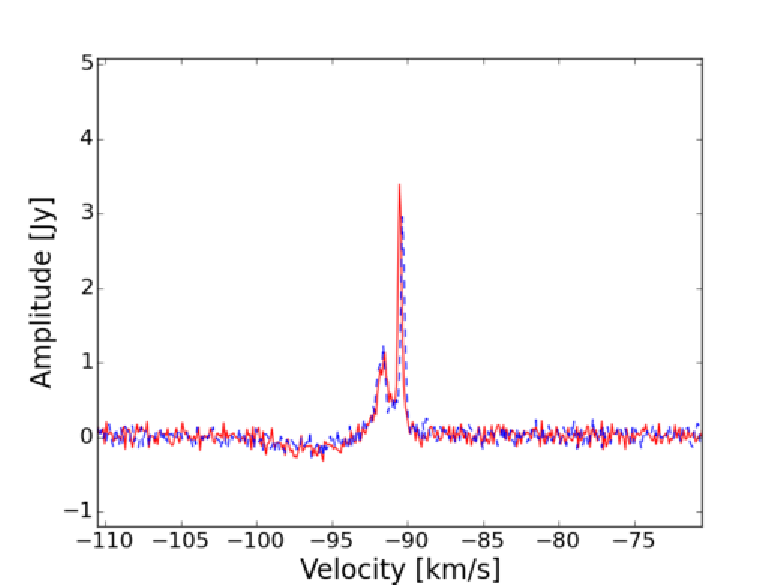}} 
	\subfloat[\textit{MMBOH-G328.808+00.633}]{\includegraphics[width=0.33\textwidth]{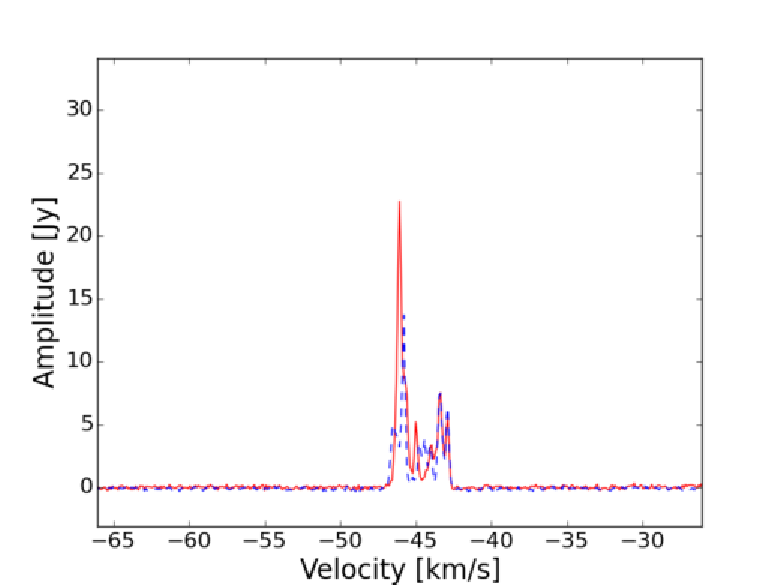}}
	\caption{(continued)}
\end{figure}
\renewcommand\thefigure{11}
\begin{figure}
\ContinuedFloat
\centering
	\subfloat[\textit{MMBOH-G329.031-00.197}]{\includegraphics[width=0.33\textwidth]{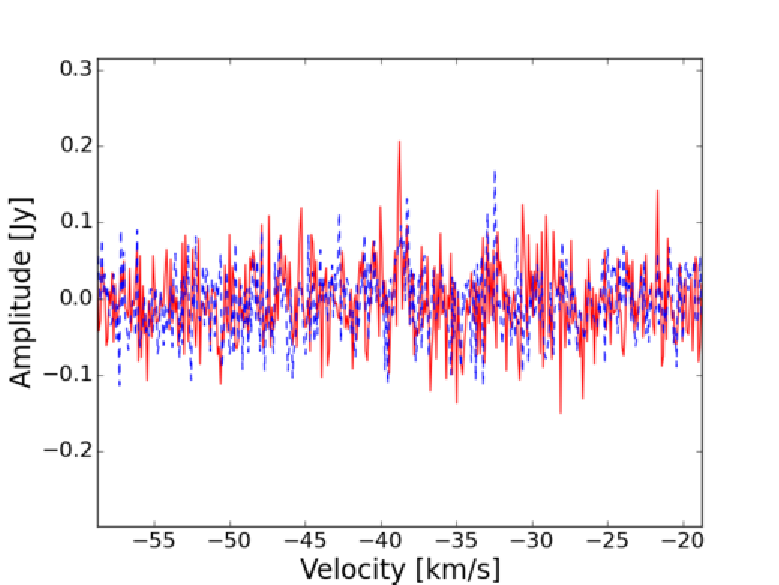}}
	\subfloat[\textit{MMBOH-G329.184-00.314}]{\includegraphics[width=0.33\textwidth]{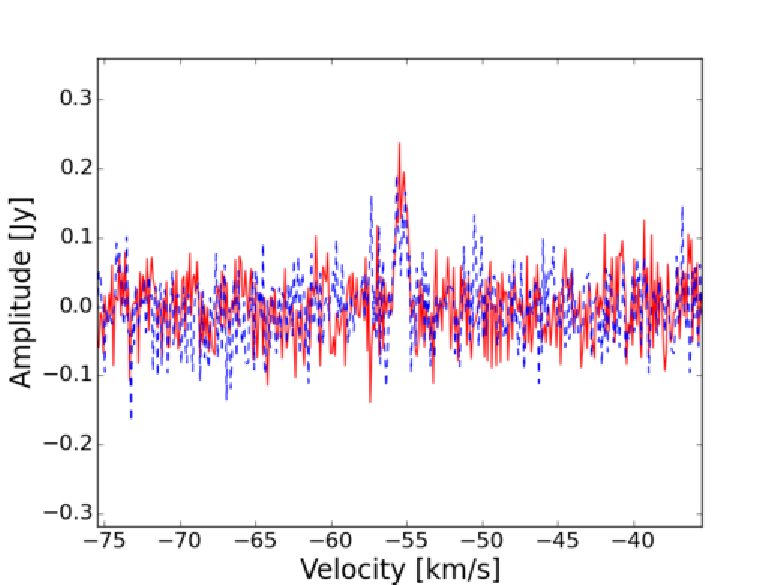}}
	\subfloat[\textit{MMBOH-G329.339+00.148}]{\includegraphics[width=0.33\textwidth]{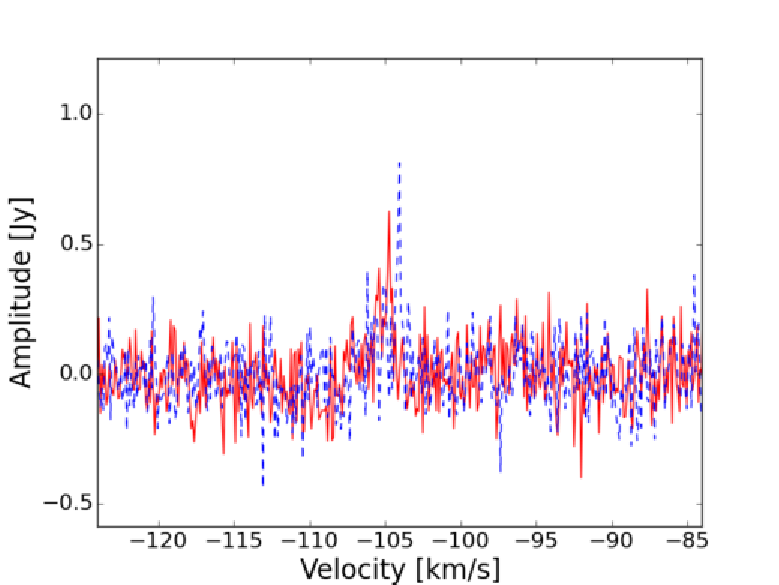}}
\qquad
	\subfloat[\textit{MMBOH-G329.405-00.459}]{\includegraphics[width=0.33\textwidth]{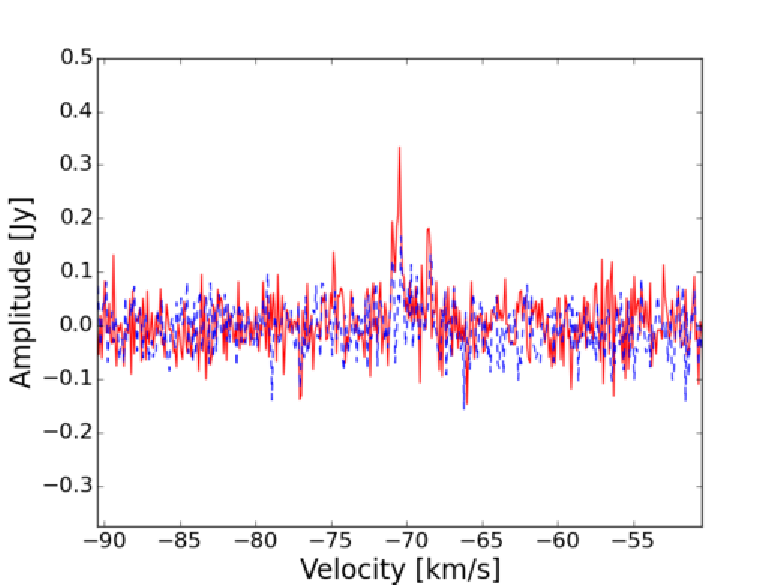}}
	\subfloat[\textit{MMBOH-G330.953-00.182}]{\includegraphics[width=0.33\textwidth]{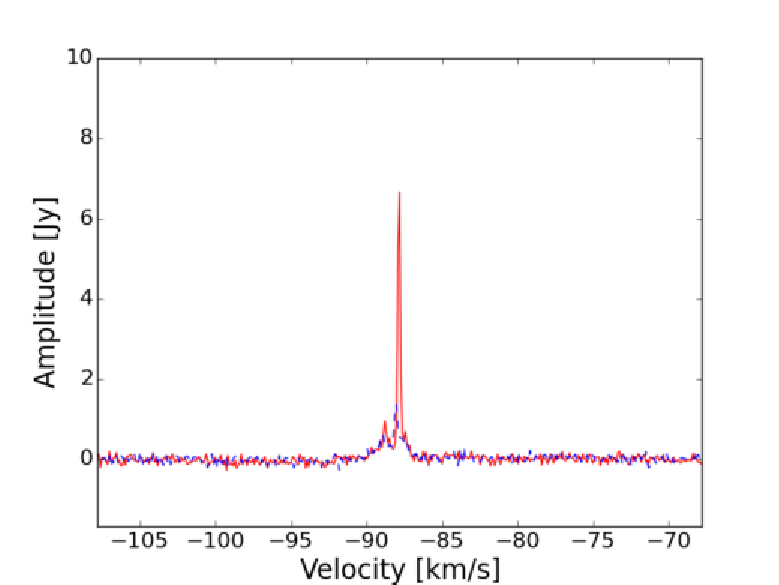}}
	\subfloat[\textit{MMBOH-G331.512-00.102}]{\includegraphics[width=0.33\textwidth]{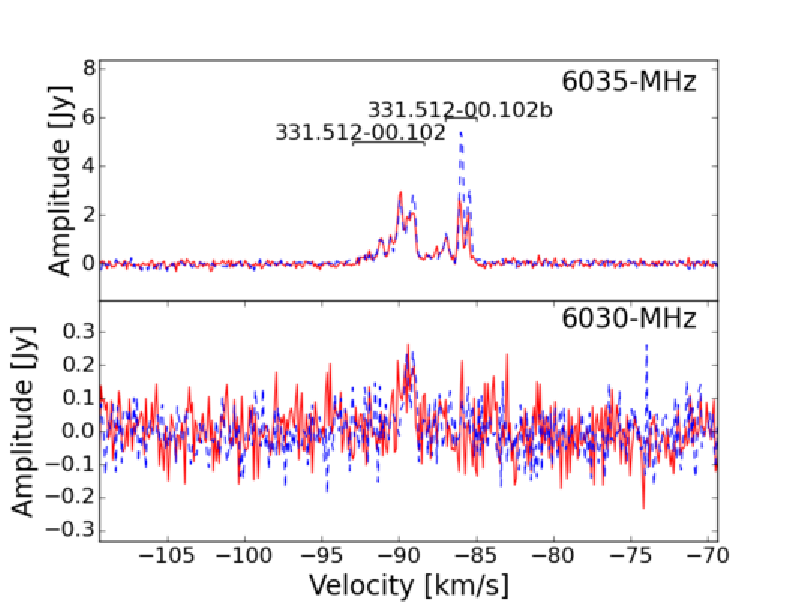}}
\qquad
	\subfloat[\textit{MMBOH-G331.512-00.102b}]{\includegraphics[width=0.33\textwidth]{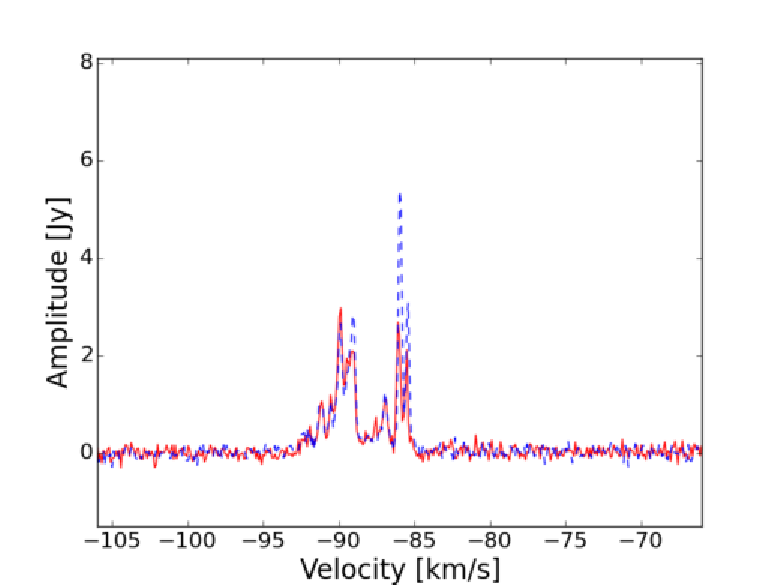}}
	\subfloat[\textit{MMBOH-G331.542-00.066}]{\includegraphics[width=0.33\textwidth]{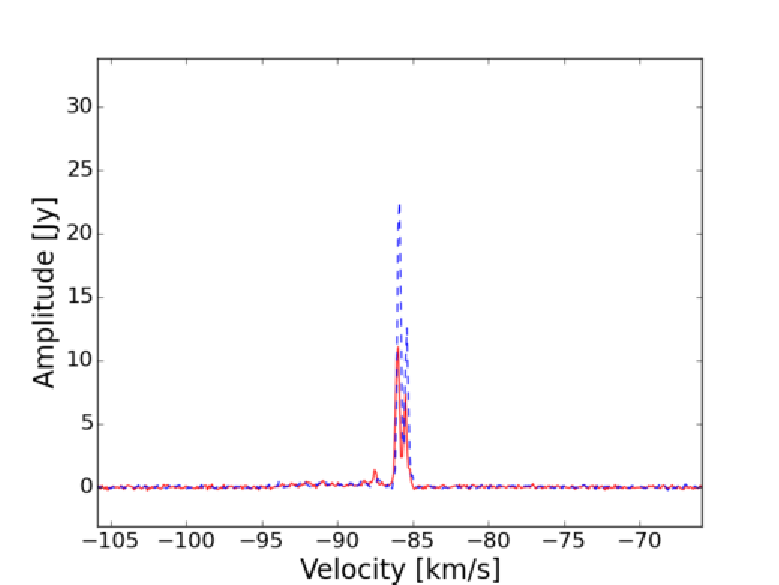}}
	\subfloat[\textit{MMBOH-G332.824-00.548}]{\includegraphics[width=0.33\textwidth]{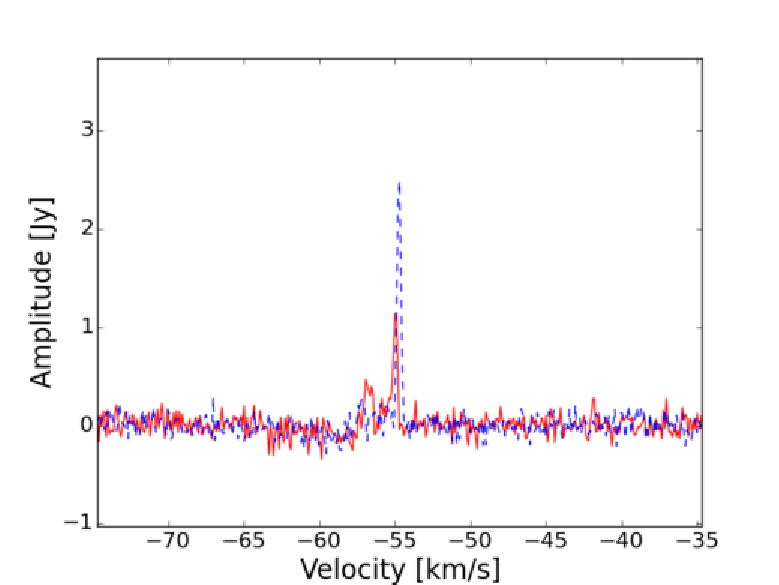}}
\qquad
	\subfloat[\textit{MMBOH-G332.964-00.679}]{\includegraphics[width=0.33\textwidth]{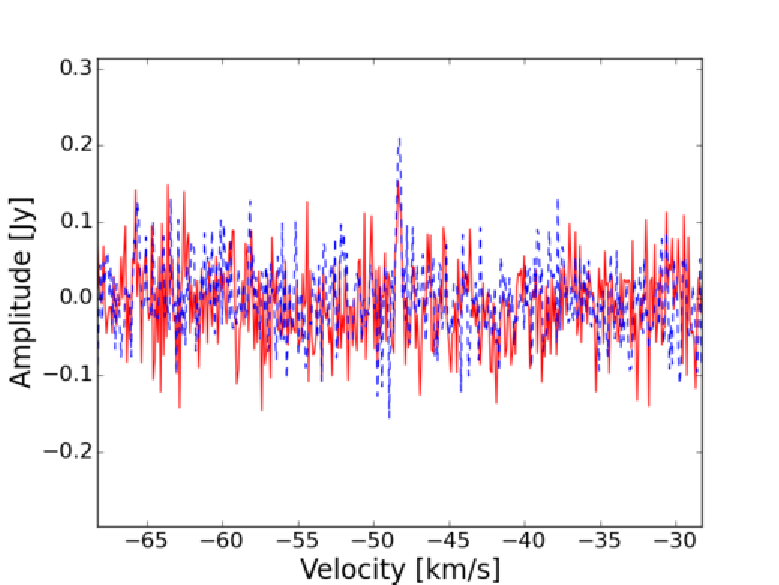}}
	\subfloat[\textit{MMBOH-G333.068-00.447}]{\includegraphics[width=0.33\textwidth]{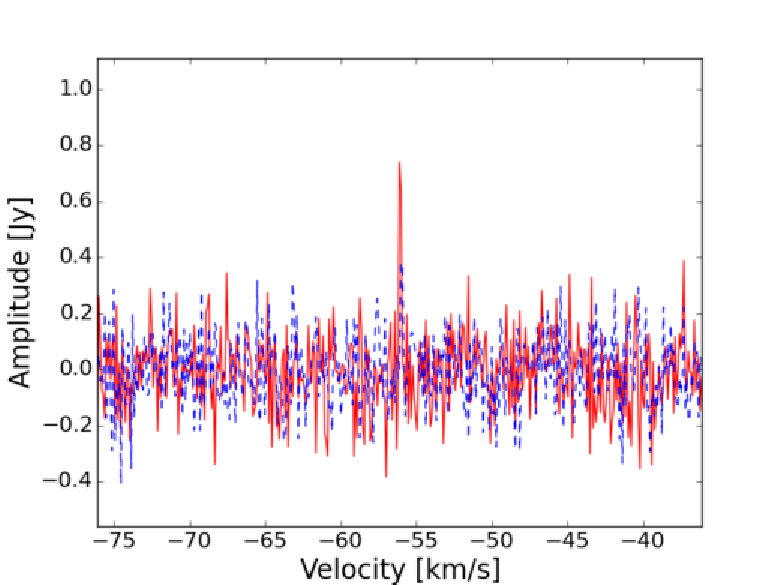}} 
	\subfloat[\textit{MMBOH-G333.135-00.431}]{\includegraphics[width=0.33\textwidth]{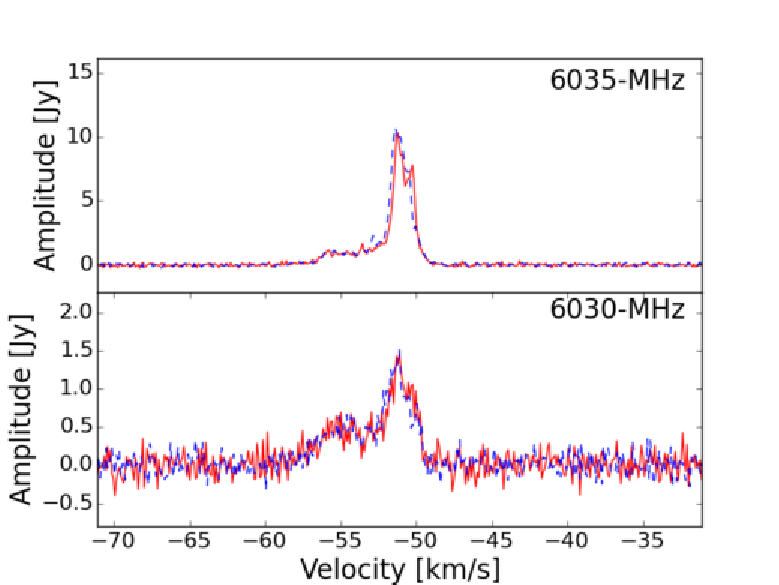}}
	\caption{(continued)}
\end{figure}
\renewcommand\thefigure{11}
\begin{figure}
\ContinuedFloat
\centering
	\subfloat[\textit{MMBOH-G333.136-00.432}]{\includegraphics[width=0.33\textwidth]{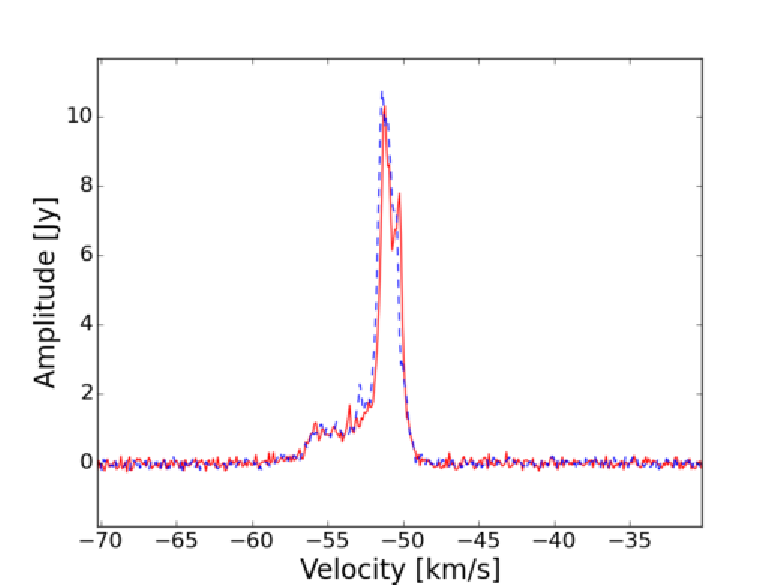}}
	\subfloat[\textit{MMBOH-G333.228-00.055}]{\includegraphics[width=0.33\textwidth]{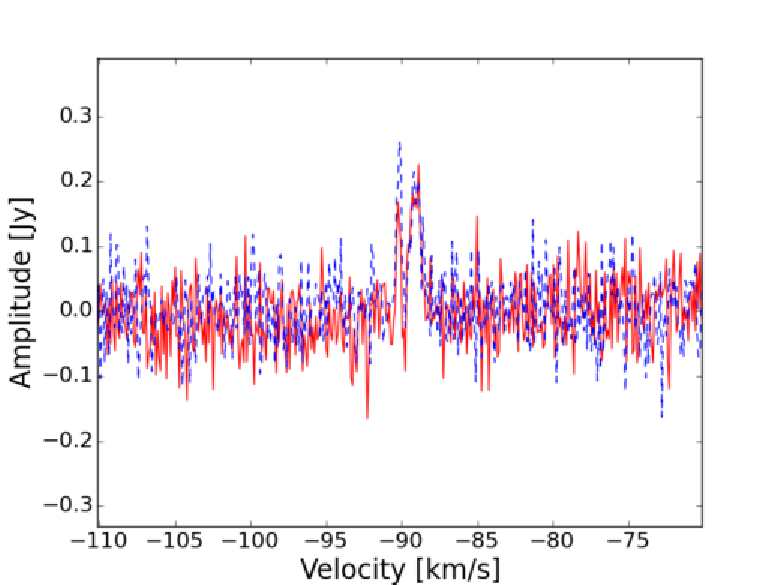}}
	\subfloat[\textit{MMBOH-G333.608-00.215}]{\includegraphics[width=0.33\textwidth]{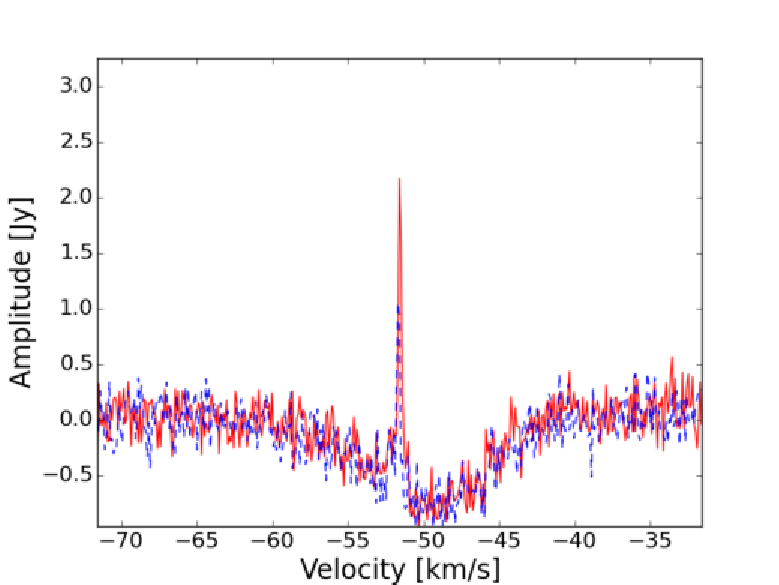}}
\qquad
	\subfloat[\textit{MMBOH-G336.822+00.028}]{\includegraphics[width=0.33\textwidth]{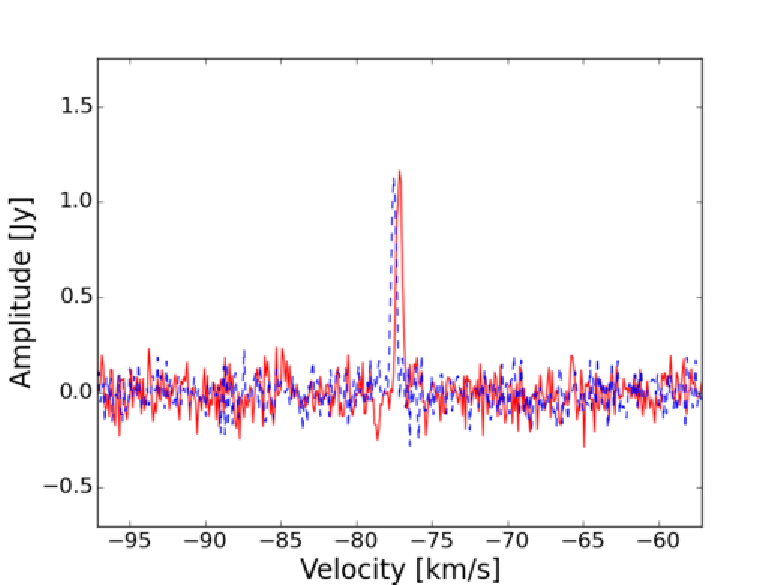}}
	\subfloat[\textit{MMBOH-G336.941-00.156}]{\includegraphics[width=0.33\textwidth]{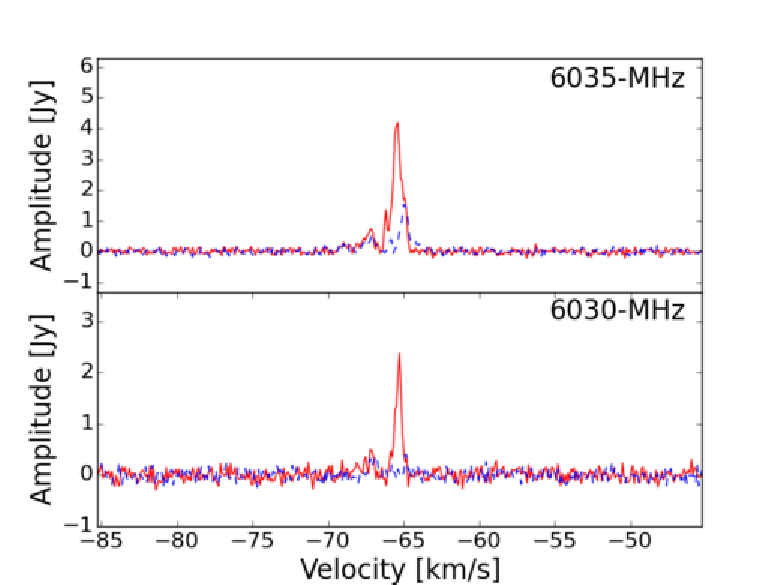}}
	\subfloat[\textit{MMBOH-G336.983-00.183}]{\includegraphics[width=0.33\textwidth]{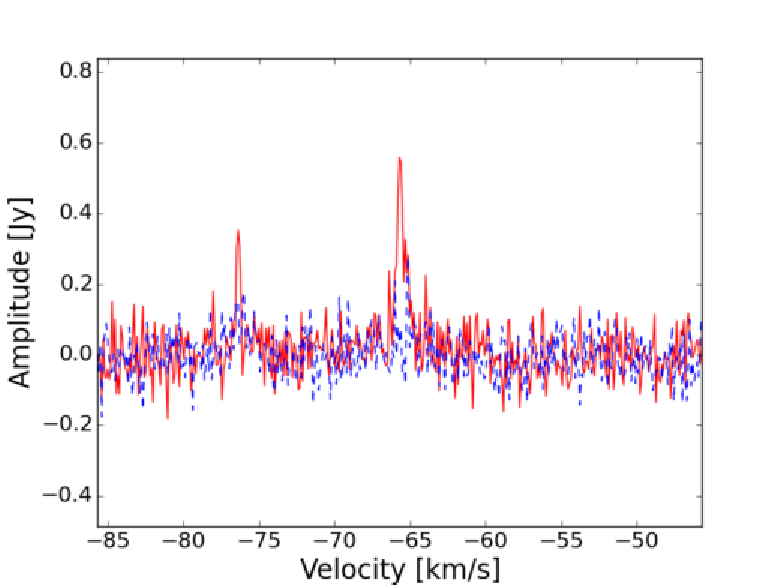}}
\qquad
	\subfloat[\textit{MMBOH-G337.098-00.928}]{\includegraphics[width=0.33\textwidth]{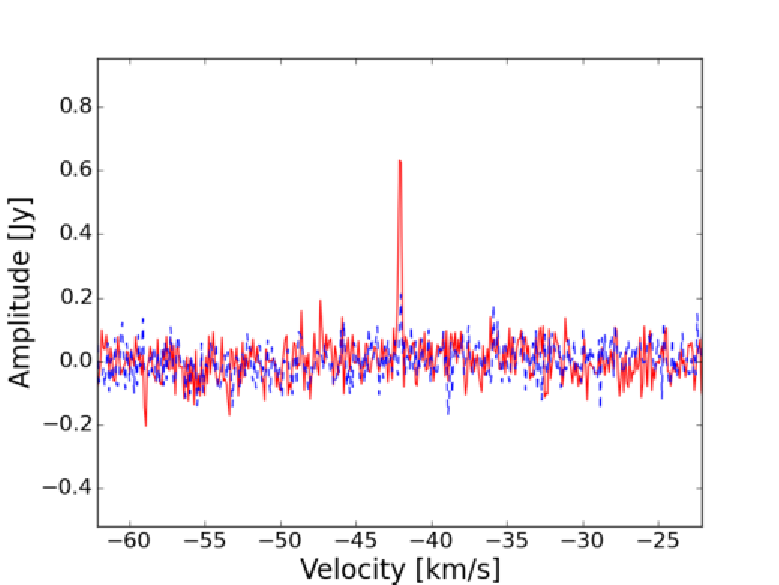}}
	\subfloat[\textit{MMBOH-G337.404-00.402}]{\includegraphics[width=0.33\textwidth]{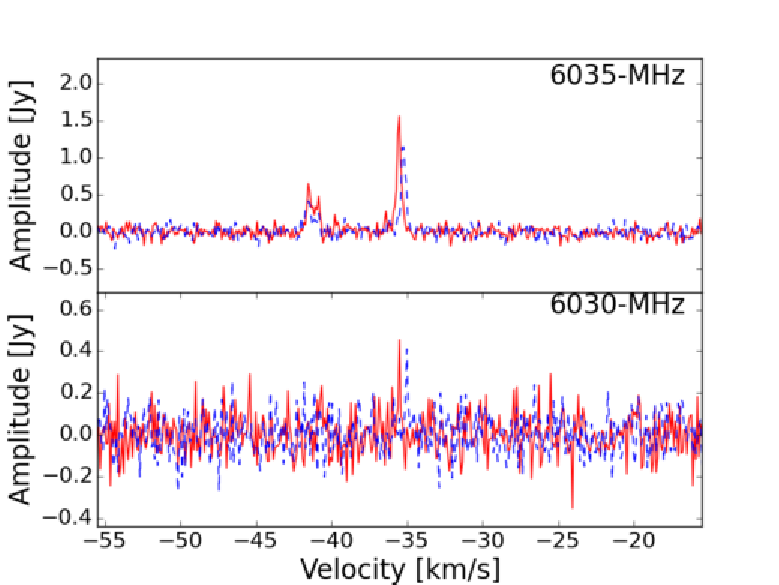}}
	\subfloat[\textit{MMBOH-G337.606-00.052}]{\includegraphics[width=0.33\textwidth]{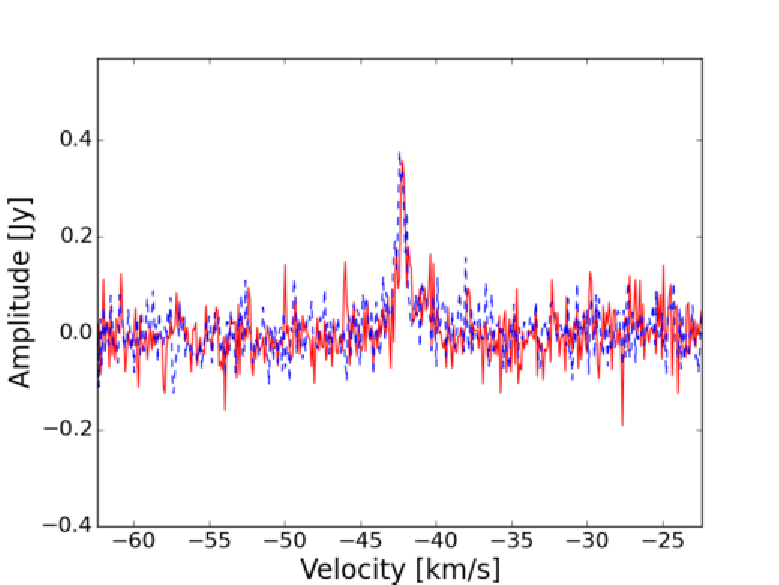}}
\qquad
	\subfloat[\textit{MMBOH-G337.705-00.053}]{\includegraphics[width=0.33\textwidth]{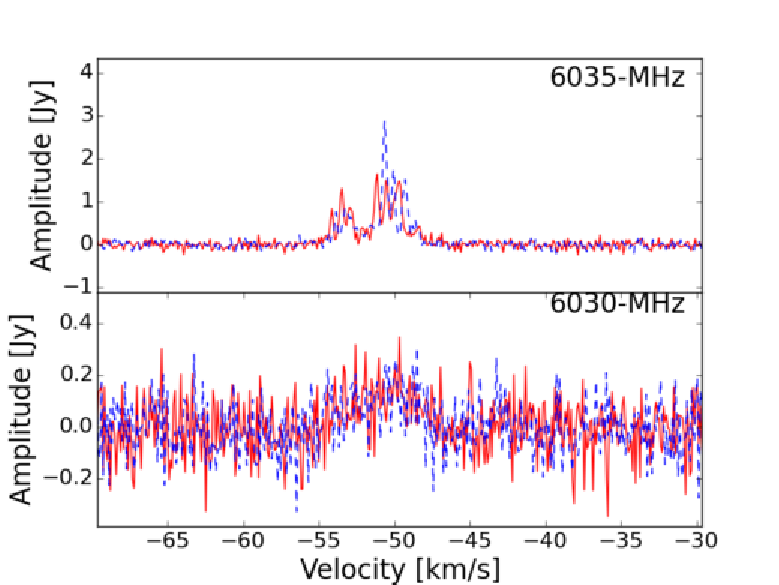}}
	\subfloat[\textit{MMBOH-G337.844-00.374}]{\includegraphics[width=0.33\textwidth]{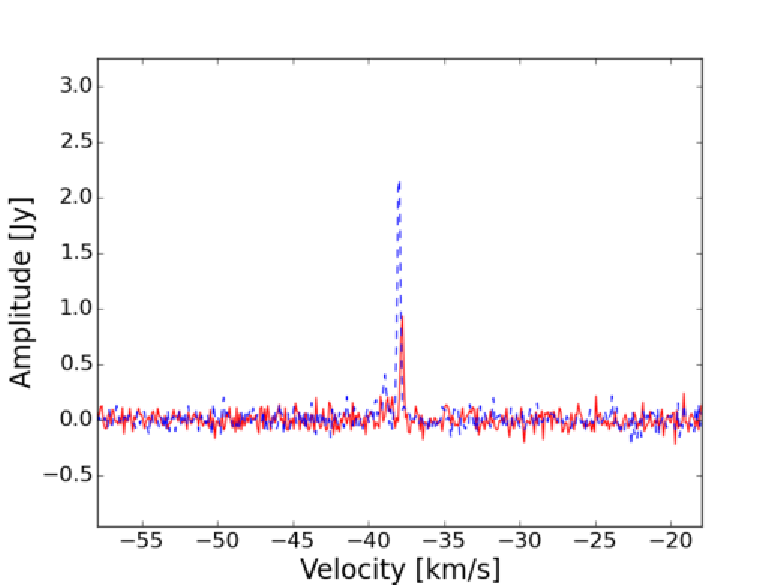}} 
	\subfloat[\textit{MMBOH-G338.925+00.557}]{\includegraphics[width=0.33\textwidth]{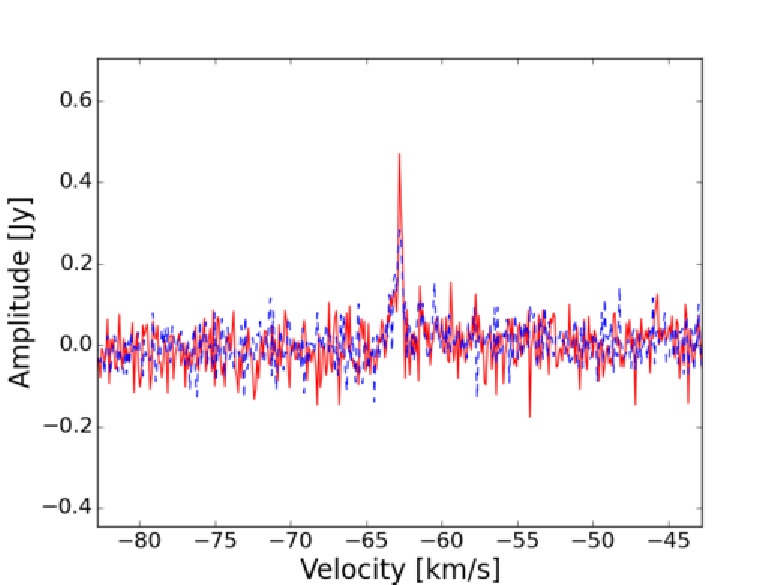}}
	\caption{(continued)}
\end{figure}
\renewcommand\thefigure{11}
\begin{figure}
\ContinuedFloat
\centering
	\subfloat[\textit{MMBOH-G339.053-00.315}]{\includegraphics[width=0.33\textwidth]{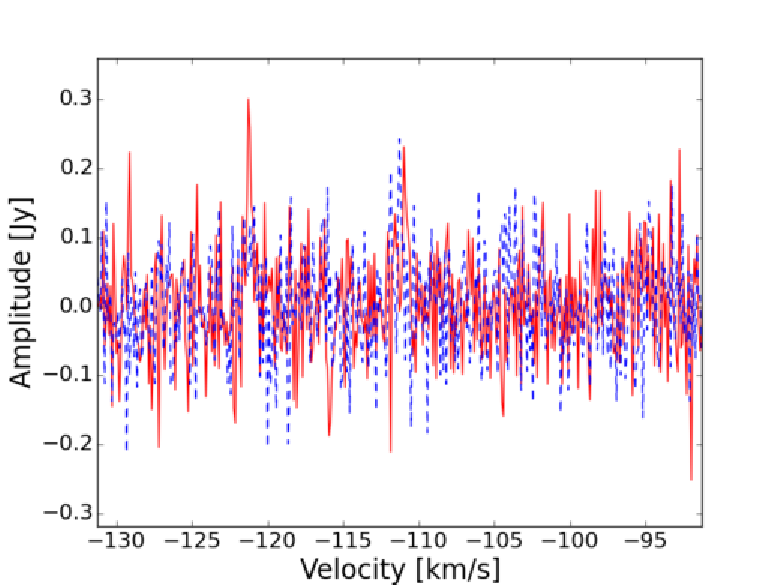}}
	\subfloat[\textit{MMBOH-G339.282+00.136}]{\includegraphics[width=0.33\textwidth]{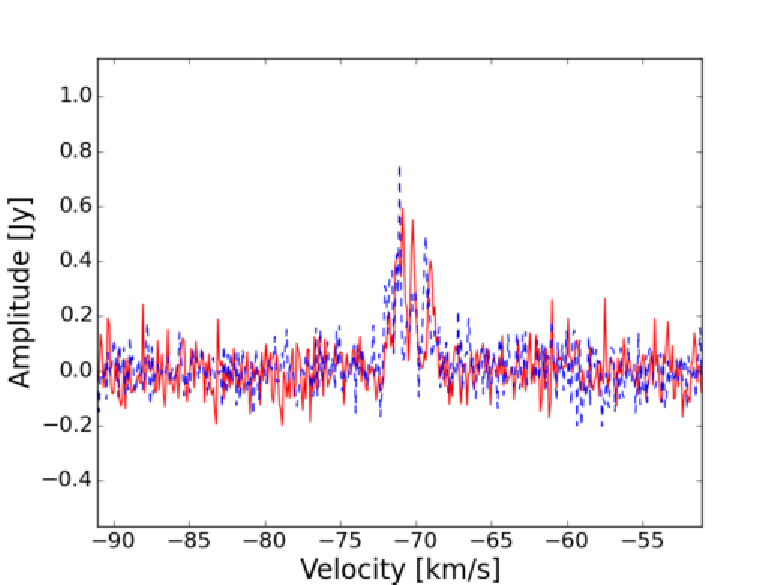}}
	\subfloat[\textit{MMBOH-G339.622-00.121}]{\includegraphics[width=0.33\textwidth]{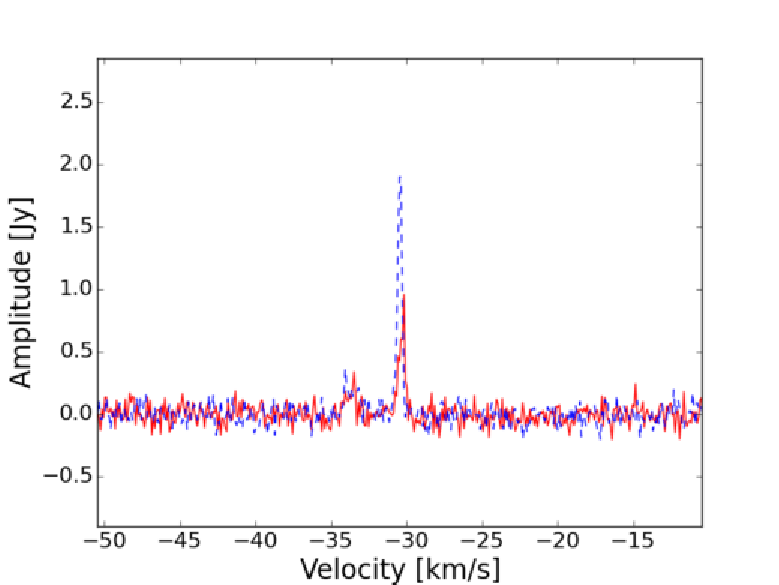}}
\qquad
	\subfloat[\textit{MMBOH-G339.884-01.259}]{\includegraphics[width=0.33\textwidth]{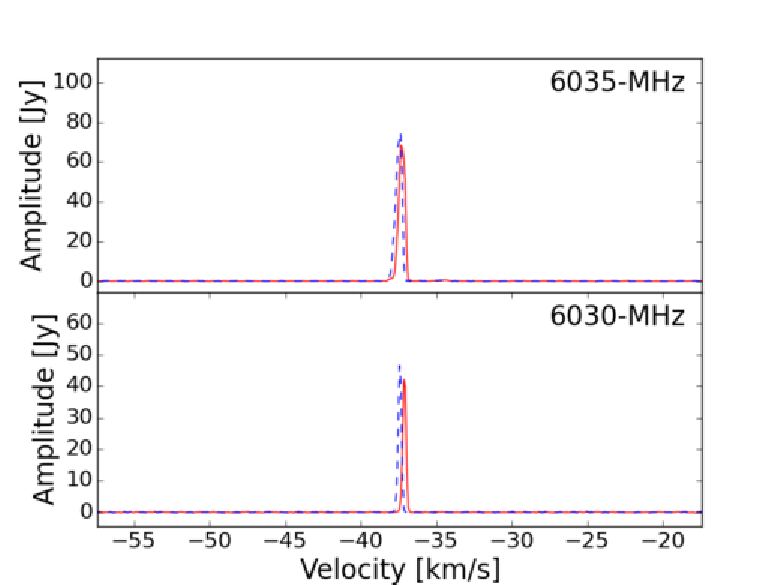}}
	\subfloat[\textit{MMBOH-G339.980-00.539}]{\includegraphics[width=0.33\textwidth]{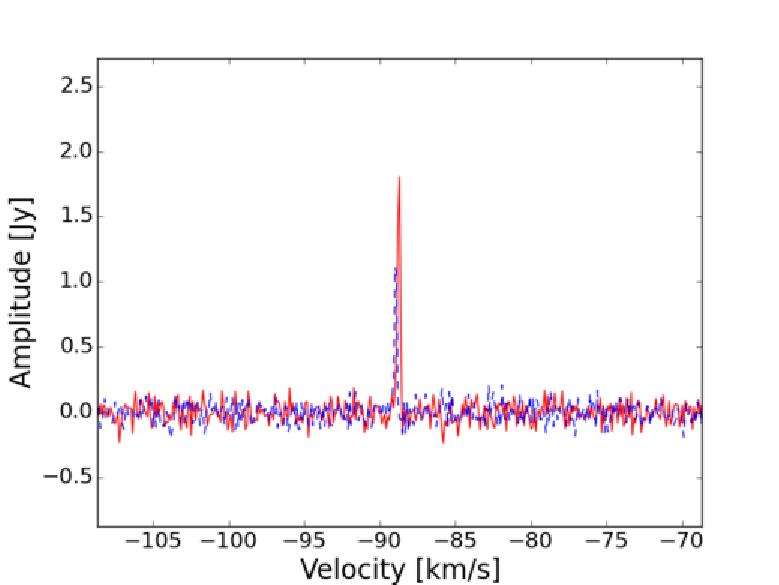}}
	\subfloat[\textit{MMBOH-G340.785-00.096}]{\includegraphics[width=0.33\textwidth]{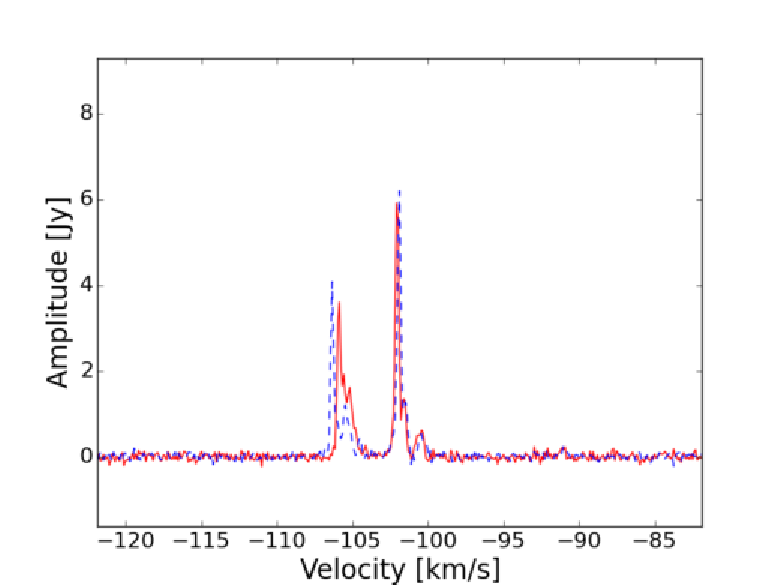}}
\qquad
	\subfloat[\textit{MMBOH-G341.974+00.225}]{\includegraphics[width=0.33\textwidth]{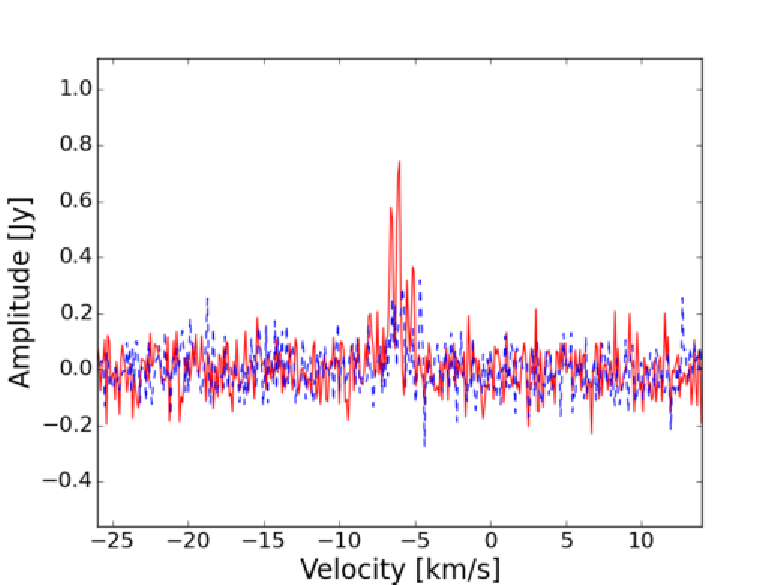}}
	\subfloat[\textit{MMBOH-G343.354-00.067}]{\includegraphics[width=0.33\textwidth]{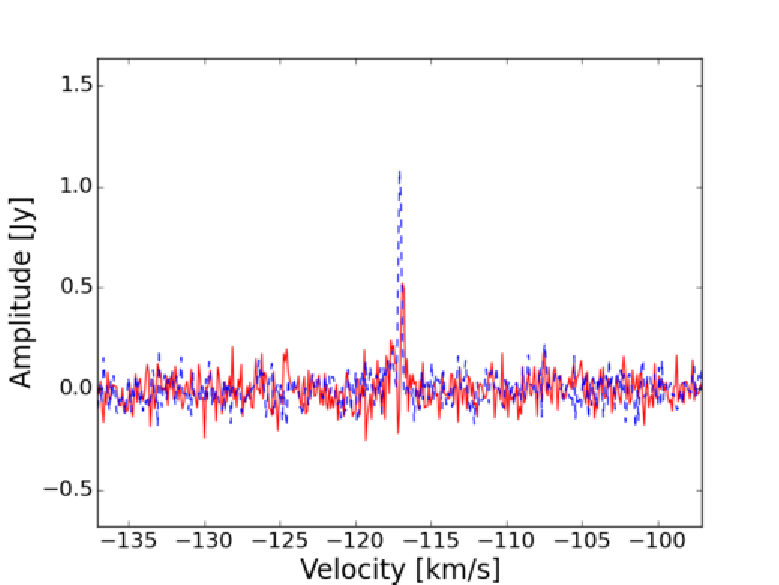}}
	\subfloat[\textit{MMBOH-G343.929+00.125}]{\includegraphics[width=0.33\textwidth]{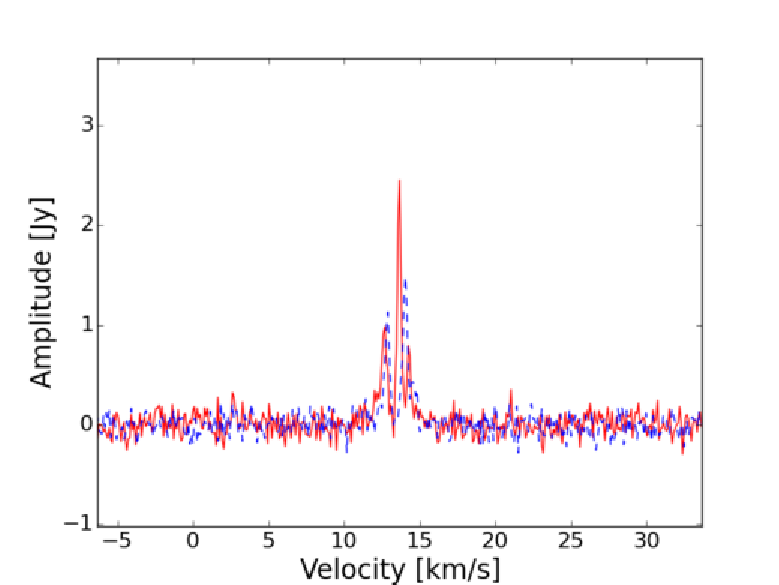}}
\qquad
	\subfloat[\textit{MMBOH-G344.419+00.044}]{\includegraphics[width=0.33\textwidth]{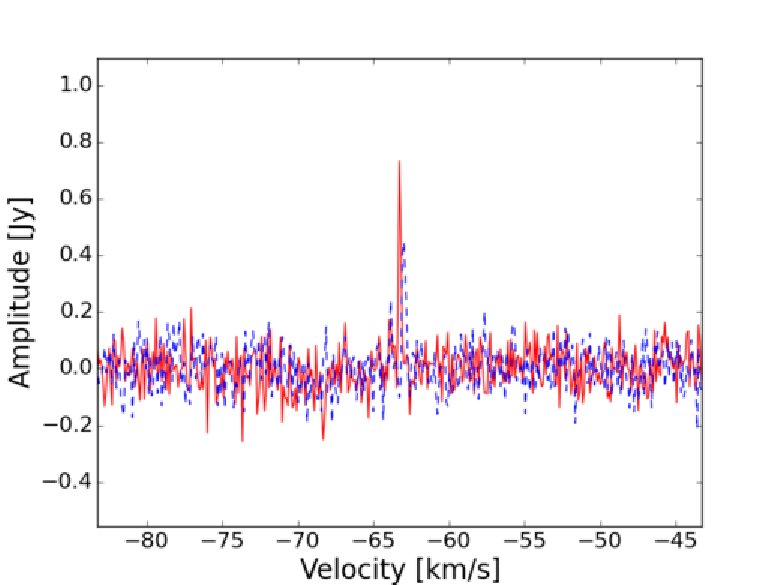}}
	\subfloat[\textit{MMBOH-G345.003-00.224}]{\includegraphics[width=0.33\textwidth]{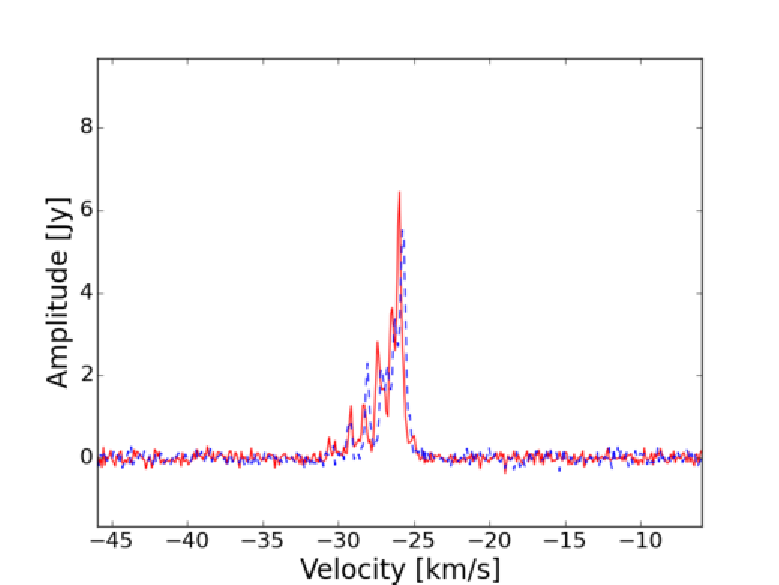}} 
	\subfloat[\textit{MMBOH-G345.010+01.792}]{\includegraphics[width=0.33\textwidth]{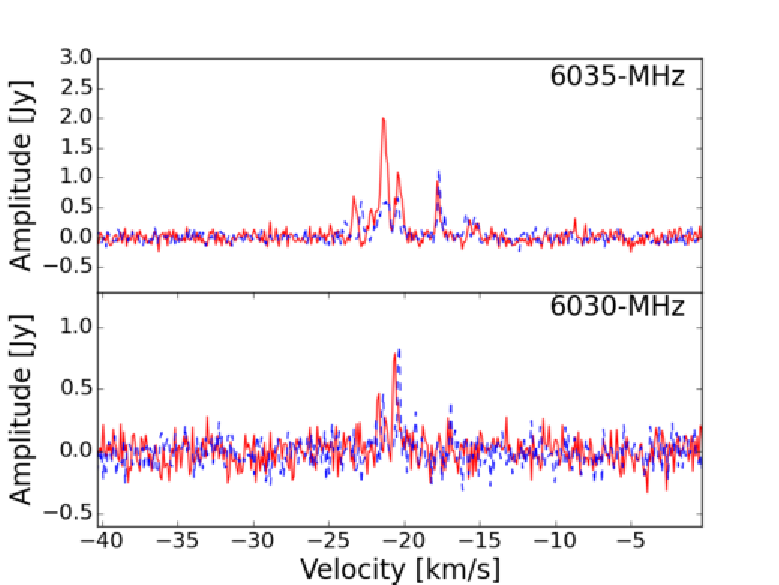}}
	\caption{(continued)}
\end{figure}
\renewcommand\thefigure{11}
\begin{figure}
\ContinuedFloat
\centering
	\subfloat[\textit{MMBOH-G345.407-00.951}]{\includegraphics[width=0.33\textwidth]{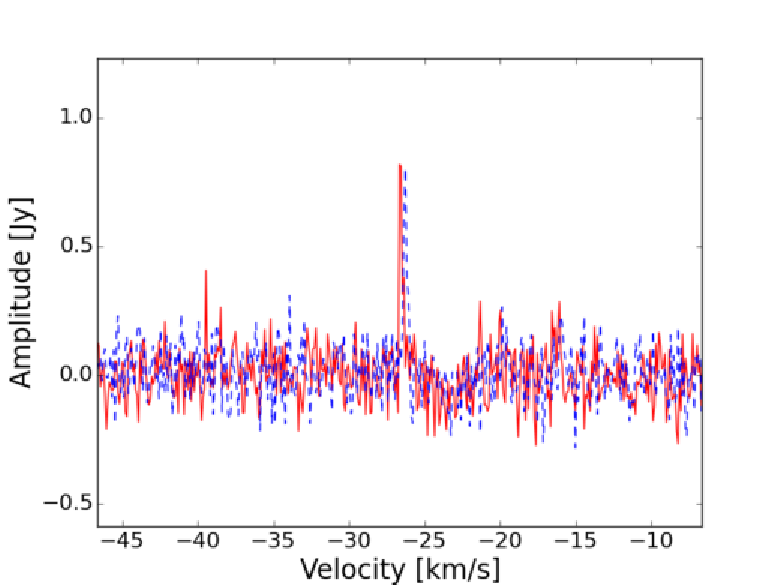}}
	\subfloat[\textit{MMBOH-G345.487+00.314}]{\includegraphics[width=0.33\textwidth]{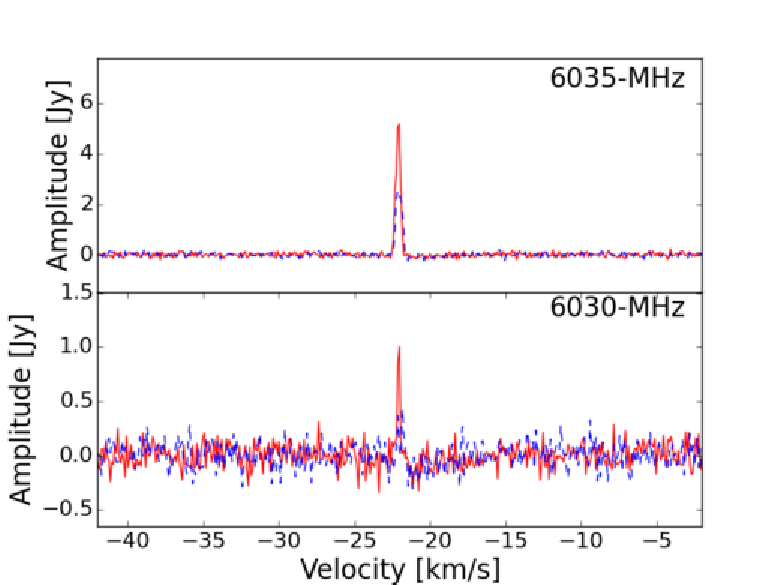}}
	\subfloat[\textit{MMBOH-G345.495+01.469}]{\includegraphics[width=0.33\textwidth]{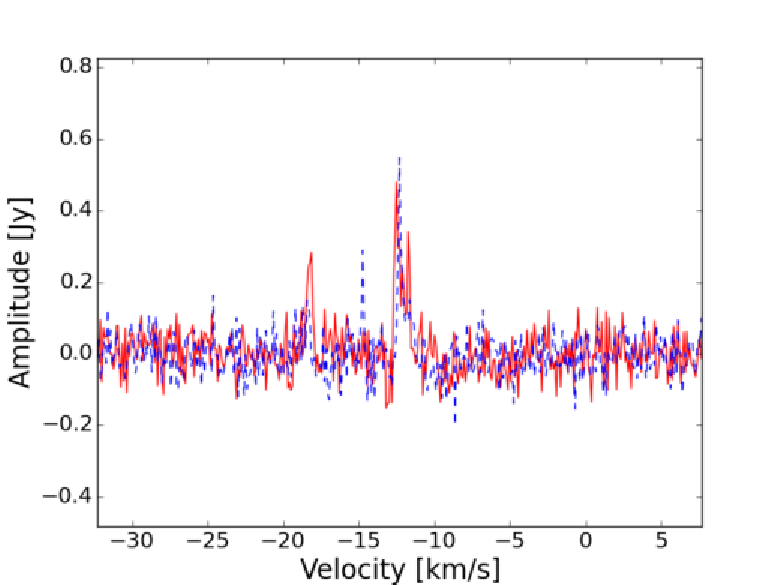}}
\qquad
	\subfloat[\textit{MMBOH-G345.698-00.090}]{\includegraphics[width=0.33\textwidth]{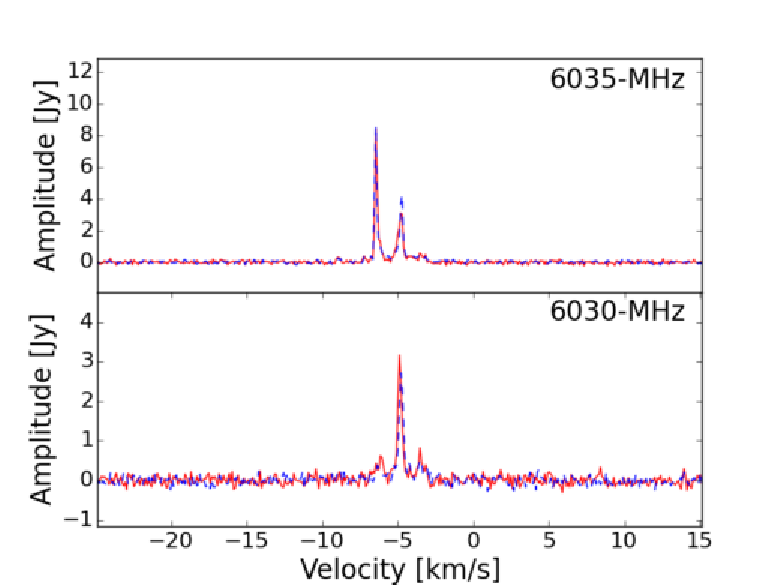}}
	\subfloat[\textit{MMBOH-G347.628+00.149}]{\includegraphics[width=0.33\textwidth]{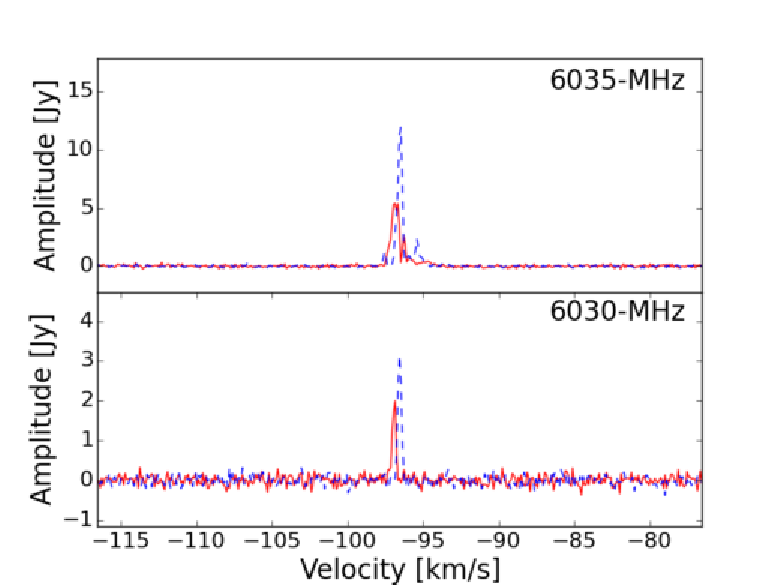}}
	\subfloat[\textit{MMBOH-G348.698-01.027}]{\includegraphics[width=0.33\textwidth]{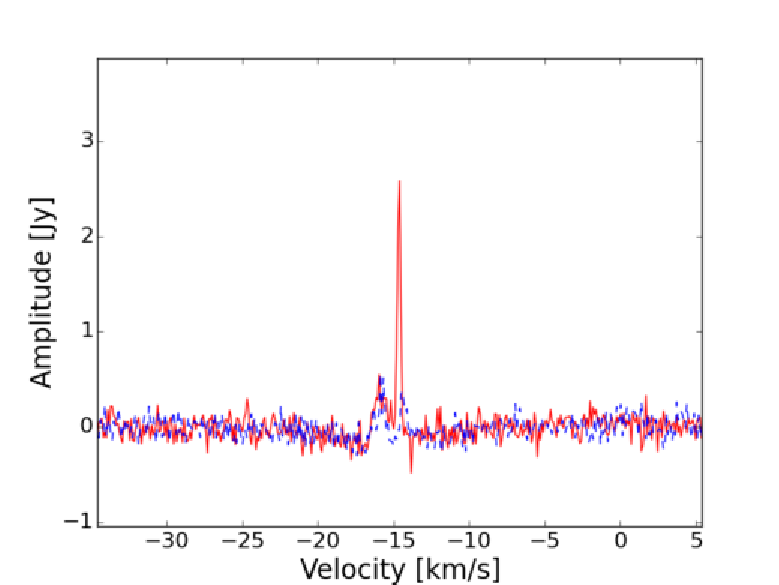}}
\qquad
	\subfloat[\textit{MMBOH-G350.014+00.434}]{\includegraphics[width=0.33\textwidth]{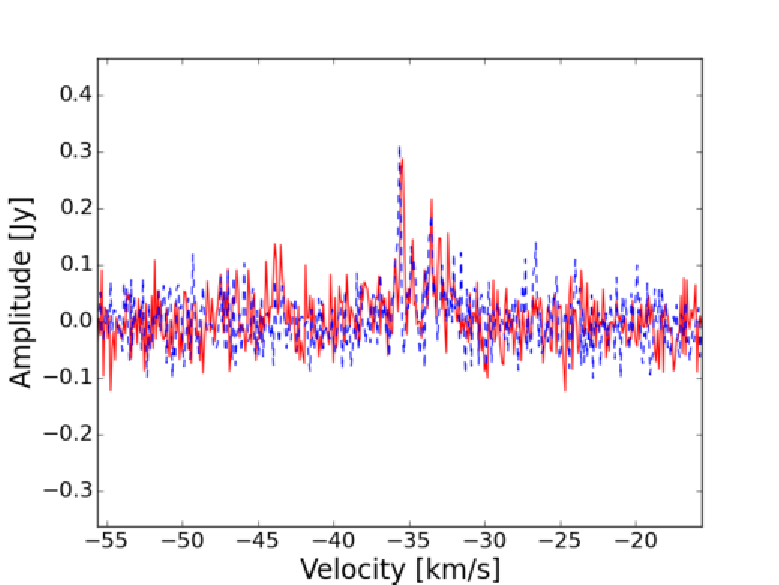}}
	\subfloat[\textit{MMBOH-G350.113+00.095}]{\includegraphics[width=0.33\textwidth]{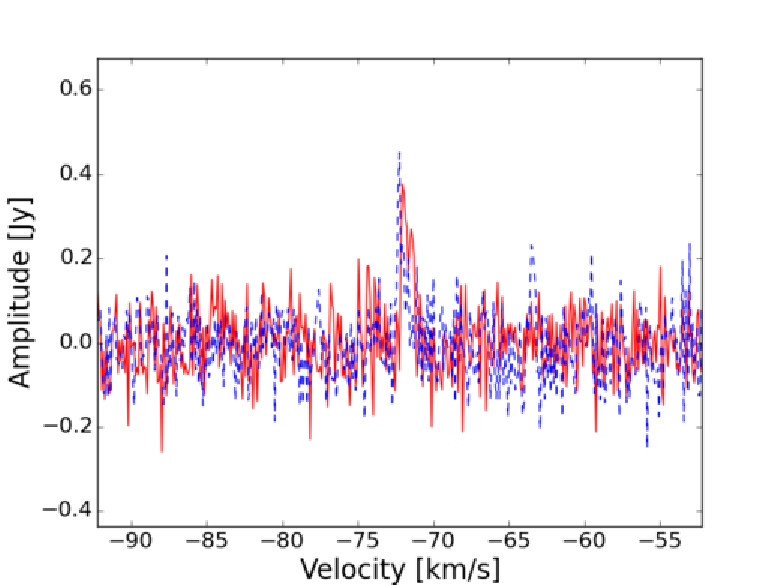}}
	\subfloat[\textit{MMBOH-G350.686-00.491}]{\includegraphics[width=0.33\textwidth]{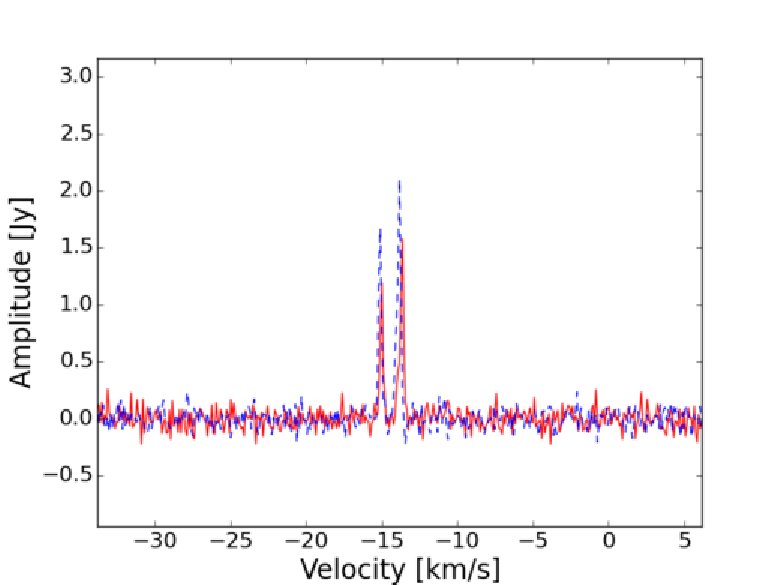}}
\qquad
	\subfloat[\textit{MMBOH-G351.417+00.645}]{\includegraphics[width=0.33\textwidth]{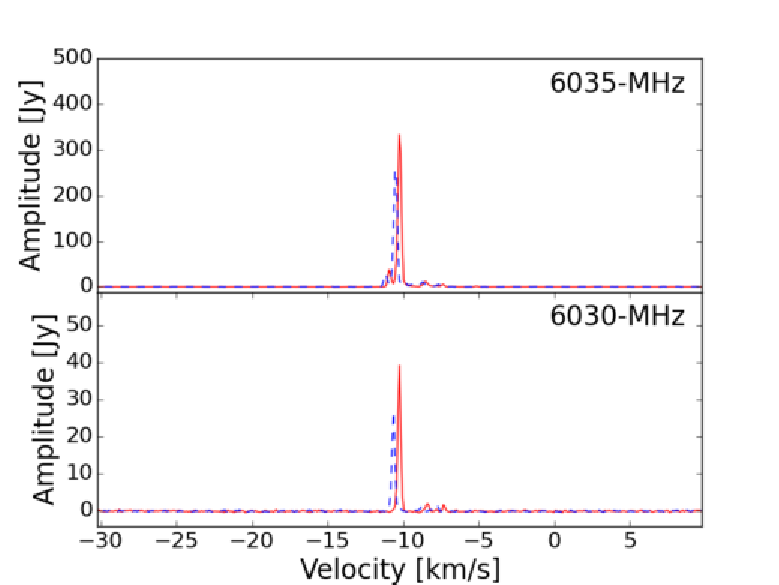}}
	\subfloat[\textit{MMBOH-G351.581-00.353}]{\includegraphics[width=0.33\textwidth]{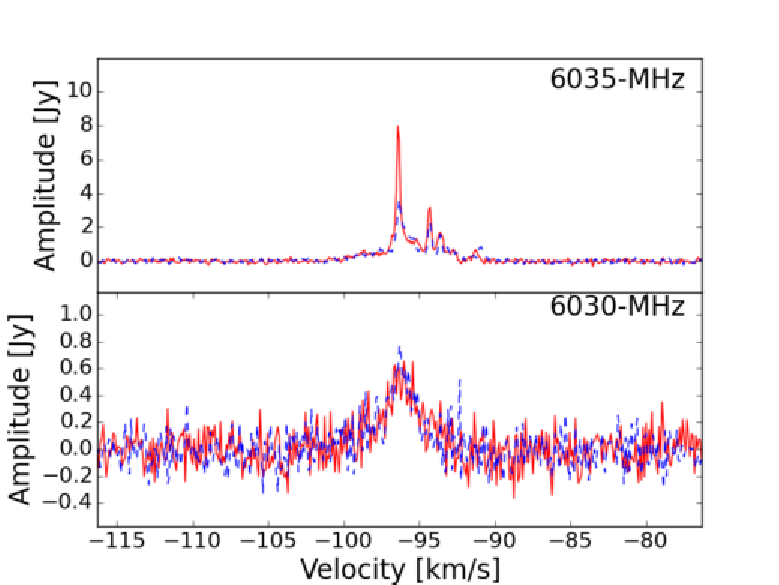}} 
	\subfloat[\textit{MMBOH-G351.775-00.536}]{\includegraphics[width=0.33\textwidth]{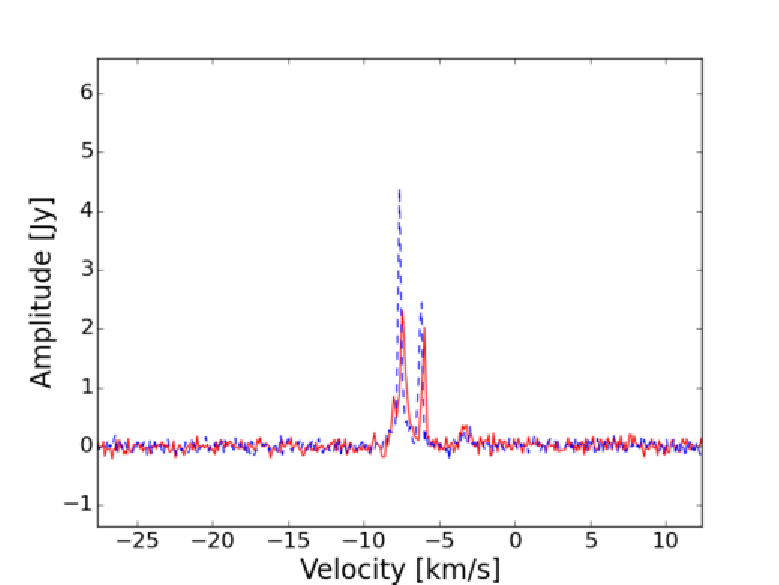}}
	\caption{(continued)}
\end{figure}
\renewcommand\thefigure{11}
\begin{figure}
\ContinuedFloat
\centering
	\subfloat[\textit{MMBOH-G353.410-00.360}]{\includegraphics[width=0.33\textwidth]{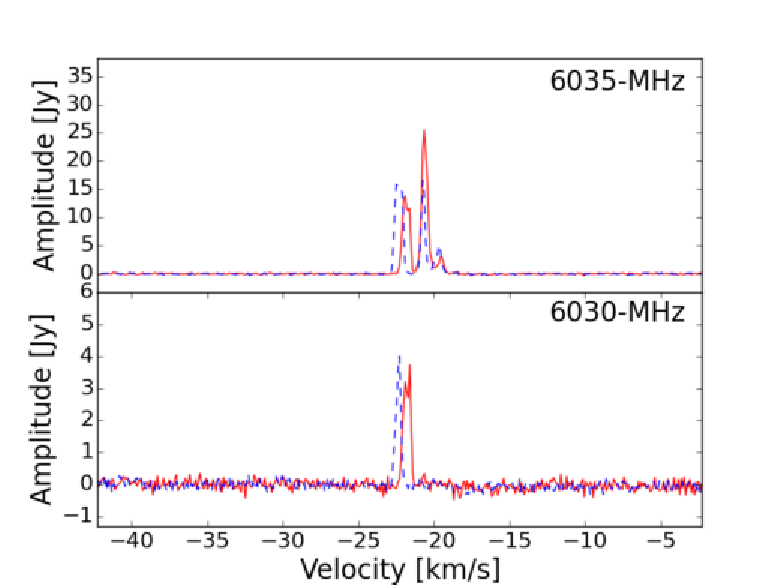}}	
	\subfloat[\textit{MMBOH-G354.725+00.299}]{\includegraphics[width=0.33\textwidth]{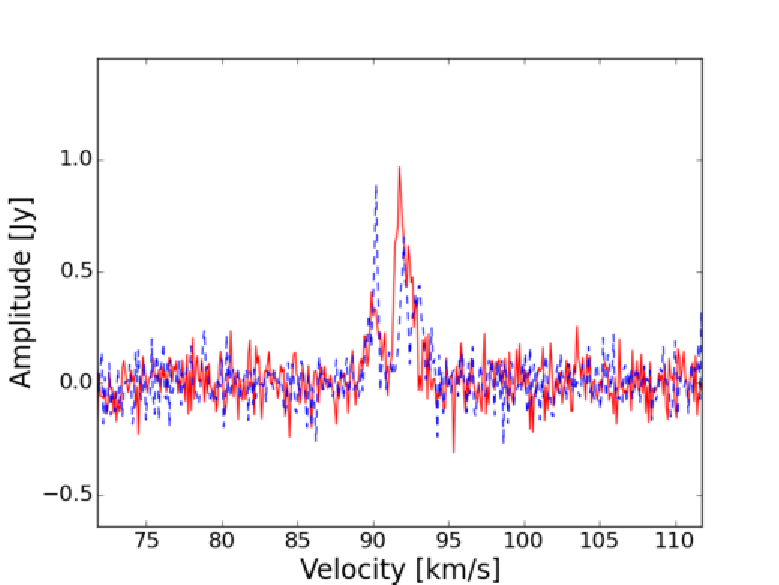}}
	\subfloat[\textit{MMBOH-G355.344+00.147}]{\includegraphics[width=0.33\textwidth]{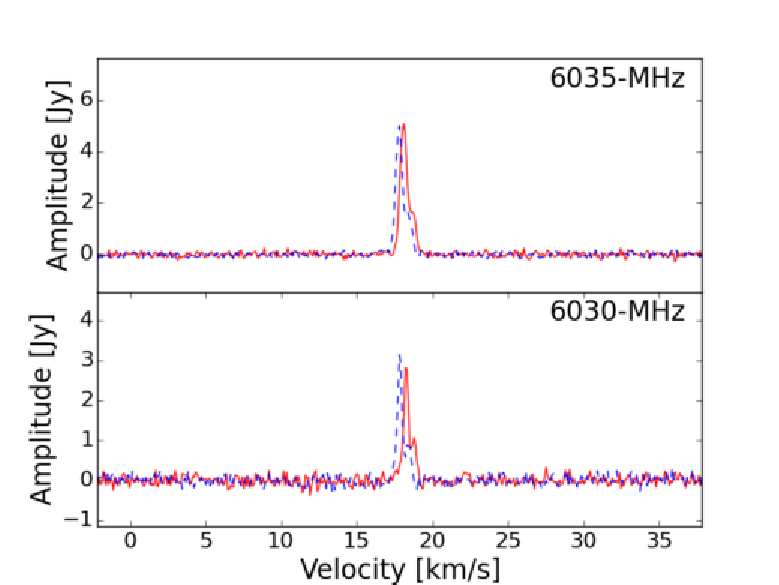}}
\qquad
	\subfloat[\textit{MMBOH-G357.924-00.338}]{\includegraphics[width=0.33\textwidth]{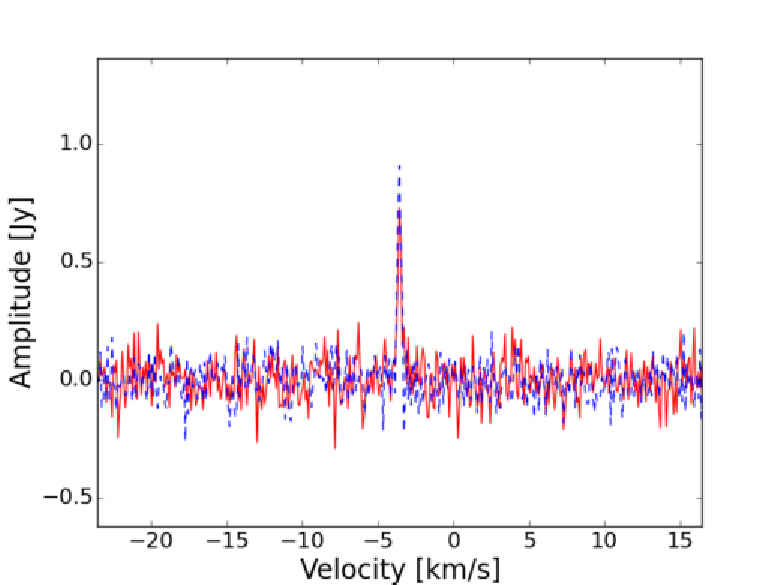}}
	\subfloat[\textit{MMBOH-G359.137+00.031}]{\includegraphics[width=0.33\textwidth]{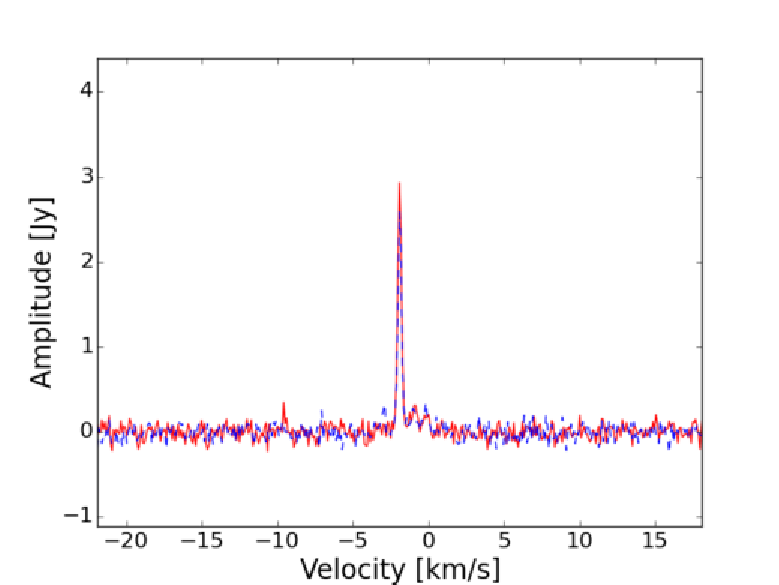}}
	\caption{(continued)}
\end{figure}


\end{document}